\documentclass[10pt,aps,prl,twocolumn,amsfont,amssymb,amsmath,superscriptaddress,floatfix]{revtex4-2}
\usepackage{graphicx}
\usepackage[normalem]{ulem}
\usepackage{mathtools}
\usepackage{dcolumn}
\usepackage{bm}
\usepackage{hhline}
\usepackage[dvipsnames]{xcolor}
\usepackage{physics}
\usepackage{fix-cm}
\usepackage{hyperref}
\usepackage{amsmath}
\hypersetup{
    colorlinks,
    linkcolor={blue!70!black},
    citecolor={blue!70!black},
    urlcolor={blue!70!black}
}
\newcommand{\mvec}[1]{\mathbf{#1}}
\newcommand{\sgrad}{\nabla}

\newcommand{\lap}{\nabla^2}

\newcommand{\del}{\partial}
\newcommand{\mat}[1]{\underline #1}

\newcommand{\pdc}[1]{\textcolor{cyan}{#1}}
\newenvironment{geqn}{\begin{equation}\begin{gathered}}{\end{gathered}\end{equation}}

\usepackage{float}

\begin{document}

    \title{Symmetry-protected phases in a 1D active solid with mechanochemical feedback}	
    \author{Soumyadeep Mondal} 
    \email{msoumyadeep@iisc.ac.in}
    \author{Phanindra Dewan}
    \email{phanindrad@iisc.ac.in}
    \author{Lakshman Santhosh Kumar}
    \email{lakshmans@iisc.ac.in}
    \author{Sumantra Sarkar}
    \email{sumantra@iisc.ac.in}
    \thanks{Everyone contributed equally.}
	\affiliation{Department of Physics, Indian Institute of Science, Bengaluru, Karnataka, India, 560012}
	\date{\today}
	\begin{abstract}

    We present a framework for mechanochemical self-organization in active solids where elasticity is reciprocally coupled to Hopf oscillators. Our model reveals a rich landscape of symmetry-protected phases, identified through amplitude equations and group-theoretic analysis. We uncover a universal transition to compression-driven oscillation death (COD), providing a physical basis for localized signaling dampening in biological tissues that resolves inconsistencies in previous models. Our work demonstrates that complex self-organization in active solids can be classified purely through symmetry arguments.
     \end{abstract} 
\maketitle

Mechanochemical feedback loops (MCFLs) emerge from the coupling between a material’s mechanical deformation and its internal chemical state. While these feedbacks are recognized as fundamental drivers of collective behavior and homeostasis in biological active matter ~\cite{gross2017active,bruckner_tissue_2024,ioratim-uba_mechanochemical_2023,lan2015biomechanical,banerjee2015propagating,sknepnek2023generating,dierkes2014spontaneous,staddon2019mechanosensitive,cavanaugh2020adaptive,etournay2015interplay,noll2017active,saw2017topological,streichan2018global,ibrahimi2023deforming,clement2017viscoelastic,khalilgharibi2019stress,wurthner2023geometry,banerjee2024hydrodynamics,kaouri2017mathematical,sartori2019effect,rautu2024active,dhanuka2025excitability,hannezo_mechanochemical_2019,de2022interplay,lim2014cell}, the physical principles governing the resulting spatiotemporal patterns remain poorly understood. In this Letter, we show that the dynamics of such feedback-driven active solids are dictated by the underlying symmetries of the system. This approach provides a universal framework for classifying mechanochemical patterns, independent of specific biochemical details.

A prototypical manifestation of MCFLs occurs in epithelial tissues~\cite{serra2012mechanical,boocock_theory_2021,perez-verdugo_excitable_2024,parmar2025spontaneous,shankaran_rapid_2009,moreno_competition_2019,hino2020erk,yin_emergence_2024, andrenvsek2023wrinkling,lin2017activation}, where traveling waves of extracellular signal-regulated kinase (ERK) chemistry coupled to actin cytoskeleton mechanics drive collective migration~\cite{serra2012mechanical,boocock_interplay_2023,boocock_theory_2021,hino2020erk}. At low cell densities, ERK exhibits characteristic Hopf oscillations~\cite{shankaran_rapid_2009,kholodenko2010signalling,arkun2018dynamics,muthukrishnan2025glassy}, which are suppressed as compressive stresses increase during tissue crowding or competition~\cite{shankaran_rapid_2009,moreno_competition_2019,muthukrishnan2025glassy}. This transition suggests a mechanically-driven bifurcation. However, two fundamental questions remain unresolved. First, is this suppression an idiosyncratic artifact of ERK’s specific biochemistry, or a generic dynamical phenomenon inherent to all mechanochemically coupled (MCC) Hopf oscillators? And second, whether this dampening originates from Amplitude Death (AD)—the stabilization of a pre-existing steady state—or Compression-driven Oscillation Death (COD)—the creation of a new, spatially localized steady state—remains a point of contention.

Previous frameworks have suggested that the tissues undergo AD driven by increased cell motility~\cite{boocock_interplay_2023}, yet this prediction struggles to reconcile with experimental observations of localized dampening in regions of low motility and persistent oscillations in regions of high motility~\cite{moreno_competition_2019,muthukrishnan2025glassy}. We argue that this discrepancy arises from neglecting the intrinsic Hopf nature of ERK. By explicitly incorporating the physics of coupled Hopf oscillators, we resolve these inconsistencies and uncover a rich landscape of nonequilibrium phases.

\begin{figure*}[htbp]
    \centering
    \includegraphics[width=1\linewidth]{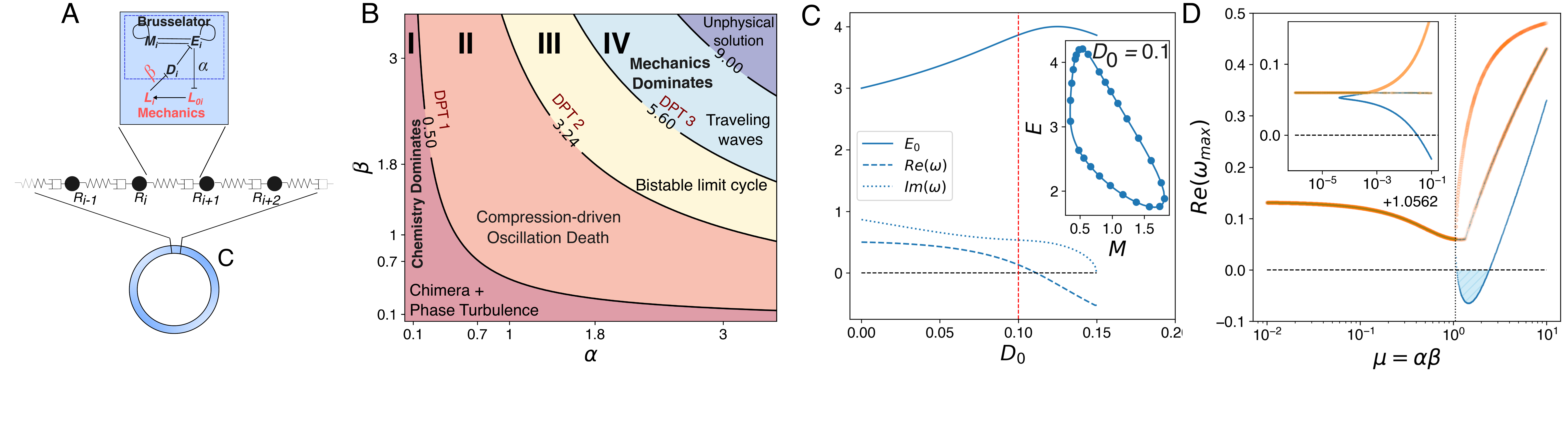}

    \caption{(A) A schematic of the 1D mechanochemically coupled vertex model. (B) A phase diagram of the system shows four distinct phases of patterns. (C)  $E$ and the real and imaginary parts of the eigenvalue with the maximum real part as a function of $D_0$. Inset: $E$ vs $M$ for $D_0 = 0.1$. The blue circles represent the initial conditions used to find the fixed points. (D) The maximum real part of the eigenvalue as a function of $\mu$ ($\tau_D = 1.1, \tau_L = 1.2$). The system undergoes a bifurcation at $\mu_c \approx 1.0566$ (inset D), which splits a single branch into three branches, one of which leads to compression-driven ($L <1$) oscillation death ($\mbox{Re}(\omega_{max}) < 0$, blue shaded area in D). }
    \label{fig:0D}
\end{figure*}

To explore these dynamics, we propose a minimal model: the Harmonic Hopf Solid (HHS). We consider a 1D periodic chain of springs where each node is coupled to a Brusselator—a canonical Hopf oscillator—via an MCFL of strength $\mu$. Biologically, the model describes a 1D vertex model of epithelial tissues, where each spring represents a cell and where a Brusselator governs each cell's chemical state. Our analysis reveals that as $\mu$ increases, the system undergoes a series of dynamical phase transitions (DPTs) that separates the patterns into different classes. At low $\mu$, the system maintains chemistry-dominated patterns, whereas at high $\mu$ the patterns originate principally from the spatial variations of mechanical deformations. Importantly, at intermediate $\mu$, mechanics and chemistry conspire together to admit a phase where mechanical compression drives spatially localized oscillation damping, i.e. COD (Fig.~\ref{fig:0D}B).

Critically, we show that these transitions are not artifacts of specific biochemical kinetics. By replacing the Brusselator with Fitzhugh-Nagumo oscillators and applying group-theoretic analysis of coupled oscillators in $D_n$ symmetric rings~\cite{collins_coupled_1993,collins_group-theoretic_1994,epstein_symmetric_1993}, we establish that these patterns are topologically and symmetrically protected. This symmetry-based approach provides a powerful tool for classifying the ``zoo" of patterns observed in active matter, proving that complex biological morphology can emerge from simple, universal rules of mechanochemical coupling.

\paragraph*{Single cell model:} A single cell, in the absence of any MCC, is described by the following equations:
\begin{geqn}
    \del_t M = a -(b+1)M + cM^2E \\
    \del_t E = bM - cM^2E - DE\\ 
    \tau_D\del_t D = -(D-D_0)
    \label{Eq:Bruss1}
\end{geqn}
Here, $a,b,c$ are the parameters of the canonical Brusselator~\cite{epstein1998introduction,strogatz2024nonlinear} and $E$ and $M$ are the oscillating species, which represent ERK and its upstream regulators, respectively~\cite{kholodenko2010signalling}. Because the level of ERK is strongly correlated with the cellular actin level~\cite{logue2015erk,de2022interplay,muthukrishnan2025glassy},  the concentration of $E$ serves as a proxy for the cellular actin level in our model. $D$ is a degrader of $E$, which we have added based on the experimental observation that ERK is degraded by various regulator proteins~\cite{moreno_competition_2019,shankaran_rapid_2009}. We assume that in the absence of any stimulus, its level is maintained at the homeostatic value $D_0$ with a characteristic recovery time $\tau_D$. The generic criterion for Hopf bifurcation in the presence of $D$ is complicated and not useful. However, some specific cases are of importance. Without $D$, this system undergoes a Hopf bifurcation to an oscillatory state when $b > 1 + a^2c$~\cite{epstein1998introduction}. Because ERK oscillation is widely observed in isolated cells~\cite{muthukrishnan2025glassy}, we choose $b = 3$, and $a = c = 1$ to ensure an oscillatory steady state. In the presence of $D$, when $\tau_D \rightarrow 0$, the Hopf bifurcation happens at $D_0 = D_c \approx 0.111422$ (Fig.~\ref{fig:0D}C). Here, we use $D_0 = 0.1$ for all our simulations to ensure that a single cell stays close to the bifurcation point~\cite{mora2011biological,munoz2018colloquium}. The corresponding limit cycle is shown in Fig.~\ref{fig:0D}C.  

\paragraph*{Mechanochemical coupling:} ERK activates cellular pathways that upregulates actomyosin level in the cell~\cite{logue2015erk,lavoie2020erk,hino2020erk,mendoza2011erk,mendoza2015erk,tanimura2017erk,barros2005activation}. Thus, increased ERK level (level of $E$ in the model) leads to cell contractions, which decreases the cell volume \cite{boocock_theory_2021,boocock_interplay_2023}. In our model, the decreased cell volume generates mechanical stress that upregulates $D$, which, in turn, downregulates $E$, leading to cellular expansion. This phenomenon can be captured by modifying Eqn.~\ref{Eq:Bruss1} as follows.
\begin{geqn}
\tau_D\del_t D = -(D-D_0) - \beta D(L-1)\\
\tau_L \del_t L_0 = -(L_0-1) - \alpha (E-E_0)\\
\tau_R \del_t L = -(L-L_0) + f(L,t)
\label{Eq:MCFL}
\end{geqn}
Here $L$ is the instantaneous length of the cell and $L_0$ is the homeostatic (rest) length of the cell, whereas $E_0$ is the $E$ concentration at the fixed point of Eqn.~\ref{Eq:Bruss1}. $\alpha$ determines how rapidly changes in $E$ modifies $L_0$ and $\beta$ determines how the instantaneous volume, $L$, changes the level of $D$. Together, $\mu = \alpha\beta$ determines the strength of the mechanochemical coupling \cite{boocock_theory_2021}. $f(L,t)$ is a forcing term that can arise due to external driving or interaction between cells in a network. For example, in the continuum limit, the equations can be written as follows~\cite{boocock_theory_2021} (\cite{SI}-III). \nocite{Ipsen1999, synchronizationArkady2001, JanakiRangarajan}   
\begin{geqn}
    \del_t M = a -(b+1)M + cM^2E \\
    \del_t E = bM - cM^2E - DE\\ 
    \tau_D\del_t \delta = -(\delta + \delta^3) - \beta D(\del_xR - 1)\\
    \tau_L \del_t l = -(l + l^3) - \alpha(E-E_0) \\
    \tau_R \del_t R = \del^2_xR - \del_xL_0
    \label{Eq:contvertex1D}
\end{geqn}
Here $R(x)$ is the location of the vertex and $\del_xR$ is the continuum equivalent of the cell length, $L$. $\tau_D$, $\tau_L$ are relaxation timescales, and $\tau_R = 1$ is the frictional timescale. Moreover, $\delta = D - D_0$ and $l = L_0 - 1 $. 
 
In the rest of the paper, we assume that there was no MCC prior to $t = 0$. Hence, we choose initial conditions from the uncoupled Brusselator limit cycle (Fig.~\ref{fig:0D}C). MCC is turned on at $t = 0$, resulting in changes in the stability of the uncoupled fixed point $\mvec{u_0} = \{M_0,E_0,D_0,1,1\}$, as the phase space becomes $5$-dimensional. Eqn.~\ref{Eq:contvertex1D} is solved using a central spatial discretization and a fourth-order Runge–Kutta time integration scheme under periodic boundary conditions. 

\begin{figure}[htbp]
    \centering
    \includegraphics[width=\linewidth]{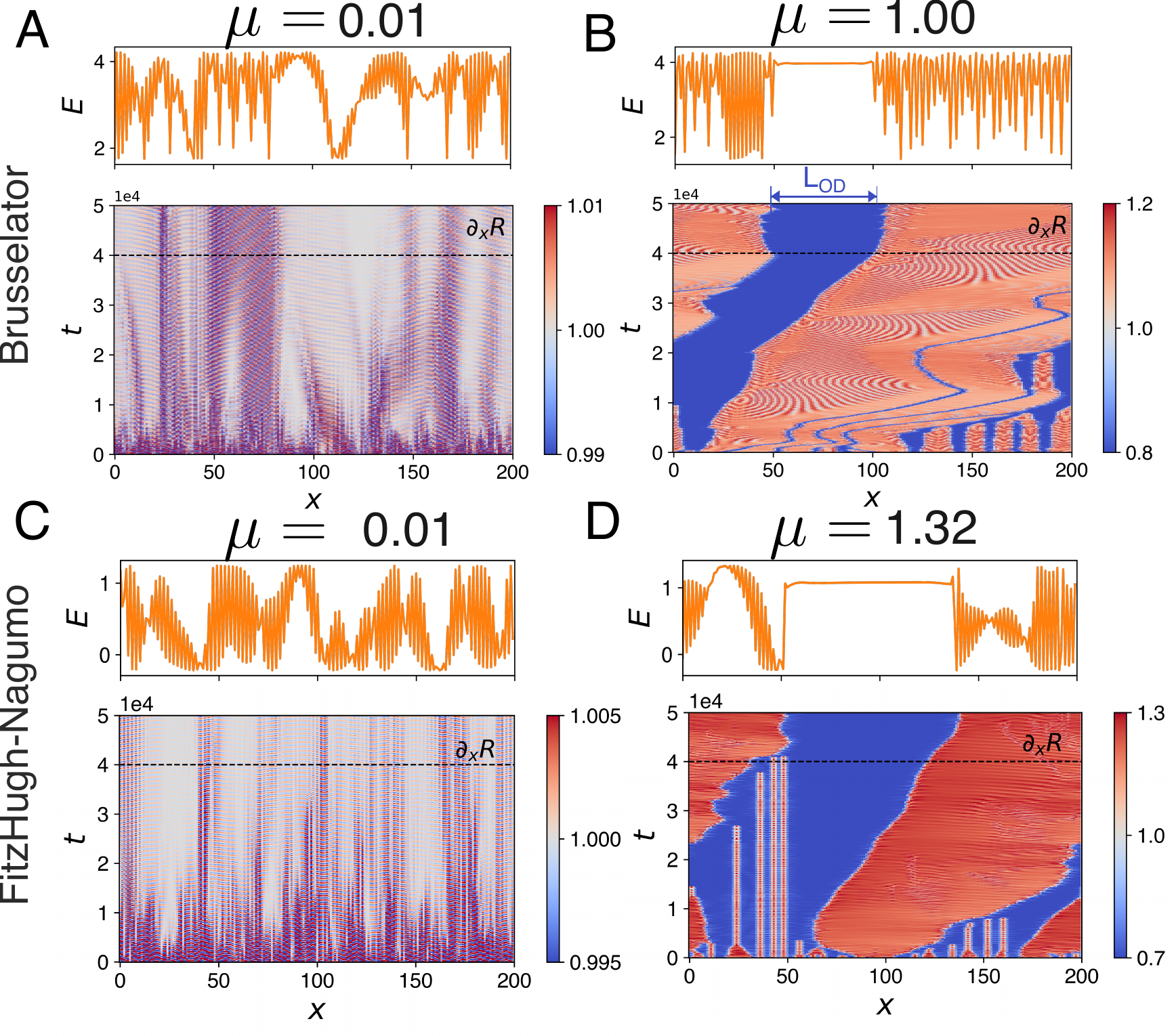}
    \caption{\textbf{COD:} Kymographs of $\partial_x R$ reveal a transition from spatially incoherent phase turbulence state to a state with compression-driven oscillation death as $\mu$ is increased. Qualitatively identical results are observed for the (A,B) Brusselator and (C,D) FitzHugh--Nagumo oscillator, which demonstrate the generality of the transition.}
    \label{fig: generality of patterns}
\end{figure}

\paragraph*{Compression-driven oscillation death:} The MCFL-coupled single cell undergoes multiple bifurcations as $\mu$ is varied (Fig.~\ref{fig:0D}D). Below a $\tau_{D,L}$-independent threshold $\mu_c \approx 1.0566$, the cell displays canonical limit-cycle oscillations of Brusselators. At $\mu_c$, the determinant of the Jacobian matrix becomes zero (\cite{SI}-II.B), and a fixed point suddenly appears, which splits into two saddle points as $\mu$ is increased (Fig.~\ref{fig:0D}D inset); a situation analogous to saddle-node bifurcation~\cite{strogatz2024nonlinear}. Closer inspection reveals that one of them favors compression ($L < 1$) and the other extension ($L \geq 1$, Fig~\ref{fig:0D}D). Furthermore, the fixed point corresponding to the compression branch becomes linearly stable for a range of $\mu$ values, which kills any oscillation in the system. We find that this transition from an oscillatory to nonoscillatory state through compression is a generic feature of the single cell and does not depend on $\tau_D$ and $\tau_L$ (\cite{SI},Fig.~S20). Coupling the single cells in a 1D periodic network reveals a far richer phenomenology. At low $\mu$, the cells in the network display phase turbulence such that phase coherence is lost quite rapidly (Fig.~\ref{fig: generality of patterns}A). However, pockets of synchronous oscillations exist, which point to the existence of chimera states in this limit (\cite{SI},Fig.~S21-23). For $\mu > 0.5$, the network phase separates into a mechanically compressed and an extended phase, signifying a dynamical phase transition at $\mu \approx 0.5$. Interestingly, oscillation is dampened in the compressed region, whereas phase turbulence~\cite{kuramoto1984chemical}(\cite{SI}-V) persists in the extended region, signifying the onset of COD through mechanochemical feedback in the spatially extended system (Fig.~\ref{fig: generality of patterns}B).

\paragraph*{Generality of the transition:} The existence of COD in the HHS with Brusselator is reassuring, but the generality of the result remains unclear. Fortunately, for coupled Hopf oscillators in a 1D periodic ring, we can use the so-called equivariant Hopf bifurcation theorems~\cite{collins_coupled_1993,collins_group-theoretic_1994,epstein_symmetric_1993} to establish the generality. These powerful theorems state that the symmetry group of the oscillator network dictates the possible patterns and, crucially, the specific details of the Hopf oscillators \textit{does not matter}, implying that the transitions predicted using the Brusselator should hold universally across all Hopf oscillators as long as the symmetry of the interaction and the network topology is preserved! To test this prediction, we replaced Brusselators with Fitzhugh-Nagumo oscillators~\cite{Sherwood2013} in the HHS, which recovered exactly the same phenomenology as $\mu$ was varied (Fig.~\ref{fig: generality of patterns}C-D). Crucially, these results prove that complex biological morphologies and patterns can arise purely from the symmetries of the underlying interactions and does not require specific biochemical details. Hence, the results that we derive here for Brusselators should apply to more complex Hopf oscillators, such as ERK. 

\paragraph*{Origin of the patterns:} To understand the origin of the patterns, we perform a linear stability analysis (LSA) of Eqn.~\ref{Eq:contvertex1D} around the fixed point $\mvec{u_0}$ (End Matter). However, LSA fails to account for the observed spatiotemporal structures. For example, at weak coupling ($\alpha=\beta=0.1$, \cite{SI},Fig.~S7), the unstable modes are independent of the wave number $q$, which would suggest the absence of spatial pattern formation. In contrast, numerical simulations clearly reveal nontrivial spatiotemporal dynamics (Fig.~\ref{fig: generality of patterns}). This apparent failure of LSA arises from the non-Hermitian nature of the linear stability matrix, which makes the eigenvalue spectrum alone insufficient to predict the dynamics. Closer inspection reveals that coupling between mechanical and chemical components generates a $\mathcal{PT}$-symmetry–broken regime bounded by exceptional points, whose structure depends on the relative timescales $\tau_D$ and $\tau_L$ (End Matter). We suspect that there is a deep connection between the generated patterns and the spontaneously broken $\mathcal{PT}$-symmetry, which we will explore in a future work. Presently, the failure of the LSA to predict the patterns warrants a nonlinear analysis of the equations. Specifically, we observe that the mechanically compressed regions are also regions with elevated levels of $D$, which suggests that the transition to the COD state can be explained by analyzing the behavior of the system near the Hopf-fold bifurcation point $D_c$ (Fig.~\ref{fig:0D}C). 

\paragraph*{Amplitude equations:} 
Near $D_c$, the linearized dynamics includes two purely imaginary and one zero eigenvalues (\cite{SI},Fig.~S9). These are the slow modes of the system which comprise the \textit{center subspace}~\cite{Ipsen2000,PhysRevE.66.026102,Carr1981}. To understand the collective behaviors, we performed a center manifold reduction by projecting Eq.~\ref{Eq:contvertex1D} on the center subspace. Retaining the leading nonlinearities in $\gamma = D_0 - D_c$ yields two coupled amplitude equations for a complex mode $w$ arising from the Hopf bifurcation and a real mode $z$ that arises from the conservation of the total length $L$ in a periodic network and describes the local mechanical strain (\cite{SI}-IV).

\begin{geqn}
\partial_t w =d_w \lap w + a_0\gamma w + a_1 w z + a_2 w z^2 + a_3 w z^3 + a_4 |w|^2 w \\
\partial_t z = d_z \lap z + d_{ww}\lap |w|^2
\end{geqn}
The $w$ mode has local and non-linear saturation term, while $z$ evolves diffusively, implying a separation of timescales. Adiabatic elimination of $w$ gives
\begin{equation}
\partial_t z = \lap \left[ b_1 \gamma + b_2 z + b_3 z^2 + b_4 z^3 \right],
\end{equation}
which describes a conserved gradient flow $\mathbf{J} = -\sgrad \frac{\delta \mathcal{L}}{\delta z}$ with Lyapunov functional,
\begin{equation}
\mathcal{L}[z] = \int l(z)dx=\int dx \left[b_1 \gamma z + \frac{b_2}{2} z^2 + \frac{b_3}{3} z^3 + \frac{b_4}{4} z^4\right]
\end{equation}
The detailed discussion of the parameters are given in \cite{SI}.

\begin{figure}
    \centering
    \includegraphics[width=\columnwidth]{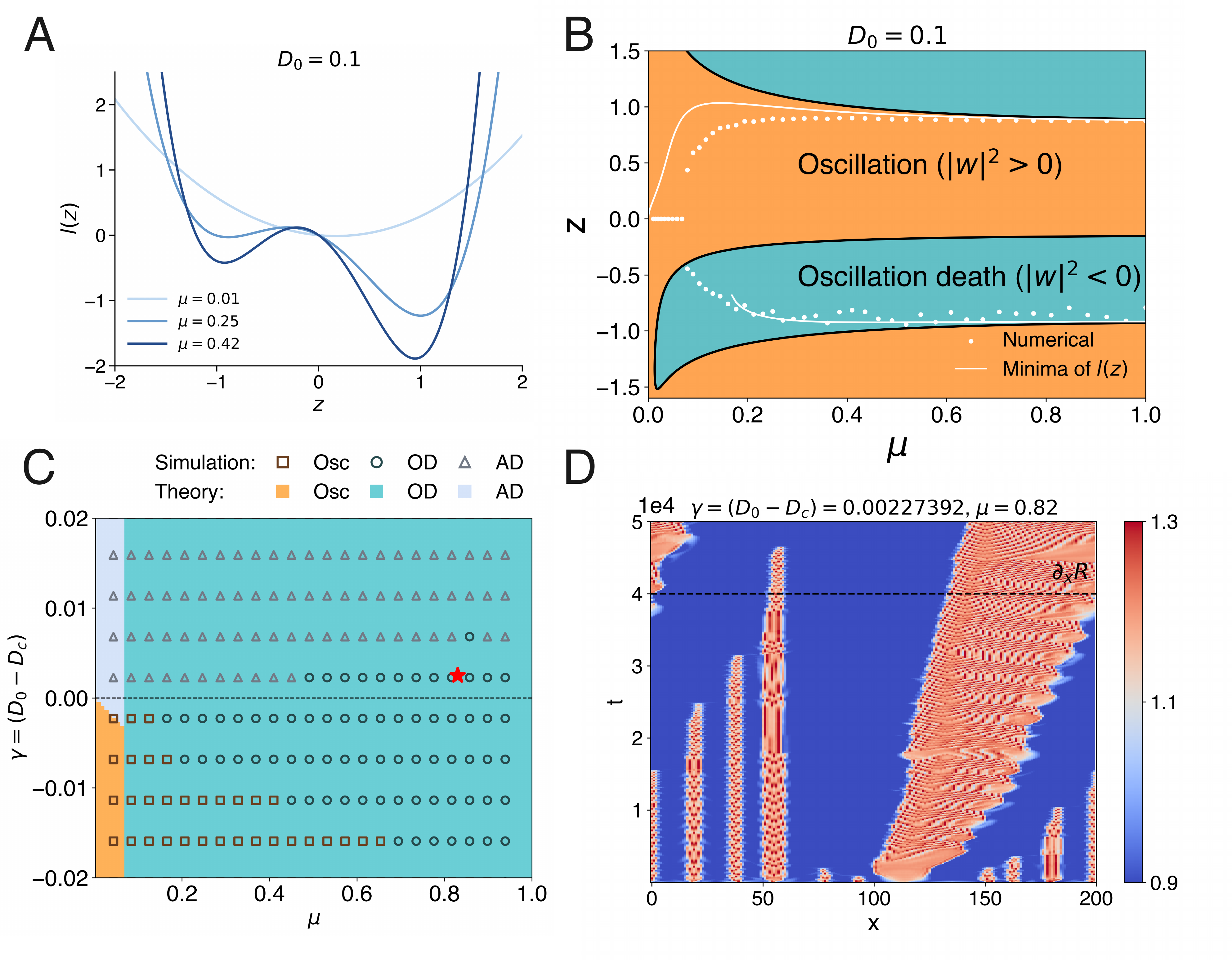}
    \caption{
(A) The effective Lyapunov density $l(z)$ plotted as a function of $z$, derived from the amplitude 
equation for $z$. (B) Two distinct regions are identified: an oscillatory regime ($|w|^2>0$) and an 
oscillation death regime ($|w|^2<0$). Numerical solutions (scatter points) of the $z$ equation reveal 
that the $z=0$ state bifurcates into two branches, one of which enters the oscillation 
death region. The solid line represents the analytical minima of $l(z)$. 
(C) Comparison between the analytical prediction (heatmap) obtained from the amplitude equation 
and the corresponding simulation results (markers). (D) A kymograph illustrating the presence 
of an oscillation death region for small positive values of $\gamma$ and $\mu$.}

    \label{fig:center_manifold_reduction}
\end{figure}

The landscape of $l(z)$  governs stability and phase structure (Fig.~\ref{fig:center_manifold_reduction}(A)). The homogeneous state $z=0$ bifurcates at finite $\mu$ into two branches (Fig.~\ref{fig:center_manifold_reduction}(B)), corresponding to opposite signs of $|w|^2$. 

Branches with $|w|^2>0$ yield sustained oscillations, whereas $|w|^2<0$ admits no oscillatory solution, resulting in oscillation death. In fact, the $|w|^2<0$ branch occurs where $z < 0$, i.e. in regions with compression, revealing the origin of COD in our system.  The phase diagram obtained from the amplitude equations qualitatively captures the boundaries between the phases as observed in simulations (Fig.~\ref{fig:center_manifold_reduction}(C)). Surprisingly, the equations predict a phase that is the counterpoint of COD. For $D_0 > D_c$ and low $\mu$, the system admits only nonoscillating solutions, i.e. AD. However, as $\mu$ is increased, $z$ separates into positive and negative branches and admits a spatially localized oscillating solution, which we indeed find from simulations, albeit at a $\mu$ value much higher than predicted (Fig.~\ref{fig:center_manifold_reduction}(D)). This is expected, as we neglect higher order nonlinearities in the amplitude equations, which can qualitatively change the phase boundaries. The equations provide another surprising insight: although the chemicals themselves are restricted to each individual cell, the fact that the cells are mechanically coupled to each other allows for an effective diffusive transport. This observation provides a justification for using reaction-diffusion equations to model morphogenesis or tissue pattern formation. \cite{bailles2022mechanochemical,turing1952chemical,roshan2023multiscale}. We postulate that this effective diffusion may allow faster information transmission than real space diffusion of chemicals, a potential biological design principle that requires further verification.  

\begin{figure}
    \centering
    \includegraphics[width=\linewidth]{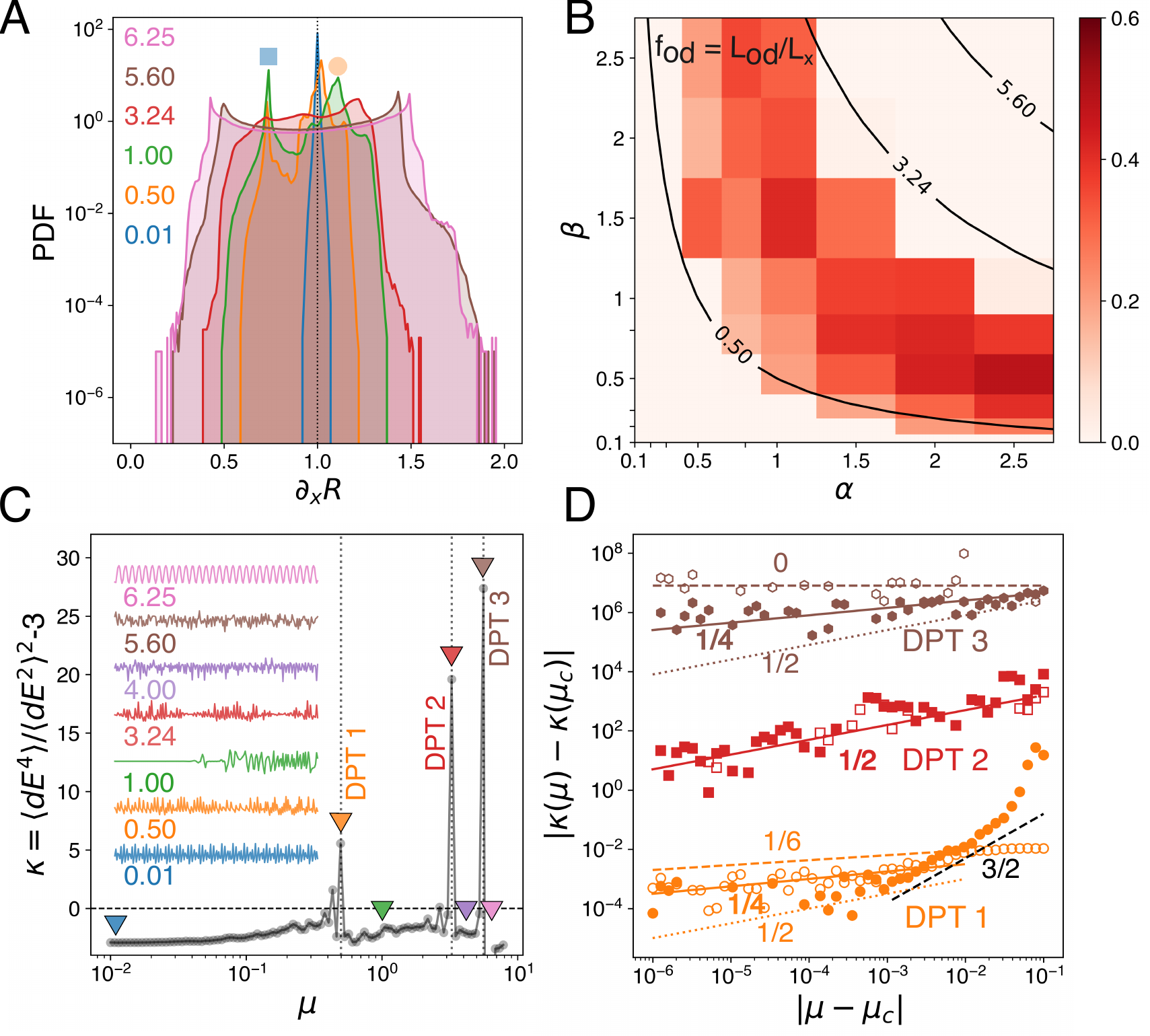}
    \caption{\textbf{Dynamic phase transitions and phase diagram:} (A) Probability density of springs' length, $\partial_xR$, transitions from unimodal to bimodal distributions as $\mu$ (legend) is increased. (B) Normalized total size of COD, $f_{OD} = L_{OD}/L_x$ (see Fig.~\ref{fig: generality of patterns}B) for different $\alpha$, $\beta$. The contours show values of $\mu$ that mark boundaries between different regions. (C) At these $\mu$, state variables, e.g. $E$, show intermittent fluctuations, captured by the excess kurtosis, $\kappa$. $\kappa > 0$ near the boundaries, suggesting fat-tailed distributions and dynamic phase transitions (DPT). Inset: time-trace of $dE = E(t+\tau_{lag}) - E(t)$ for different $\mu$ (legend). (D) $|\kappa(\mu) - \kappa(\mu_c)|$ scales with $|\mu - \mu_c|$, where $\mu_c = 0.5, 3.24, 5.60$ are the locations of the DPTs. The estimated power law exponents are shown as rational fractions with the best fit in bold. Open and closed symbols: $\mu < \mu_c$ and $>\mu_c$. Data is shifted along y-axis for clarity. }
   
    \label{fig:distro}
\end{figure}

\paragraph*{Dynamic phase transitions:} The amplitude equations fail to capture the dynamics beyond $\mu = 1$ due to the truncation of the nonlinear modes. Therefore, we probed this strongly nonlinear regime through numerical solutions of Eq.~\ref{Eq:contvertex1D}. Consistent with the predictions of the amplitude equations, the $\partial_xR$ distribution at $\tau_D = 1.1$ and $\tau_L = 1.2$ (for other $\tau_{D,L}$ values see \cite{SI},Fig.~S26-27) remains bimodal above $\mu \approx 0.5$ (Fig.~\ref{fig:distro}A), the location of DPT-1 (Fig.~\ref{fig:0D}B). In fact, we observed that the COD regions and the oscillatory regions phase separate for $0.5\leq \mu \leq 3.24$, such that the COD regions remain in one coherent cluster. Total size of these clusters, $L_{OD}$, is comparable to the system size, $L_x$, for a range of parameters (Fig.~\ref{fig:distro}B). Importantly, the disapperance of these clusters at $\mu \approx 3.24$ marks the onset of another phase transition. Indeed, we observe that at these transition points, state variables, such as $E$ shows intermittent fluctuations, which can be quantified through excess kurtosis $\kappa$ (Fig.~\ref{fig:distro}C). $\kappa$ becomes positive at $\mu_c \approx 0.5, 3.24,$ and $5.60$, where $E$ shows highly intermittent fluctuations. In contrast, for in-between $\mu$ values, $\kappa$ remains negative, suggesting that the fluctuations are highly suppressed. In fact, we find that $\kappa$ diverges as a power law as $\mu \rightarrow \mu_c$ from both directions (Fig.~\ref{fig:distro}D), signifying a dynamic phase transition (DPT) at each $\mu_c$~\cite{international_centre_for_theoretical_sciences_dynamical_2019,nyawo2017minimal}. The origin of intermittency is different at different DPTs. At $\mu_c = 0.5$, the available phase space expands and the limit cycle becomes diffused. At $\mu_c = 3.24$, A new limit cycle, corresponding to the mechanical traveling waves, appears inside the Brusselator limit cycle, leading to bistability in oscillation, akin to canard explosion~\cite{wechselberger2007canards}. Finally, at $\mu = 5.60$, the Brusselator limit cycle disappears and only the mechanical limit cycle remains (\cite{SI} movies).

\paragraph*{Phase diagram:} From these observations, we propose a phase diagram of the HHS (Fig.~\ref{fig:0D}B). It has four realizable phases. In phase I ($\mu < 0.5$), chemistry dominates over mechanics, and the collective behavior of HHS is governed mainly by the Brusselator's chemistry. We observe chimera states~\cite{kuramoto2002coexistence,abrams2004chimera,martens_chimera_2013,das2024order}, and phase turbulence. In phase IV, mechanics dominates, and we observe spatially-extended traveling waves, as predicted before~\cite{boocock_theory_2021,boocock_interplay_2023}. The Hopf nature of ERK oscillation becomes important in phases II and III, where patterns originate from the interplay of mechanics and chemistry. Specifically, in III, we see bistable limit cycle oscillations. Finally, in II, we observe spatiotemporal chaos, and most importantly, phase separation of the system into COD and phase turbulent phase (\cite{SI} movies). The traveling waves become evanescent at the boundaries of COD. The location of these phases varies with $\tau_D$ and $\tau_L$, but the qualitative features remain invariant (\cite{SI},Fig.~S28-29).

\paragraph{Discussion:}  In this Letter, we have resolved a fundamental ambiguity in the collective dynamics of active epithelial tissues: that mechanochemical feedback triggers COD.

By investigating a 1D periodic chain of coupled Hopf oscillators, we demonstrate that these patterns are not idiosyncratic to epithelial signaling. Instead, they emerge from the universal properties of mechanochemically coupled systems near a Hopf bifurcation, fundamentally constrained by the symmetry and topology of the interaction network. Crucially, our results suggest that the phase architecture predicted by our minimal model constitutes a universality class that governs complex biological networks regardless of specific biochemical details~\cite{guru2022microtubule,moreno_competition_2019,muthukrishnan2025glassy}. This robustness, which we have verified in a companion study~\cite{muthukrishnan2025glassy}, implies that the underlying dynamical transitions are protected against local microscopic fluctuations (see End Matter) and the topology and the dimensionality of the system~\cite{muthukrishnan2025glassy}. Beyond epithelial monolayers~\cite{parmar2025spontaneous,boocock_theory_2021,ioratim-uba_mechanochemical_2023,banerjee2015propagating,perez-verdugo_excitable_2024,boocock_interplay_2023,moreno_competition_2019,shankaran_rapid_2009,muthukrishnan2025glassy}, our framework provides a predictive lens for exploring spatiotemporal self-organization across a broad spectrum of synthetic active matter, from chemically active droplets to adaptive metamaterials~\cite{roshan2023multiscale, tompkins2014testing,kumar2024emergent, zhai2004amplitude, baconnier2022selective,arora2024shape}. 

The search for universal principles in biological pattern formation remains a cornerstone of nonequilibrium physics~\cite{bailles2022mechanochemical,turing1952chemical, cross1993pattern}. Our findings establish that the ``zoo" of biological patterns can be systematically classified into universal classes through the underlying symmetry and the topology of the active solids. 

\begin{acknowledgments}
The authors would like to thank Aayush Dasgupta, Navneet Roshan, Tanmoy Das, Manish Jain, Medhavi Vishwakarma, Sindhu Muthukrishnan, Subroto Mukerjee, Nairita Pal, and Sriram Ramaswamy for helpful discussions and Sumilan Banerjee and Anindya Das for critically reading the manuscripts. The authors gratefully acknowledge Medhavi Vishwakarma and Sindhu Muthukrishnan for providing access to their experimental data. SS would like to thank Axis Bank Centre for Mathematics and Computing, SERB-DST (SERB-22-0223), MOE-STARS (2023-0850), ANRF (ANRF/ARG/2025/003123/PS) and IISc for financial support. PD thanks IISc, and SM acknowledges support from the PMRF (0203001) for Ph.D. fellowships.
\end{acknowledgments}
\newpage
\section*{End Matter}
\paragraph{Complexity of the model (why five variables?):}
Our model solves the coupled dynamics of five variables: $M,E,D,L_0,R$. $L_0$ and $R$ are the parameters of the spring and cannot be removed. We require a feedback loop between mechanics and chemistry. Hence, we need $E$ and $D$. We could have gotten a feedback loop with just $E$, but it would have given us amplitude death, not oscillation death~\cite{boocock_theory_2021}. Moreover, experimental evidence~\cite{moreno_competition_2019} shows that degraders of ERK are present in compressed cells. Hence, $E$ and $D$ are essential. $M$ seems to be the only variable that we could spare. However, $E$-oscillation then has to be put in by hand. Hence, our model, with five variables, is the minimal model that captures an essential biological fact that ERK and proteins like it are Hopf oscillators that undergo bifurcation through the actions of degrader molecules. Ignoring this fact simplifies the model~\cite{boocock_theory_2021,boocock_interplay_2023}, but then it fails to capture collective oscillation death and associated patterns that we observe in experimental systems~\cite{moreno_competition_2019,muthukrishnan2025glassy,guru2022microtubule}.

\paragraph{Values of $\tau_L$ and $\tau_D$:}
In our model, $\tau_R$ sets the unit of time and $\tau_L$ and $\tau_D$ are free parameters. We expect them to vary from system to system. A direct comparison with experimental system would require access to 1D epithelial tissue experiments, which we unfortunately do not have. However, we can get a sense of the numbers by comparing a 2D version of this model with experiments on MDCK cells~\cite{muthukrishnan2025glassy}. There, $\tau_R \approx 2$ min, and $\tau_D \approx \tau_L \in [100,240]$ min can capture the observed dynamics well. In contrast, for \cite{guru2022microtubule}, we expect $\tau_R \approx 0.2$ min and $\tau_L$ and $\tau_D$ to vary in a few minutes timescale.

\paragraph{Linear Stability Analysis:} 
To understand the system we at first do the linear stability analysis of the Eqn.~\ref{Eq:contvertex1D} around the steady state values of $[M_0, E_0, D_0, L_{0o},R_{0}]$. We take the ansatz, $y(t,x) = \tilde{y}(\omega,q) exp(\omega t+iqx)$ and put this ansatz to Eqn.~\ref{Eq:contvertex1D} which gives us the eigenvalue equation, $\omega \tilde{\mathbf{u}} = \mat{A_q} \mathbf{\Tilde{u}} $
where, $\mathbf{\Tilde{u}} = [\Tilde{M},\Tilde{E},\Tilde{D},\Tilde{L_0},\Tilde{R}]^T$, and
\begin{equation}
\scriptsize
\setlength{\arraycolsep}{4pt}
\mat{A_q}
=
\begin{bmatrix}
-1-b+c_1 & c_2 & 0 & 0 & 0 \\
b-c_1 & -(c_2+D_0) & -E_0 & 0 & 0 \\
0 & 0 & -\frac{1}{\tau_D} & 0 &
-\frac{i\beta D_0 q}{\tau_D} \\
0 & -\frac{\alpha}{\tau_L} & 0 &
-\frac{1}{\tau_L} & 0 \\
0 & 0 & 0 &
-\frac{iq}{\tau_R} &
-\frac{q^2}{\tau_R}
\end{bmatrix}
\label{Eq:SIAq}
\end{equation}

\begin{figure}[!htbp]
    \centering
    \includegraphics[width=\linewidth]{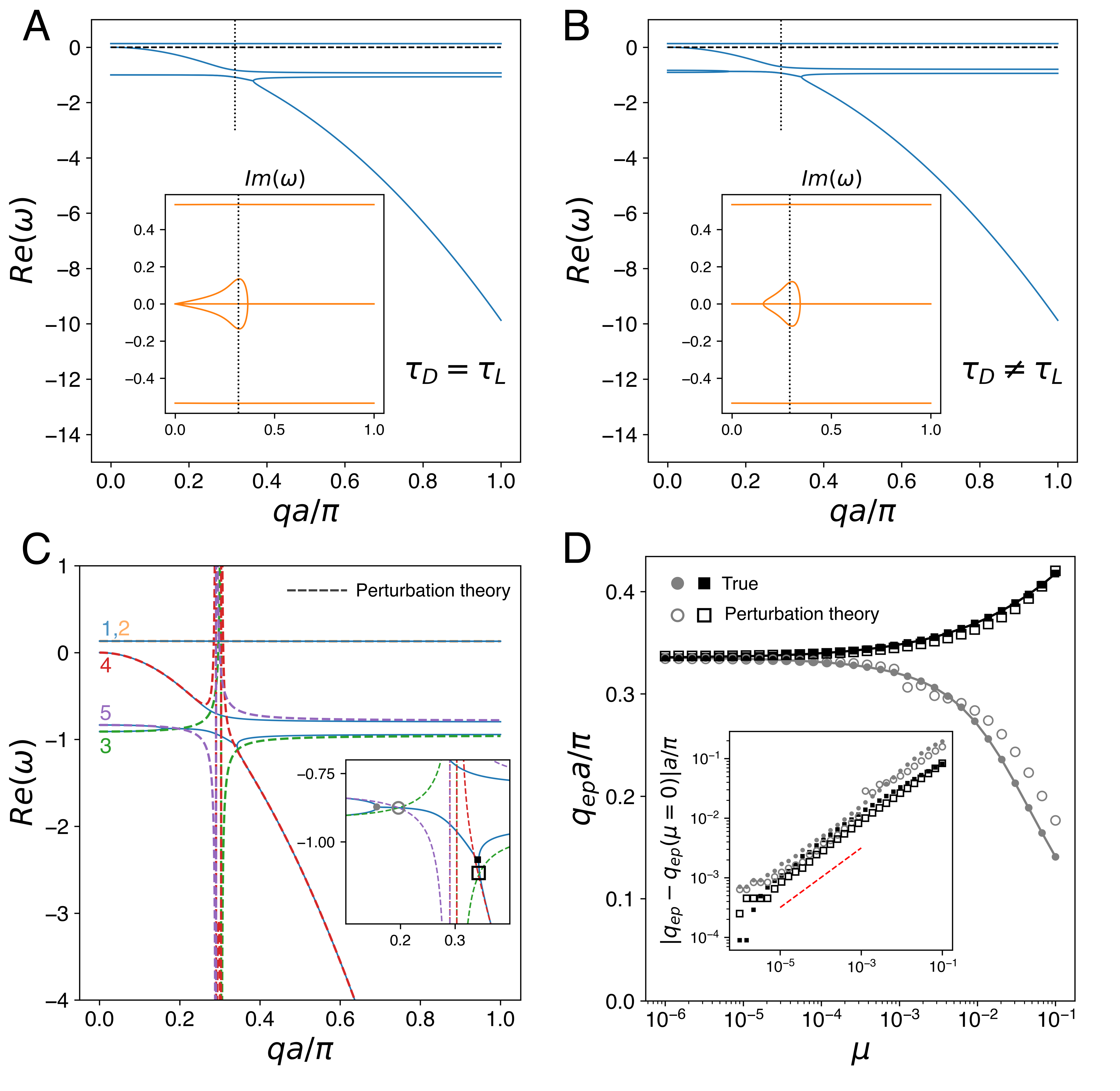}
    \caption{Eigenvalue spectra for $\alpha = \beta = 0.1$ and (A) $\tau_D = 1.1, \tau_L = 1.2$ and (B) $\tau_D = \tau_L = 1$, show a region of broken $\mathcal{PT}$-symmetry, the boundaries of which end at two exceptional points $q^{\pm}_{ep}$. $q^{\pm}_{ep} \neq 0$ for A, but $q^{-}_{ep} = 0$ for B. (C) Eigenvalues (numbered in the legend) obtained from second-order nondegenerate perturbation theory (dashed lines) match well with the spectra in A for $q$ values away from the $\mathcal{PT}$-symmetry broken region. $q^{\pm}_{ep}$ can be estimated from the failure of the perturbation theory (inset). (D) The estimated $q^{\pm}_{ep}$ matches well with their true values. Inset: $|\Delta q_{ep}| = |q_{ep} - q_{ep}(\mu = 0)|$ grows as $\mu^{1/2}$ (red dashed line) for $q_{ep}^+$ and $\mu^{3/4}$ (blue dotted line) for $q_{ep}^-$.}
    \label{fig:spectra}
  
\end{figure}

\paragraph{Broken $\mathcal{PT}$-symmetry:} The coupling between mechanical and chemical components produces a $\mathcal{PT}$-symmetry–broken region bounded by exceptional points $q_{ep}^{\pm}$ (Fig.~\ref{fig:spectra}A,B). For $\tau_D \neq \tau_L$, the spectrum is nondegenerate with $q_{ep}^{-} \neq 0$ (Fig.~\ref{fig:spectra}A), whereas for $\tau_D = \tau_L$, $q_{ep}^{-}$ merges with a diabolical point at $q=0$, yielding a Dirac-cone–like dispersion (Fig.~\ref{fig:spectra}B). Perturbation theory about the uncoupled state reproduces the spectrum away from $q_{ep}^{\pm}$ but fails near these points (Fig.~\ref{fig:spectra}C), reflecting eigenvector coalescence and providing accurate estimates of the phase boundaries (Fig.~\ref{fig:spectra}D). The separation $|q_{ep}^{+} - q_{ep}^{-}|$ increases monotonically with $\mu$ (Fig.~\ref{fig:spectra}D inset) leading to collective behaviors such as spontaneous excitability and traveling waves.

\paragraph{Noisy HHS:} In the main text, we explore HHS in the absence of noise. However, all biological systems are subjected to external and internal noise. Hence, here we assess the robustness of our results under stochastic fluctuations. To do so, we modified the $R$-equation, which describes the motion of the vertices.
\begin{equation}
\tau_R \, \partial_t R = \partial_x^2 R - \partial_x L_0 + \tilde{\eta}(x,t) 
\end{equation}
Here, $\tilde{\eta}(x,t) = \eta(x,t) - \frac{1}{L}\int^L_0 \eta(x',t) dx'$ is a conservative noise. 
Subtracting the spatial mean ensures that the total length $\int_0^{L} \partial_xR dx$ remains conserved for a 1D periodic system.  In the discrete implementation, this is imposed as 
\begin{eqnarray}
    \tau_{R} \dot{R_i} &=& (R_{i+1}+R_{i-1} -2R_i) - (L_{i} - L^0_{i-1}) + \tilde{\eta}\\ 
    \tilde{\eta} &=& \eta -\frac{1}{N}\sum^N_{j=1} \eta_j
\end{eqnarray}

where, $\eta(x,t)$ is a Gaussian white noise with, 
\begin{equation}
\langle \eta(x,t) \eta(x',t') \rangle = 2D\, \delta(x - x') \, \delta(t - t'),
\end{equation}
    where $D$  denotes the noise strength. 

\begin{figure}[htbp]
    \centering
    \includegraphics[width=0.48\textwidth]{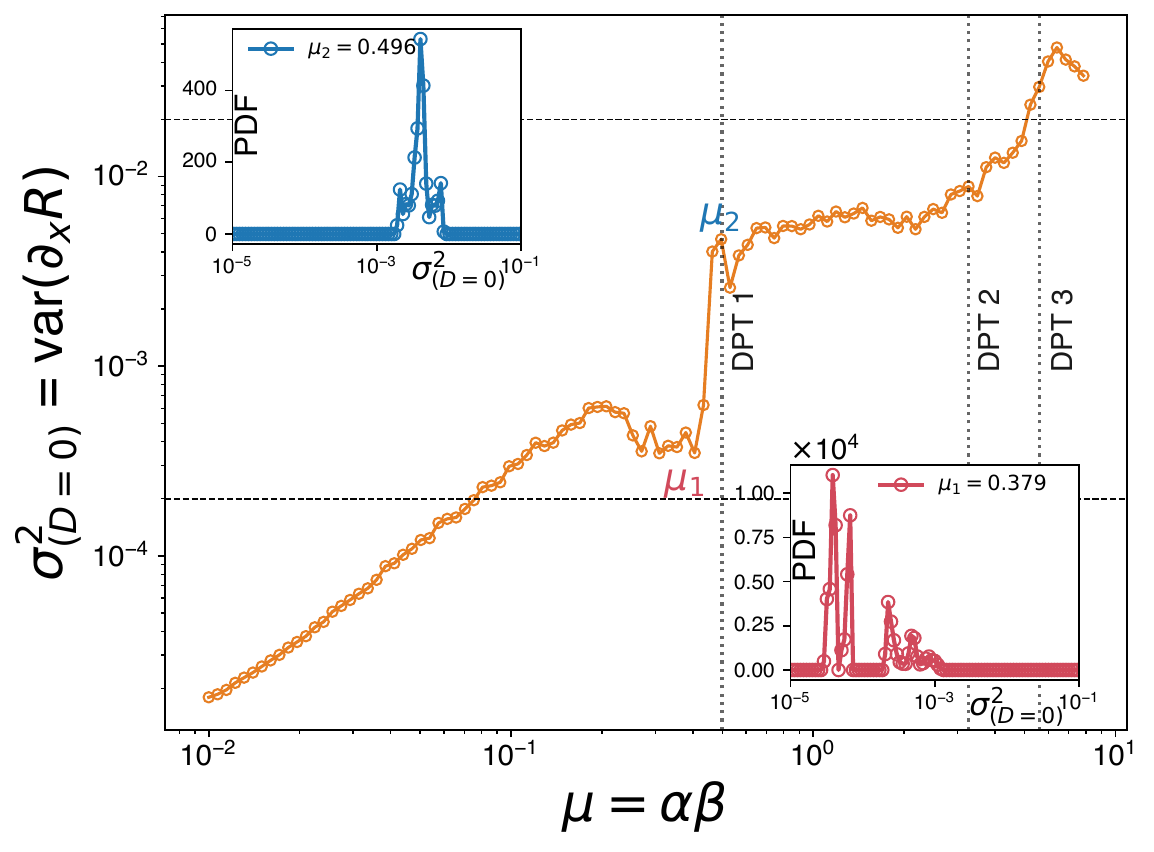}
    \caption{Figure shows how the variance $\sigma^2_{(D=0)}$ of the spring length ($\partial_x R$) changes with coupling strength ($\mu$). The discontinuous jumps around DPT1 ($\mu \sim 0.4$). The horizontal dashed lines correspond to two noise strengths shown in Fig.~\ref{fig:end_note_noise_intermittence}. The geometric mean of the variance is presented here, which is calculated from its steady state distribution at each $\mu$. The geometric standard deviation are negligible and are not shown here.}
    \label{fig:end_note_noise_variance}
\end{figure}

\paragraph{Variance of $\partial_x R$:} To determine a physically meaningful range of $D$, we first evaluated the intrinsic deterministic fluctuations of the system in the absence of noise ($D=0$). The variance of $\partial_x R$, $\sigma^2_{(D=0)}$, generally increases with $\mu$ and (Fig.~\ref{fig:end_note_noise_variance}) lies within the range $\sim 10^{-5} - 10^{-2}$, providing a reference scale for natural dynamical fluctuations. Interestingly, $\sigma^2_{(D=0)}$, shows a sharp increase near DPT1, implying a discontinuous transition from a solid-like to a fluid-like state, since $\sigma^2_{(D=0)}$ should be inversely related to the elastic modulus of the network. Indeed, this jump corresponds to an abrupt increase in strain variability and marks the onset of large-amplitude oscillations. The corresponding change in the distribution of $\partial_x R$ further highlights the qualitative difference between the two dynamical regimes (Fig.~\ref{fig:end_note_noise_variance} inset). No such sharp transition is observed near DPT2 and 3. However, above DPT2, the variance starts increasing faster than below. In contrast, variance seems to decrease above DPT3 (Fig.~\ref{fig:end_note_noise_variance}).

\begin{figure}[!htbp]
    \centering
    \includegraphics[width=0.5\textwidth]{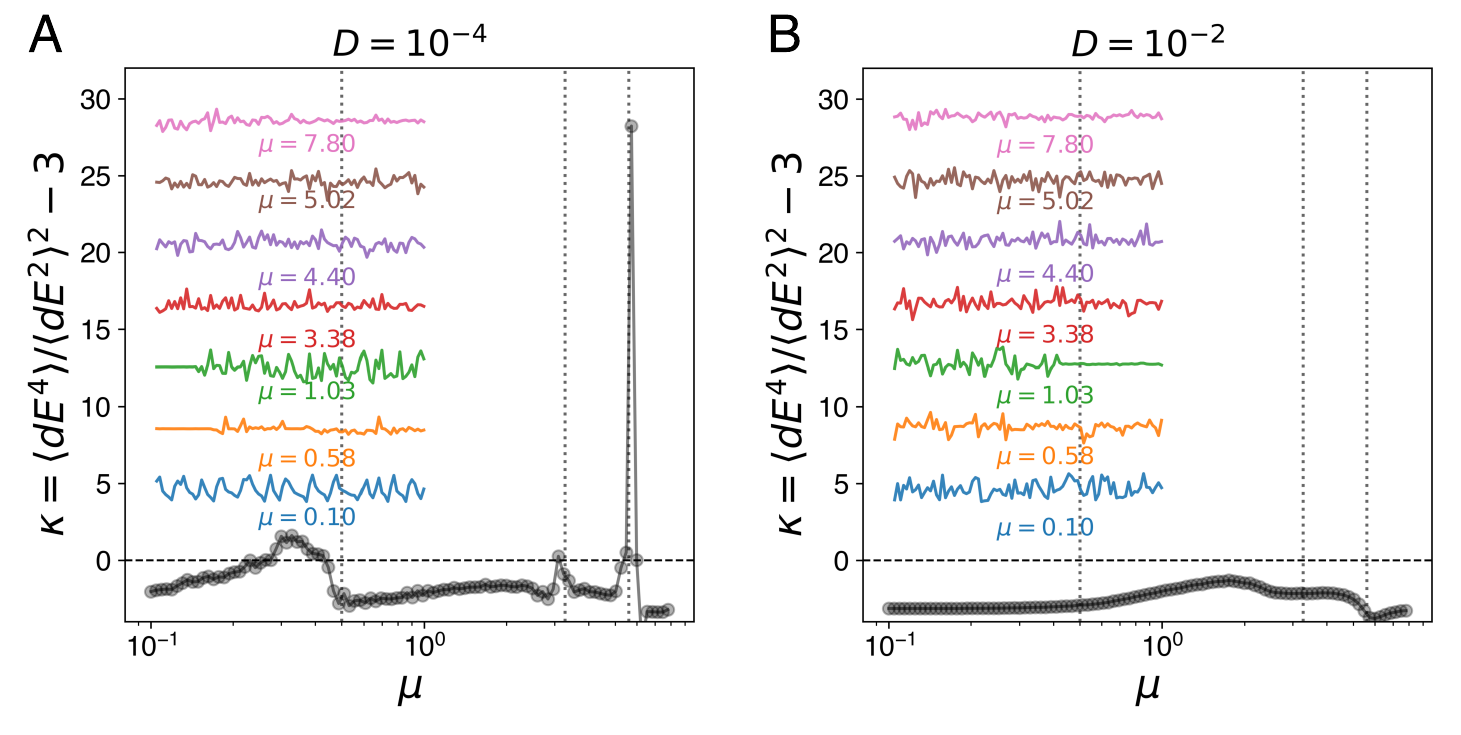}
    \caption{Effect of noise in the system with varying noise strength: (A) $D=10^{-4}$ and  (B) $D=10^{-2}$. Intermittent fluctuations persist at low noise levels and are suppressed as the noise increases.}
    \label{fig:end_note_noise_intermittence}
\end{figure}

\paragraph{Effect of noise:} Next, we test the effect of noise for two extreme cases. When, $D = 10^{-4}$, the noise strength is weaker than $\sigma^2_{(D=0)}$ for $\mu \gtrsim 0.1$. Hence, in this weak noise limit, we expect the system to show changed behavior only for the lowest values of $\mu$. Indeed, as shown in Fig.~\ref{fig:end_note_noise_intermittence}A, only the location of DPT1 is affected at this noise strength. In contrast, for $D = 10^{-2}$, the noise is strong enough to affect all $\mu$ values considered here, which is what we observe in Fig.~\ref{fig:end_note_noise_intermittence}B. At large noise amplitudes ($D = 10^{-2}$), stochastic forcing completely dominates the dynamics, and the system loses coherent structure entirely (Fig.~\ref{fig:end_note_noise_intermittence}B). Interestingly, although the intermittency is lost, the patterns survive even at such high level of noise (\cite{SI}), which implies that the dynamic phase transitions do happen, albeit the divergence of $\kappa$ may occur within a much narrower range of $\mu$.

This systematic exploration demonstrates that the distinct dynamical phases reported in our work are robust to biologically relevant levels of active stochastic fluctuations, while coherence is naturally lost only when noise overwhelms the deterministic driving forces.

\paragraph{On the use of the term ``symmetry-protected phases":}
In this work, the term symmetry-protected phases refers to phases whose existence is determined by the symmetry of the mechanochemical interactions together with the translational invariance of the periodic ring. Because of the equivariant Hopf theorem~\cite{collins_coupled_1993,collins_group-theoretic_1994,epstein_symmetric_1993}, these symmetries fix the admissible phase structure independently of the specific microscopic oscillator model. Consequently, the same phase diagram is obtained for both the Brusselator and FitzHugh–Nagumo models. Thus, the phase organization depends only on the underlying symmetry constraints and the topology of the oscillator network, and not on the detailed chemistry of the Hopf-oscillator dynamics. 
It is in this sense, we use the term ``symmetry-protected phases", which is different from the term commonly used in quantum condensed matter~\cite{Senthil2015}.

\clearpage
\onecolumngrid
\setcounter{section}{0}
\setcounter{equation}{0}
\setcounter{figure}{0}
\setcounter{table}{0}
\renewcommand{\theequation}{S\arabic{equation}}
\renewcommand{\thefigure}{S\arabic{figure}}
\renewcommand{\thetable}{S\Roman{table}}
\renewcommand{\lap}{\mvec{\nabla}^2}
\renewcommand{\pdc}[1]{\textcolor{blue}{#1}}

\vspace{1.5\baselineskip}
\begin{center}
  {\large\bfseries Supplementary Information:\\ Symmetry protected phases in a 1D active solid with mechanochemical feedback}\\[1.0em]
  Soumyadeep Mondal, Phanindra Dewan, Lakshman Santhosh Kumar, and Sumantra Sarkar\\[0.4em]
  {\itshape Department of Physics, Indian Institute of Science, Bengaluru, Karnataka, India, PIN 560012}\\[0.4em]
  (Dated: \today)
\end{center}
\vspace{1.0\baselineskip}

\section{Methods}

\subsection{Initial condition}
\begin{figure}[htbp]
    \centering
    \includegraphics[width=\linewidth]{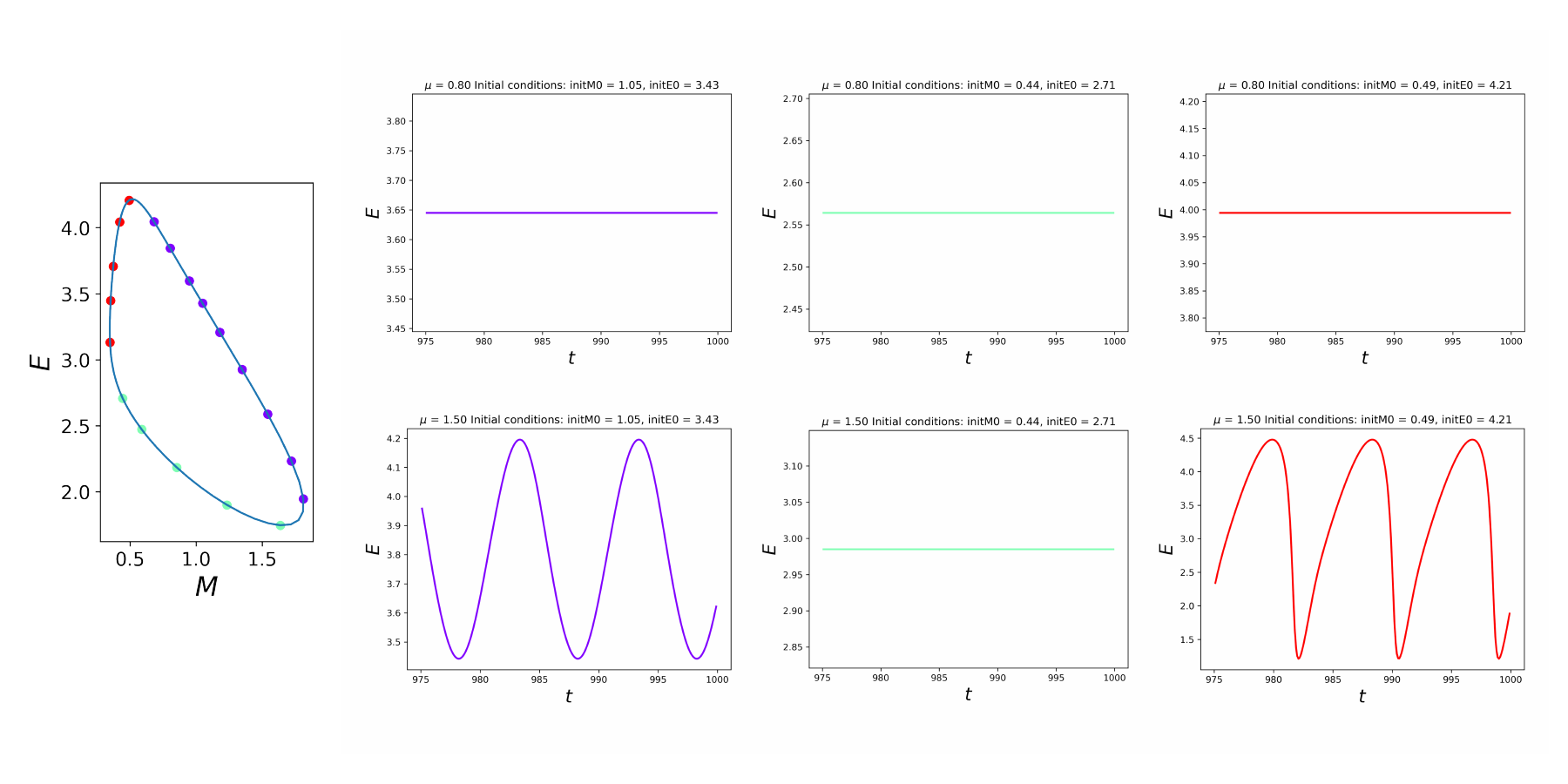}
    \caption{\textbf{Single cell model}: Time series of $E$ for different $\alpha = \beta$ values and different initial conditions $E_0$ and $M_0$.}
    \label{si:fig:SI-TimeSeries}
\end{figure}

We assume that for $t < 0$, there was no mechanochemical coupling, i.e., $\alpha = \beta = 0$. Hence, the brusselators were oscillating independently, and they were in some random location in the uncoupled Brusselator limit cycle (Fig.~\ref{si:fig:SI-TimeSeries}). For both single cell and 1D system, the coupling was turned on at $t = 0$, which suddenly changed the dimension of the phase space from 3 to 5. Hence, depending on the initial location in the limit cycle, the system moves to different fixed points, as illustrated in Fig.~\ref{si:fig:SI-TimeSeries}. 

To show this, in the single-cell model, we uniformly sampled the limit cycle and solved the steady-state single-cell equations to obtain the fixed points. We then used these fixed point solutions to compute the eigenvalues of the Jacobian matrix for different $\mu = \alpha\beta$ values, which were plotted in Fig 1C,D.

In the continuum model, we assumed that each ``cell" is in a random location of the limit cycle. Hence, we randomly sampled the limit cycle for $M_0$ and $E_0$ and chose $D_0 = 0.1$, $L_0(t = 0) = 1$, and $\partial_xR(x,t=0) = 1$. For Fig. 4, we used 10-20 different initial conditions to compute the averages and the distributions.   

\subsection{Numerical solution of continuum model}
To simulate the coupled 1D Partial Differential Equations (PDE)~\ref{si:Eq:SIcontvertex1D}, we have implemented the central in-space method with Runge-Kutta time marching. We use spatial and temporal discretization with $a = \Delta x = 1$ and $\Delta t = 0.01$, satisfying the Courant–Friedrichs–Lewy (CFL) condition to ensure the stability of the numerical solutions. Here are the simulation details: 

\begin{enumerate}
    \item $a = 1$ and $\tau_R = 1$ for all simulations. They are the unit of length and time in our model. $\tau_D = 1.1$ and $\tau_L = 1.2$, unless otherwise stated. 
    
    \item For Fig 3A-C, we have simulated a model with $L_x = 200a$ and $t_{max} = 5\times10^4\tau_R$. A single seed value is shown.

    \item For Fig 4A-B, we have chosen $\alpha = \beta = \sqrt{\mu}$, where $\mu \in [1e-2,9]$. $L_x = 1000a$, $t_{max} = 10^5\tau_R$. The distributions were obtained from datapoints from last $t = 9.9\times10^4-10^5$ and averaged over $10$ random initial conditions. All together, $10^3\times10^3\times10 = 10^7$ samples were used to compute the distributions. 
 
    \item For Fig 4A-B and Fig.~\ref{si:fig:SI-total-size},~\ref{si:fig:SI-total-number}, we have chosen $\alpha, \beta \in \{0.1,0.2,0.3,0.5,0.8,1,1.5,2,2.5,3\}$. For each of the 100 pairs of $\alpha,\beta$ values and 9 pairs of $\tau_D,\tau_L$ values, we used 20 random initial conditions. 
\end{enumerate}

\begin{figure}[htbp]
    \centering
    \includegraphics[width=\linewidth]{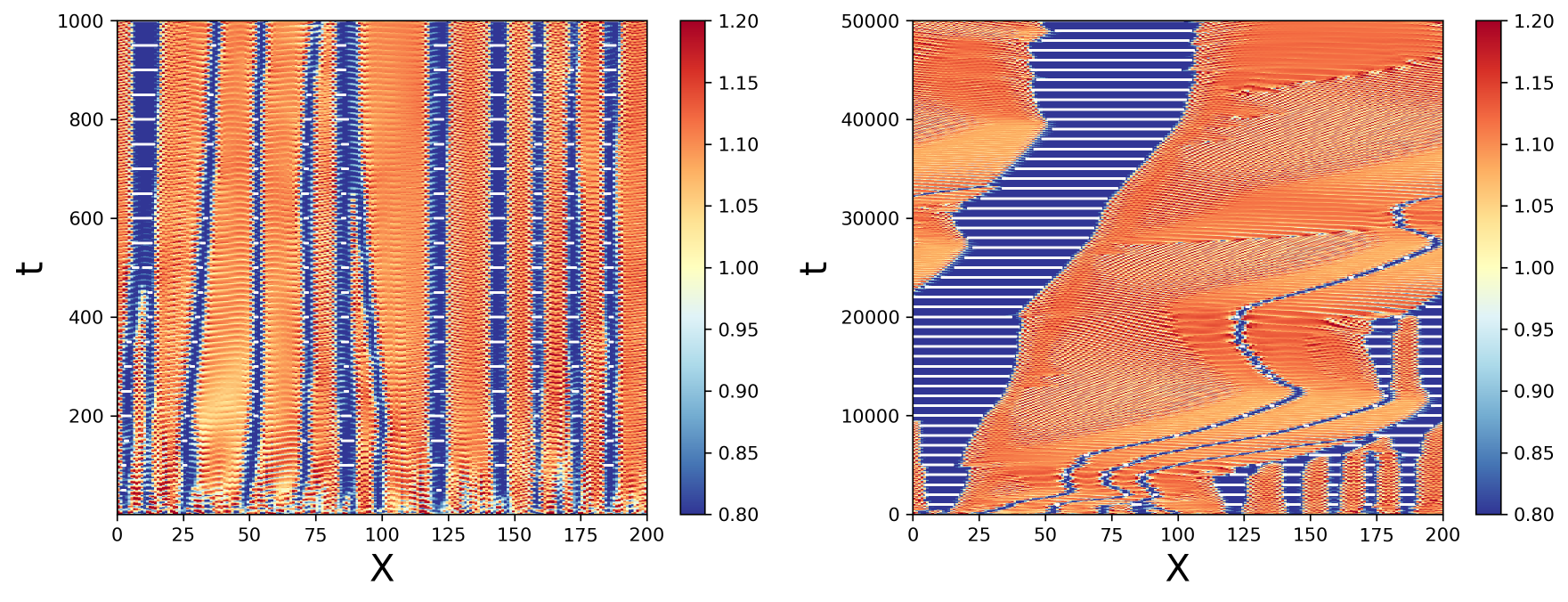}
    \caption{Numerical estimate of $L_{OD}$ at short (left) and long timescales (right). The white lines show our estimate. The color map indicates the value of $\partial_xR$.}
    \label{si:fig:static-algo}
\end{figure}

\subsection{Numerical estimation of $L_{OD}$}
We observed that $E$ does not change appreciably in the COD regions. Hence, the variance of $E$ over some sampling time in COD regions is low compared to the surrounding regions. By using a threshold of $10^{-2}$ and a sampling time of $100$ time steps, we were able to get good estimates of $L_{OD}$ (Fig.~\ref{si:fig:static-algo}). The size does not depend sensitively on the choices of these hyperparameters, as long as they are chosen reasonably.

\subsection{Excess kurtosis}
Near the dynamic phase transition boundaries, the state variables, such as $E$, show intermittent fluctuations. To quantify it, we use excess kurtosis as a measure~\cite{davidson2015turbulence}. To do so, we define $dE = E (t+\tau_{lag}) - E(t)$. We use the long-lag time values of $dE$, with $\tau_{lag} = 100\tau_R, 1000\tau_R,$ and $10000\tau_R$. 

Kurtosis or any measure of dynamic fluctuations depends on the timescale. For this analysis, we excluded all data below $t<10000\tau_R$. Above this limit, the memories of the initial condition are washed out by the dynamics. For the last lag time, we did not have sufficient statistics. So, we did not use it. The other two lag times show similar behaviors. We define excess kurtosis as follows ($\langle dE \rangle = 0$): 
\begin{equation}
    \kappa = \langle dE^4\rangle / \langle dE^2 \rangle^2 - 3.
\end{equation}
$\kappa = 0$ for a Gaussian distribution. Distributions with $\kappa > 0$, such as near the dynamic phase transitions, have fat tails, whereas for $\kappa < 0$, the underlying distribution has thinner tails. $dE$ fluctuation within each region has this property.

\section{Single-cell Model}

The single-cell model with mechanochemical feedback is written as:

\begin{geqn}
    \del_t M = a -(b+1)M + cM^2E \\
    \del_t E = bM - cM^2E - DE\\ 
    \tau_D\del_t D = -(D-D_0) - \beta D(L-1)\\
\tau_L \del_t L_0 = -(L_0-1) - \alpha (E-E_0)\\
\tau_R \del_t L = -(L-L_0) + f(L,t)
    \label{si:Eq:Bruss1}
\end{geqn}

Let us take $f(L,t) =0$ for simplicity. We can linearize the equations above to write:

\begin{geqn}
    \del_t \delta M = [-(b+1) + 2cM_0 E_0] \delta M + cM_0^2 \delta E\\
    \del_t \delta E = (b - 2cM_0 E_0)\delta M + (-c M_0^2 - D_0) \delta E - E_0 \delta D\\ 
    \tau_D\del_t \delta D = -\delta D - \beta D_0 \delta L\\
\tau_L \del_t \delta L_0 = - \alpha \delta E -\delta L_0 \\
\tau_R \del_t \delta L = \delta L_0 -\delta L
    \label{si:Eq:Bruss2}
\end{geqn}

The above equation can be written as:

\begin{equation}
    \del_t \begin{bmatrix}
        \delta M \\
        \delta E \\
        \delta D \\
        \delta L_0 \\
        \delta L
    \end{bmatrix}
    = \mat{J} \begin{bmatrix}
        \delta M \\
        \delta E \\
        \delta D \\
        \delta L_0 \\
        \delta L
    \end{bmatrix}
\end{equation}

where the Jacobian of the above linearized system of equations is given by:

\begin{equation}
    \mat{J} = \begin{bmatrix}
-(b+1) + 2cM_0 E_0 & cM_0^2 & 0 & 0 & 0\\
b - 2cM_0 E_0 & -c M_0^2 - D_0 & -E_0 & 0 & 0\\
0 & 0 & -1/\tau_D & 0 & -\beta D_0 / \tau_D \\
0 & -\alpha/\tau_L & 0 & -1/\tau_L & 0\\
0 & 0 & 0 & 1/\tau_R & -1/\tau_R
\end{bmatrix}
\label{si:brussljacobian}
\end{equation}

\subsection{Eigenvalues}

Using SciPy, we computed the eigenvalues of the Jacobian given in \ref{si:brussljacobian}. We numerically solved for the fixed points using scipy.optimize for various initial conditions chosen from the limit cycle of uncoupled Brusselator (Fig 1B in the main text and Fig.~\ref{si:fig:SI-TimeSeries}. For $\mu > \mu_c$ (see next subsection), it took the system to three fixed points, one of which is a stable node. The system is compressed ($L < 1$) in this node and there are no oscillations in the system. This is illustrated in Fig.~\ref{si:fig:SI-maxEV},  where the real part of the eigenvalues with the largest real parts are plotted for different $\tau_D$ and $\tau_L$. It shows that compaction-driven ($L<1$) oscillation death arises generically in the single-cell model. 

\subsection{Critical $\mu_c$}

The determinant of the Jacobian in \ref{si:brussljacobian} is given below:

\begin{equation}
    \det \mat{J} = \begin{vmatrix}
-(b+1) + 2cM_0 E_0 & cM_0^2 & 0 & 0 & 0\\
b - 2cM_0 E_0 & -c M_0^2 - D_0 & -E_0 & 0 & 0\\
0 & 0 & -1/\tau_D & 0 & -\beta D_0 / \tau_D \\
0 & -\alpha/\tau_L & 0 & -1/\tau_L & 0\\
0 & 0 & 0 & 1/\tau_R & -1/\tau_R
\end{vmatrix}
\end{equation}

The critical value of the product $\alpha \beta $ where $\det \mat{J} = 0$ is where the fixed point appears. This can be calculated for a given set of parameters as given below:

\begin{align}
    \det \mat{J} = [-(b+1) + 2cM_0 E_0] \frac{cM_0^2 + D_0 + (\alpha \beta)_c E_0 D_0}{\tau_D \tau_L \tau_R} - c M_0^2 \frac{b - 2c M_0 E_0}{\tau_D \tau_L \tau_R} &= 0 \\
    \implies [-(b+1) + 2cM_0 E_0](cM_0^2 + D_0 + \mu_c E_0 D_0) - c M_0^2 (b - 2c M_0 E_0) &= 0
\end{align}

where $\mu_c = (\alpha \beta)_c$. The value of $\mu_c$ does not depend on the parameters $\tau_D, \tau_L, \tau_R$ of the model.

\subsection{Routh Hurwitz Stability Criterion for $5\times5$ Matrix}

\subsubsection{Construction of the Routh-Hurwitz table}
A $5\times5$ Matrix has a 5th-order characteristic equation, as shown below. 
\begin{equation*}
    a_5 x^5 + a_4 x^4 + a_3 x^3 + a_2 x^2 + a_1 x + a_0 = 0
\end{equation*}
Here, $\{a_0,\ldots,a_5\}$ are the coefficients of the characteristic equations, from which we can construct the Routh-Hurwitz Table.  
\begin{equation*}
    \begin{matrix}
        a_5 & a_3 & a_1\\
        a_4 & a_2 & a_0\\
        b_1 & b_2 & 0 \\
        c_1 & c_2 & 0\\
        d_1 & 0 \\
        e_1 \\
    \end{matrix}
\end{equation*}

This table contains the Routh-Hurwitz coefficients, $\{b_1,b_2,c_1,c_2,d_1,e_1\}$, which are obtained from the coefficients of the characteristic equation, as below. 

\begin{equation*}
    b_1 = -\frac{1}{a_4}\biggl|\begin{bmatrix}
        a_5 & a_3 \\
        a_4 & a_2
    \end{bmatrix}\biggr| = \frac{a_4a_3 - a_5a_2}{a_4}
\end{equation*}

Similarly,

\begin{equation*}
    b_2 = \frac{a_4 a_1 - a_5 a_0}{a_4}
\end{equation*}

\begin{equation*}
    c_1 = \frac{b_1 a_2 - a_4 b_2}{b_1}
\end{equation*}

\begin{equation*}
    c_2 = \frac{b_1 a_0}{b_1} = a_0
\end{equation*}

\begin{equation*}
    d_1 = \frac{c_1 b_2 - b_1 c_2}{c_1}
\end{equation*}

\begin{equation*}
    e_1 = \frac{d_1 c_2}{d_1} = c_2 = a_0
\end{equation*}

\subsubsection{Stability analysis using the Routh-Hurwitz table}
For a system to be stable the coefficients populating the first column of the Routh-Hurwitz table must have the same sign. We use this stability criterion to investigate the stability of the 0D model.  

\begin{figure}[!htbp]
    \centering
    \includegraphics[width=0.7\linewidth]{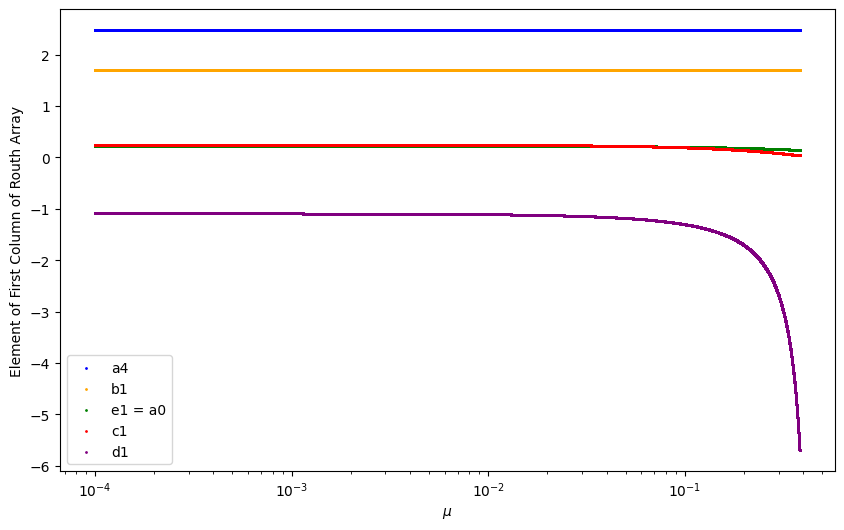}
    \caption{Routh Hurwitz Criterion for $\mu<\mu_c$.}
    \label{si:fig:RH_less}
\end{figure}

For $\mu<\mu_c$, the system has only one fixed point. The coefficient $d_1$ at this fixed point is always negative, whereas rest of the column coefficients are always positive. Hence, for $\mu < \mu_c$, the fixed point is always unstable (Fig.~\ref{si:fig:RH_less}). 

\begin{figure}[!htbp]
    \centering
    \includegraphics[width=0.7\linewidth]{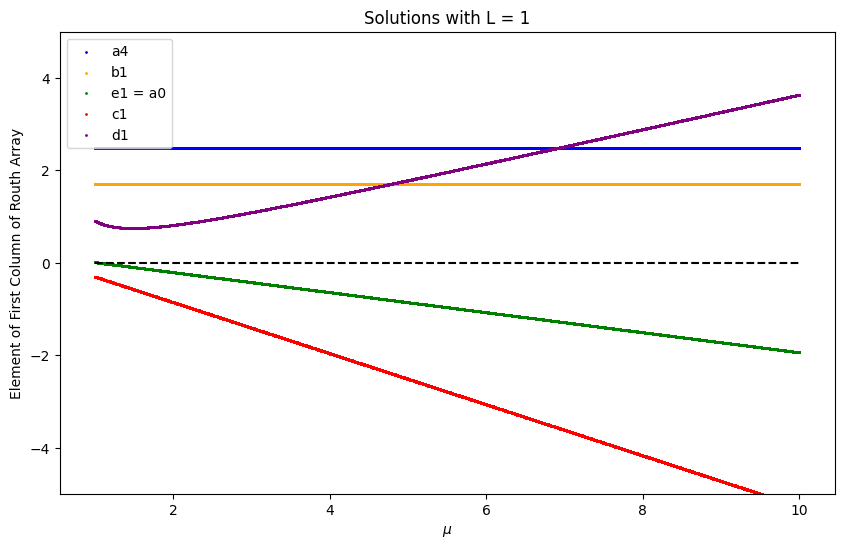}
    \caption{Routh Hurwitz Criterion for $\mu>\mu_c$ and $L = 1$.}
    \label{si:fig:RH_L=1}
\end{figure}

\begin{figure}[!htbp]
    \centering
    \includegraphics[width=0.7\linewidth]{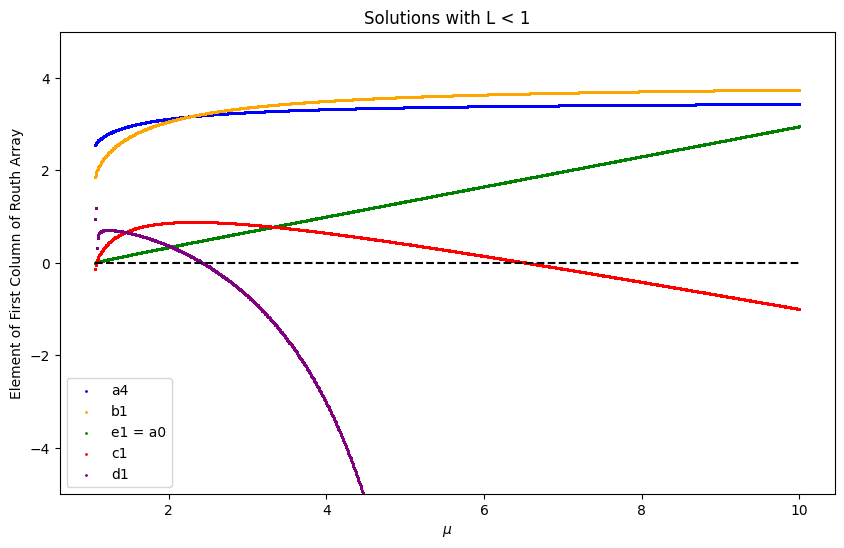}
    \caption{Routh Hurwitz Criterion for $\mu>\mu_c$ and $L < 1$ shows that for a range of $\mu$ all five coefficients are greater than zero, which implies that, in this range, this fixed point is stable.}
    \label{si:fig:RH_L<1}
\end{figure}

\begin{figure}[!htbp]
    \centering
    \includegraphics[width=0.7\linewidth]{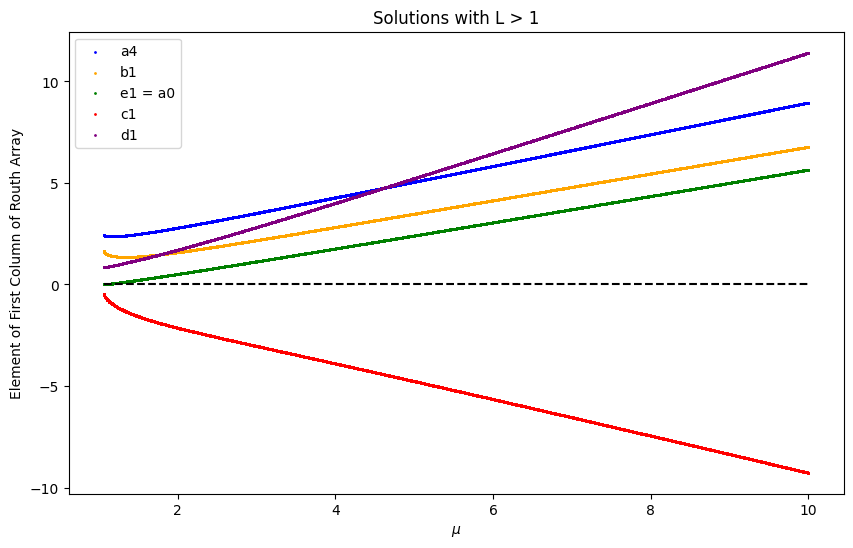}
    \caption{Routh Hurwitz Criterion for $\mu>\mu_c$ and $L > 1$.}
    \label{si:fig:RH_L>1}
\end{figure}

In contrast, for $\mu>\mu_c$, there are 3 fixed points, corresponding to a system with $L=1$, $L>1$, and $L<1$. The column coefficients for these fixed points are plotted in Fig.~\ref{si:fig:RH_L=1},~\ref{si:fig:RH_L<1}, and \ref{si:fig:RH_L>1}. As expected, for $\mu$ in a certain range, all coefficients are positive for the fixed point corresponding to $L<1$. Hence, in this range of $\mu$, the fixed point is stable. For the other fixed points, $c_1$ or $d_1$ is consistently negative. Hence, they are always unstable.   

\section{1D Continuum Model}

To model a linear chain of cells, we consider each cell to be a spring whose rest lengths $L^0_i(t)$ can be coupled to the chemistry within it. If each cell is bounded by two vertices $R_i$ and $R_{i+1}$, then the equation of motion for the vertices is given by:

\begin{equation}
    \zeta (R_i(t+\Delta t) - R_i(t)) = k_i (R_{i+1}(t) - R_i(t) - L^0_i(t)) - k_{i-1}(R_{i}(t) - R_{i-1}(t) - L^0_{i-1}(t))
\end{equation}

If we take homogeneous stiffness constants, $k_i = k \; \; \forall \; \; i$, then in the continuum limit, we obtain the equation:

\begin{equation}
    \tau_R \del_t R = \del_{xx}R - \del_x L_0,
\end{equation}
where $R(x,t)$ is the vertex position and $L_0(x,t)$ is the preferred length of the springs in the continuum limit. This is equivalent to the single-cell model, which can be seen by taking a spatial derivative on both sides of the equation. 
\begin{equation}
    \tau_R \del_t (\partial_x R) = \del_{xx}(\partial_x R - L_0)
\end{equation}
This equation is a continuity equation of $\del_xR$ and arises from the conservation of the total length. This is a novel form of conservation law that arises in confluent tissues. 

The time scale of relaxation, $\tau_R = \zeta/k$ depends on the friction coefficient of the medium and the spring constant of the cells~\cite{boocock_theory_2021}. This is the continuum version of a 1D vertex model. This derivation is similar to the one in \cite{boocock_theory_2021}.

As a minimal model of the chemistry, we consider a Brusselator model for the ERK pathways \cite{epstein1998introduction}. The Brusselator equations are:

\begin{geqn}
    \del_t M = a -(b+1)M + cM^2E \\
    \del_t E = bM - cM^2E - DE\\ 
    \tau_D\del_t D = -(D-D_0)
    \label{si:Eq:Bruss1b}
\end{geqn}

The species $E$ and $M$ are the oscillating species in the usual Brusselator. These two species are analogous to ERK and its regulatory protein MEK. Here, an extra term $-DE$ is present because we consider a degrader $D$, which degrades $E$. $D$ follows relaxation dynamics to $D_0$ with a timescale of relaxation, $\tau_D$.

Now if we consider that the instantaneous lengths of the cells $\del_x R$ affects the degrader $D$ and the ERK level affects in the rest lengths $L_0$ in the following way:

\begin{geqn}
    E \ \raisebox{-.2ex}{\rotatebox{90}{\scalebox{1}[1.2]{$\bot$}}}\ L_0 \\
    L \ \raisebox{-.2ex}{\rotatebox{90}{\scalebox{1}[1.2]{$\bot$}}}\ D
\end{geqn}

then the equations for the chemical dynamics become:

\begin{geqn}
    \del_t M = a -(b+1)M + cM^2E \\
    \del_t E = bM - cM^2E - DE\\ 
    \tau_D\del_t D = -(D-D_0) - \beta D (\del_x R -1)
\end{geqn}

The equations for the mechanics will also be modified and are written as:

\begin{geqn}
    \tau_L \del_t L_0 = - (L_0 - L_0^{eq}) - \alpha (E - E_0)\\
    \tau_R \del_t R = \del_{xx}R - \del_x L_0
\end{geqn}

Finally, the 1D continuum vertex model is given by,
\begin{geqn}
    \del_t M = a -(b+1)M + cM^2E \\
    \del_t E = bM - cM^2E - DE\\ 
    \tau_D\del_t \delta = -(\delta + \delta^3) - \beta D(\del_xR - 1)\\
    \tau_L \del_t l = -(l + l^3) - \alpha (\epsilon + \epsilon^3)\\
    \tau_R \del_t R = \del^2_xR - \del_xL_0
    \label{si:Eq:SIcontvertex1D}
\end{geqn}
Here, $a$, $b$, and $c$ are the parameters of the canonical Brusselator, and E and M are the oscillating species. Here, $R(x)$ is the location of the cell vertex, and $\partial_x R$ is the continuum equivalent of the cell length, $L$. $\tau_D, \tau_L$ and $\tau_R$ are the relaxation timescales. Moreover, $\delta = D - D_0$, $\epsilon = E-E_0$ and $l=L_0 -1$ (taking $L_0^{eq} =1$). The cubic terms are included due to the numerical stability and do not change the general behavior of the system. 
\subsection{Cubic term in the $\delta$ and $l$ equation}
The cubic term in Eqn.~\ref{si:Eq:SIcontvertex1D} is introduced as the minimal non-linear saturation needed to keep the amplitudes of the variables finite in non-linear simulations. In the absence of such terms, the linear equations correctly capture the onset of oscillatory or unstable modes, but the amplitudes of $\delta$ and $l$ can grow without limit under sustained coupling. Adding a cubic contribution, e.g. $-(\delta + \delta^3)$ is the simplest symmetric regularization that (i) leaves the linear stability unchanged, (ii) corresponds to the gradient of the potential function, $V(\delta) = \frac{1}{2} \delta^2 + \frac{1}{4} \delta^4$. This choice follows standard practice in amplitude-equation theory (Landau expansion) and in related mechanochemical models \cite{hannezo_mechanochemical_2019}, where cubic saturation terms are used to regularize ERK dynamics. Furthermore, the nonlinear term in the potential function affects the quantitative location of the phase boundaries.

\subsection{Linear Stability Analysis} 
To understand the system we at first do the linear stability analysis of the Eq. \ref{si:Eq:SIcontvertex1D} around the steady state values of $[M_0, E_0, D_0, L_{0o},R_{0}]$. We take the ansatz, $y(t,x) = \tilde{y}(\omega,q) exp(\omega t+iqx)$ and put this ansatz to Eq.~\ref{si:Eq:SIcontvertex1D} which gives us the Eigen value equation, $\omega \tilde{\mathbf{u}} = \mat{A_q} \mathbf{\Tilde{u}} $
where, $\mathbf{\Tilde{u}} = [\Tilde{M},\Tilde{E},\Tilde{D},\Tilde{L_0},\Tilde{R}]^T$, and
\begin{equation}
    \mat{A_q} = \begin{bmatrix}
    -1 -b + c_1 & c_2 & 0 & 0 & 0\\
     b - c_1 & -(c_2 + D_0) & -E_0& 0& 0\\
     0 & 0 & -1/\tau_D & 0 & -i\beta D_0q/\tau_D\\
     0 & -\alpha/\tau_L & 0 & -1/\tau_L & 0\\
     0 & 0 & 0& -iq/\tau_R & -q^2/\tau_R
    \end{bmatrix}
    \label{si:Eq:SIAq}  
\end{equation}
 Where, $c_1 = 2cM_{0}E_{0}$ and $c_2 = cM_{0}^2$. 
 \begin{figure}[!htbp]
\centering
\includegraphics[width=\linewidth]{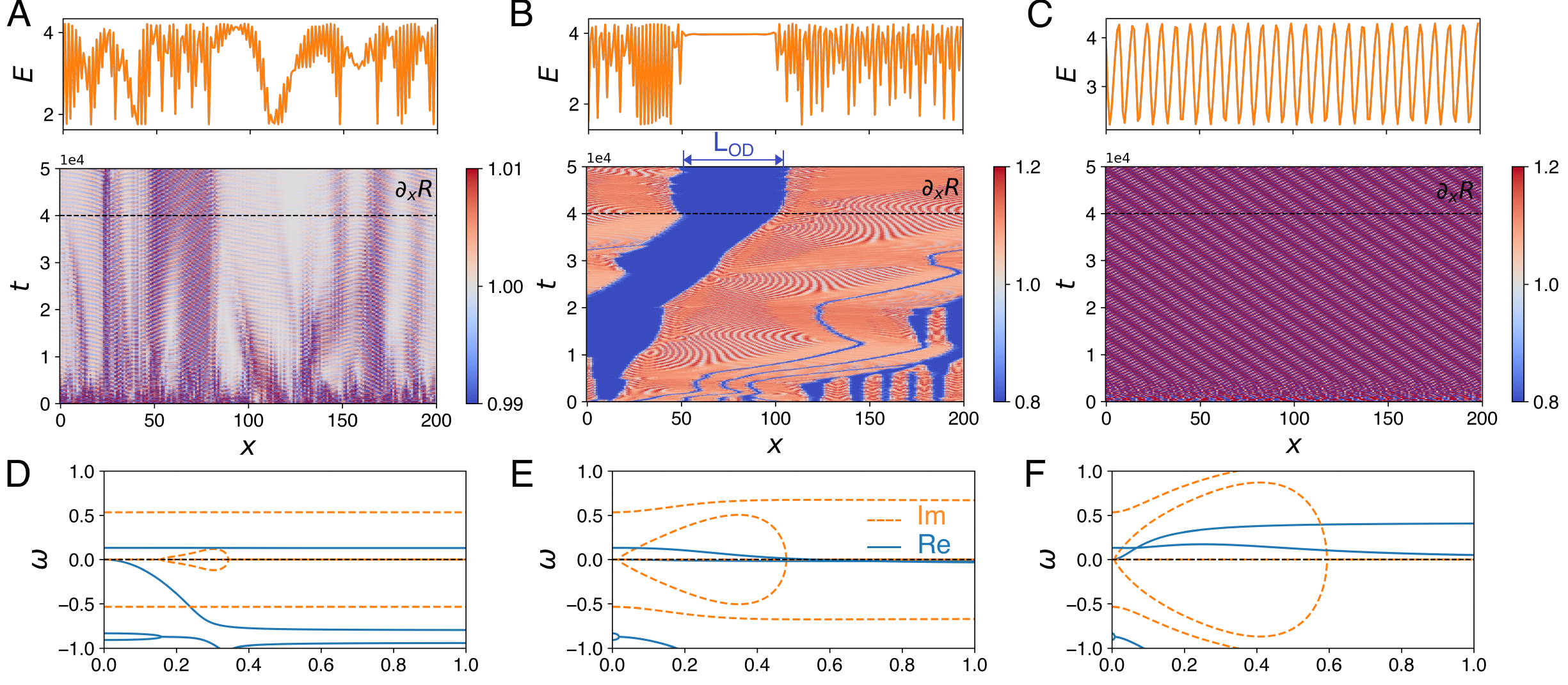}
\caption{Kymograph of $\partial_x R$ by solving Eqn.~\ref{si:Eq:SIcontvertex1D} show spatiotemporal patterns for (A) $\mu = 0.01 $, (B) $1$, and (C) $6.25$; $\tau_D = 1.1, \tau_L = 1.2$. Top: $E(x)$ at $t=40000$ (black dashed line in the Kymograph); $\alpha = \beta = \mu^{1/2}$.(D,E,F) Due to the nonorthogonality of the eigenvectors, the real (solid blue) and the imaginary (dashed orange) parts of the eigenvalues, $\omega (qa/\pi)$, do not have positive peaks at nonzero $q$ values, and, hence, does not explain the pattern forming instabilities.}

\end{figure}

\subsection{Exceptional point of the uncoupled system}
The matrix $\mat{A_q}$ is non-Hermitian, i.e., $\mat{A}_q^{\dagger} \neq \mat{A_q}$. Consequently, its left and right eigenvectors are not the same. In the absence of Mechano-Chemical Coupling (MCC) ($\mu= \alpha\beta =0$), the matrix $\mat{A_q}$ decouples into two independent blocks: a block $\mat{A^{{chem}}_q}$, which describes Brusselator dynamics and a block $\mat{A^{{mech}}_q}$, describing the mechanical part, spring network, represented as,

\begin{equation}
    \mat{A^{\text{chem}}_q} = 
    \begin{bmatrix}
    -1 - b + c_1 & c_2 & 0 \\
    b - c_1 & -(c_2 + D_0) & -E_0 \\
    0 & 0 & -1/\tau_D
    \end{bmatrix}
\end{equation}

\begin{equation}
    \mat{A^{\text{mech}}_q} = 
    \begin{bmatrix}
    -1/\tau_L & 0 \\
    -iq/\tau_R & -q^2/\tau_R
    \end{bmatrix}
\end{equation}
The eigenvalues corresponding to the mechanical part is given by,
\begin{geqn}
    \lambda_1 = -\dfrac{1}{\tau_L}\\
    \lambda_2 = -\dfrac{q^2}{\tau_R}
\end{geqn}
The corresponding right eigenvectors are given as,
\begin{geqn}
    \ket{R_1} = \dfrac{1}{\sqrt{1+\dfrac{q^2/\tau_R^2}{(\frac{1}{\tau_L}-\frac{q^2}{\tau_R})^2}}}\left[1,\dfrac{\frac{iq}{\tau_R}}{\left(\frac{1}{\tau_L}-\frac{q^2}{\tau_R}\right)} \right]^T\\
    \ket{R_2}=[0,1]^T
\end{geqn}
 Now, Exceptional points are defined as the point at which two eigenvectors and eigenvalues collapse. The condition where the two eigenvectors collapse is given by,
 $q=\sqrt{\frac{\tau_R}{\tau_L}}$. At this value of $q$, the eigenvector $\ket{R_1} = \ket{R_2}$. However, at $\mu>0$ the coupling between the mechanical and the chemical components results in a region of broken $\mathcal{PT}$-symmetry that ends at two exceptional points, $q_{ep}^{\pm}$.  
 
\subsection{Non-Hermitian Perturbation Theory}

The analysis is based on ref.~\cite{brody2013biorthogonal}.

Consider a non-Hermititan matrix, $H(\epsilon) = H_0 +\epsilon V$. Where, $H_0$ is the unperturbed non-Hermitian matrix with right eigenvectors $\ket{\phi^{(0)}_n}$, left eigenvectors $\bra{\chi^{(0)}_n}$, and eigenvalues $\lambda^{0}_n$.\\
\begin{geqn}
    H_0\ket{\phi^0_n} = \lambda^{0}_n \ket{\phi^0_n}\\
    \bra{\chi^0_m} H_0 = \lambda^{0}_m \bra{\chi^0_m}
\end{geqn}
with biorthogonality,$\bra {\chi^{(0)}_m}\ket {\phi^{(0)}_n} = \delta_{mn}$.
$V$ is the perturbation matrix, and $\epsilon$, a small parameter, denotes the perturbation strength. We expand the eigenvalues and eigenvectors in terms of $\epsilon$ as follows,
\begin{geqn}
    \lambda_n = \lambda^{0}_n + \epsilon \lambda^{(1)}_n + \epsilon^{2}  \lambda^{(2)}_n+... \\
    \ket{\phi_n} =\ket{\phi^{(0)}_n} +\epsilon \ket{\phi^{(1)}_n} +\epsilon^2 \ket{\phi^{(2)}_n}+ ...\\
    \bra{\chi_n} = \bra{\chi^{(0)}_n} + \epsilon\bra{\chi^{(1)}_n} + \epsilon^2\bra{\chi^{(2)}_n} + ...
    \label{si:SI_perturb_series}
\end{geqn}
The perturbed equation is given by,
\begin{equation}
    (H_0 + \epsilon V)\ket{\phi_n} = \lambda_n \ket{\phi_n}
    \label{si:SI_perturbation_non_degenerate_Eigen_value_equation}
\end{equation}
Substitute the perturbative expansion from Eq.\ref{si:SI_perturb_series} into Eq.\ref{si:SI_perturbation_non_degenerate_Eigen_value_equation} and equate terms order by order in $\epsilon$ we get,
\\
\begin{eqnarray}
    \text{\textbf{Zeroth Order}} ~(\epsilon^{0})&&: H_0 \ket{\phi^{(0)}_n}  =\lambda^{(0)}_n\ket{\phi_n^{(0)}} \label{si:SI_zeroth_order}\\
    \textbf{First Order ($\epsilon^{1}$)}&&: H_0 \ket{\phi^{(1)}_n} + V\ket{\phi^{(0)}_n} = \lambda_n^{(0)}\ket{\phi^{(1)}_n} + \lambda_n^{(1)}\ket{\phi^{(0)}_n}  \label{si:SI_first_order}\\
    \textbf{Second Order  ($\epsilon^{2}$)}&&: H_0 \ket{\phi^{(2)}_n} + V\ket{\phi^{(1)}_n} = \lambda_n^{(0)}\ket{\phi^{(2)}_n} + \lambda_n^{(1)}\ket{\phi^{(1)}_n} + \lambda_n^{(2)}\ket{\phi^{(0)}_n} \label{si:SI_second_order}
\end{eqnarray}
Eq.\ref{si:SI_zeroth_order} is the unperturbed eigenvectors and eigenvalues. For calculating the first order correction to the eigenvalues and eigenvectors we project $\bra{\chi^{(0)}_n}$ in Eq.\ref{si:SI_first_order} we get,
\begin{geqn}
\bra{\chi^{(0)}_n}H_0 \ket{\phi^{(1)}_n} + \bra{\chi^{(0)}_n}V\ket{\phi^{(0)}_n} = \lambda_n^{(0)}\bra{\chi^{(0)}_n}\ket{\phi^{(1)}_n} + \lambda_n^{(1)}\bra{\chi^{(0)}_n}\ket{\phi^{(0)}_n}\\
\lambda_n^{(1)} = \bra{\chi^{(0)}_n}V\ket{\phi^{(0)}_n}
\end{geqn}

The first order eigenvector can be written as, $\ket{\phi^{(1)}_n} = \sum_{m\neq n} c^{(n)}_m\ket{\phi^{(0)}_m}$. Substitute this into Eq.~\ref{si:SI_first_order} and project $\bra{\chi^{(0)}_k}$ with $k\neq n$ we get,
\begin{geqn}
   \bra{\chi^{(0)}_k} H_0 \ket{\sum_{m\neq n} c^{(n)}_m\phi^{(0)}_m} + \bra{\chi^{(0)}_k}V\ket{\phi^{(0)}_n} = \lambda_n^{(0)} \sum_{m\neq n} c^{(n)}_m\bra{\chi^{(0)}_k}\ket{\phi^{(0)}_m} + \lambda_n^{(1)}\bra{\chi^{(0)}_k}\ket{\phi^{(0)}_n}
\end{geqn}
Using biorthoigonality condition, $\bra{\chi^{(0)}_k}\ket{\phi^{(0)}_m} = \delta_{km}$ and noting that,$m\neq n$, the vanishes. Thus we obtain,
\begin{geqn}
    c^{(n)}_k(\lambda^{(0)}_k - \lambda_n^{(0)}) = - \bra{\chi^{(0)}_k}V\ket{\phi^{(0)}_n}\\
    c^{(n)}_k =\dfrac{\bra{\chi^{(0)}_k} V \ket{\phi^{(0)}_n}}{(\lambda^{(0)}_n - \lambda_k^{(0)})}
\end{geqn}

Therefore, the first-order correction of the right eigenvectors is given by,

\begin{equation}
\ket{\phi_n^{(1)}} = \sum_{m\neq n}\dfrac{\bra{\chi^{(0)}_m}V\ket{\phi^{(0)}_n}}{(\lambda^{(0)}_n - \lambda_m^{(0)})}\ket{\phi^{(0)}_m} \label{si:SI_first_oreder_eigen_vector}
\end{equation}

For second order eigenvalue correction, we project $\bra{\chi^{(0)}_n}$ in Eq.~\ref{si:SI_second_order} and substitute Eq.~\ref{si:SI_first_oreder_eigen_vector} we get the second order eigenvalue correction as,
\begin{equation}
    \lambda^{(2)}_n = \sum_{m\neq n}\dfrac{\bra{\chi^{(0)}_n}V\ket{\phi_m^{(0)}}\bra{\chi^{(0)}_m}V\ket{\phi_n^{(0)}}}{(\lambda^{(0)}_n- \lambda^{(0)}_m)}
\end{equation}

In our model, $H_0$ is the uncoupled matrix (when $\mu=0$) and given by,
\begin{equation}
     H_0 = \begin{bmatrix}
    -1 -b + c_1 & c_2 & 0 & 0 & 0\\
     b - c_1 & -(c_2 + D_0) & -E_0& 0& 0\\
     0 & 0 & -1/\tau_D & 0 & 0\\
     0 & 0 & 0 & -1/\tau_L & 0\\
     0 & 0 & 0& -iq/\tau_R & -q^2/\tau_R
    \end{bmatrix}
    \label{si:Eq:SIH0}  
\end{equation}
The coupled part in MCC acts as a perturbation matrix of the system, which is given as,
\begin{equation}
    V = \begin{bmatrix}
     0 & 0 & 0 & 0 & 0\\
     0 & 0 & 0 & 0& 0\\
     0 & 0 & 0 & 0 & -i\beta D_0q/\tau_D\\
     0 & -\alpha/\tau_L & 0 & 0 & 0\\
     0 & 0 & 0& 0 & 0
    \end{bmatrix}
    \label{si:Eq:SI_Vmat}  
\end{equation}

\subsection{The cubic term in the $\delta$ and $l$ equation}
The cubic term in Eq.(3) is introduced as the minimal non-linear saturation needed to keep the amplitudes of the variables finite in non-linear simulations. In the absence of such terms, the linear equations correctly capture the onset of oscillatory or unstable modes, but the amplitudes of $\delta$ and $l$ can grow without limit under sustained coupling. Adding a cubic contribution, e.g. $-(\delta + \delta^3)$ is the simplest symmetric regularization that (i) leaves the linear stability unchanged, (ii) corresponds to the gradient of the potential function, $V(\delta) = \frac{1}{2} \delta^2 + \frac{1}{4} \delta^4$. This choice follows standard practice in amplitude-equation theory (Landau expansion) and in related mechanochemical models \cite{hannezo_mechanochemical_2019}, where cubic saturation terms are used to regularize ERK dynamics. Furthermore, the nonlinear term in the potential function affects the quantitative location of the phase boundaries.

\section{Center Manifold Reduction of 1D HHS}

\subsection{Reformulation around the fixed point}

We start from the 1D continuum vertex model defined in Eq.~\eqref{si:Eq:SIcontvertex1D}. To make the mechanical coupling explicit, we introduce the local strain variable
\begin{equation}
L = \partial_x R,
\end{equation}
under which the mechanical equation becomes
\begin{equation}
\tau_R\partial_t L = \partial_x^2 (L - L_0).
\end{equation}

We expand the dynamics of a fixed point
$(M_0,E_0,D_0,1,1)$ and define the deviations
\begin{equation}
m = M - M_0, \quad
e = E - E_0, \quad
\delta = D - D_0, \quad
l = L_0 - 1, \quad
\Lambda = L - 1.
\end{equation}

Introducing the state vector
\begin{equation}
\mvec{X} = [m,e,\delta,l,\Lambda]^T,
\end{equation}
The fixed point corresponds to $\mvec{X} = 0$.

\begin{figure}[!t]
\centering
\includegraphics[scale=0.7]{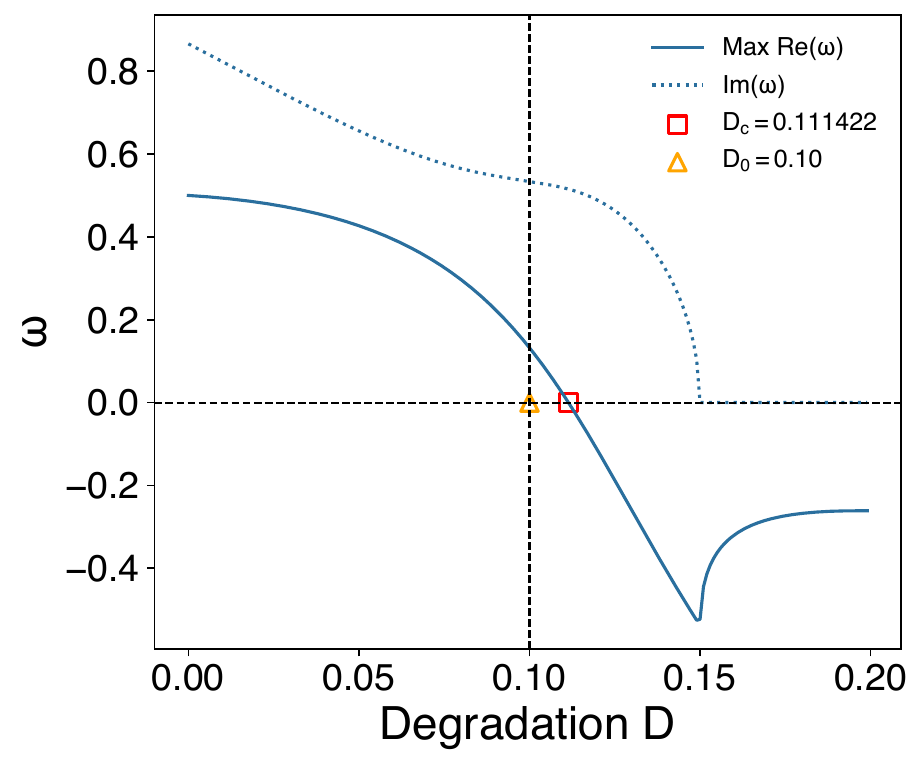}
\caption{Eigenvalues of the Jacobian of the local reaction subsystem as a function of the degradation rate $D$. The real part crosses zero at $D_c$, while the imaginary part remains finite, indicating a Hopf bifurcation. The operating point $D_0$ is also shown.}
\label{si:fig:SI_degrad_eigenspectrum}
\end{figure}

Linearizing the local reaction subsystem $(M,E,D)$ about this state,
we compute the eigenvalues of the Jacobian as a function of the operating
degradation rate $D_0$. As shown in Fig.~\ref{si:fig:SI_degrad_eigenspectrum}, the real part
of a complex conjugate pair crosses zero at a critical value $D_c$, while the imaginary part remains finite, indicating a Hopf bifurcation.

We therefore introduce the control parameter
\begin{equation}
\gamma = D_0 - D_c,
\end{equation}
which measures the distance from the Hopf point.

With these definitions, the dynamics can be written in the compact
reaction--diffusion form
\begin{equation}
\partial_t \mvec{X} = F(\mvec{X};\gamma) + D\,\partial_x^2 \mvec{X},
\end{equation}
where $F(\mvec{X};\gamma)$ collects the local nonlinear reaction terms and
$D$ encodes the mechanical diffusion coupling.

The reaction term is given by
\begin{equation}
F(\mvec{X};\gamma) =
\begin{pmatrix}
a - (b+1)(m+M_0) + c(m+M_0)^2(e+E_0) \\
b(m+M_0) - c(m+M_0)^2(e+E_0) - (D_0+\delta)(e+E_0) \\
\dfrac{1}{\tau_D}\left[-\delta - \delta^3 - \beta(D_0+\delta)\Lambda\right] \\
\dfrac{1}{\tau_L}\left[-l - l^3 - \alpha(e + e^3)\right] \\
0
\end{pmatrix}.
\end{equation}
while the diffusion operator is
\begin{equation}
D = \frac{1}{\tau_R}
\begin{pmatrix}
0 & 0 & 0 & 0 & 0 \\
0 & 0 & 0 & 0 & 0 \\
0 & 0 & 0 & 0 & 0 \\
0 & 0 & 0 & 0 & 0 \\
0 & 0 & 0 & -1 & 1
\end{pmatrix}.
\end{equation}

The operator $D$ acts only on the mechanical degree of freedom,
coupling the strain field $\Lambda$ to the rest-length variable $l$.

\subsection{Spectral Decomposition and Center Manifold}

We decompose the vector field as $F(\mvec{X},\gamma) = J\mvec{X} + f(\mvec{X},\gamma)$,
where $J = D_\mvec{X} F|_{\mvec{X}=0}$ and $f = \mathcal{O}(|\mvec{X}|^2)$.
Evaluating the Jacobian at the homogeneous fixed point $\mvec{X}=0$ and at the bifurcation point $\gamma=0$, we obtain
\begin{equation}
J =
\begin{pmatrix}
-(b+1) + 2c M_c E_c & c M_c^2 & 0 & 0 & 0 \\
b - 2c M_c E_c & -c M_c^2 - D_c & -E_c & 0 & 0 \\
0 & 0 & -\frac{1}{\tau_D} & 0 & -\frac{\beta D_c}{\tau_D} \\
0 & -\frac{\alpha}{\tau_L} & 0 & -\frac{1}{\tau_L} & 0 \\
0 & 0 & 0 & 0 & 0
\end{pmatrix}.
\end{equation}

The Jacobian has eigenvalues
\begin{equation}
  \lambda_{1,2} = \pm i\omega_c,
  \qquad
  \lambda_3 = 0,
  \qquad
  \mathrm{Re}(\lambda_{4,5}) < 0,
\end{equation}
corresponding to a Hopf pair, a neutral mode, and two stable modes.

\begin{figure}[!t]
  \centering
  \includegraphics[width=0.52\linewidth]{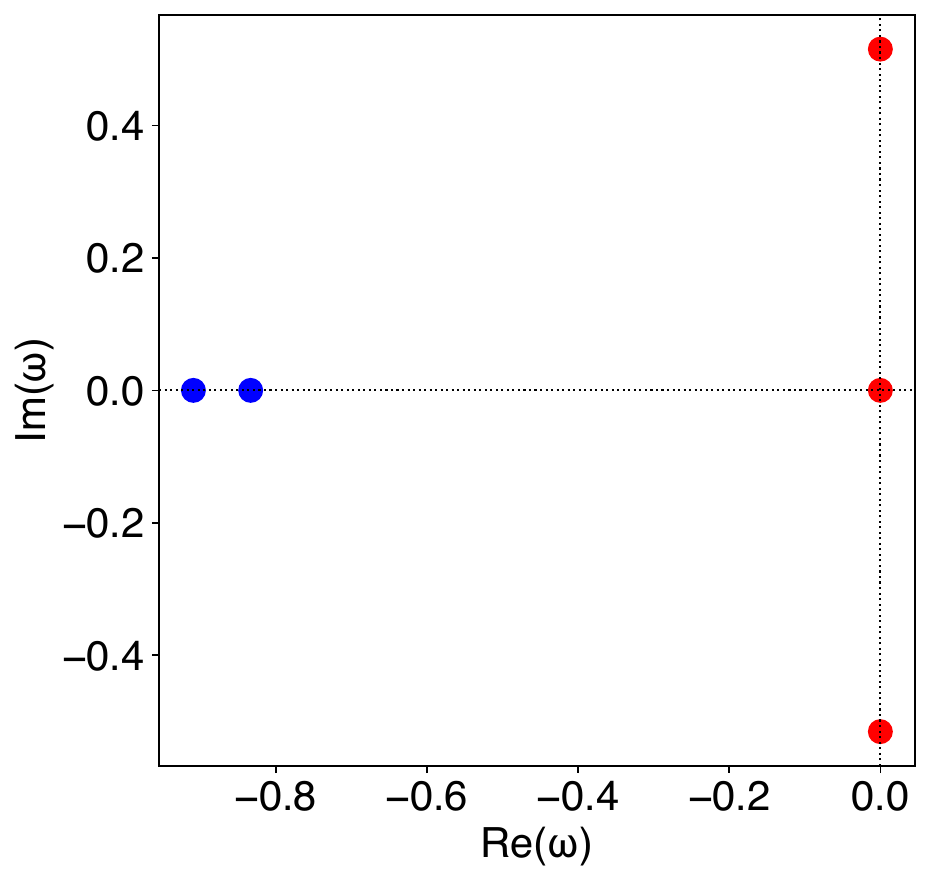}
  \caption{Eigenvalue spectrum of $J$ at $\gamma = 0$. Red:
    center eigenvalues ($\pm i\omega_c$, $0$). Blue:
    stable eigenvalues with $\mathrm{Re}\,\lambda < 0$.}
  \label{si:fig:hopf_fold_spectrum}
\end{figure}

The spectrum is shown in Fig.~\ref{si:fig:hopf_fold_spectrum}. The pair
$\lambda_{1,2} = \pm i\omega_c$ corresponds to the Hopf instability,
$\lambda_3 = 0$ is a Hopf-fold mode arising from the conservation of the
total length (or, equivalently, translational invariance of the strain
field), and the remaining eigenvalues lie in the left half-plane.

Let $|\mathbf{u}\rangle$, $|\bar{\mathbf{u}}\rangle$, and $|\mathbf{v}\rangle$ denote the right eigenvectors associated with $\lambda_{1,2,3}$, and let $\langle\mathbf{u}^*|$, $\langle\bar{\mathbf{u}}^*|$, and $\langle\mathbf{v}^*|$ denote the corresponding adjoint eigenvectors. These satisfy the biorthogonality condition
\begin{equation}
  \mathbf{u}_i^* \cdot \mathbf{u}_j = \delta_{ij}.
\end{equation}

The center subspace is
\begin{equation}
E_c = \mathrm{span}\{\mathbf{u},\bar{\mathbf{u}},\mathbf{v}\}.
\end{equation}

Any $y \in E_c$ can be written as
\begin{equation}
y = w\,\mathbf{u} + \bar{w}\,\bar{\mathbf{u}} + z\,\mathbf{v},
\end{equation}
where $w \in \mathbb{C}$ is the Hopf amplitude and $z \in \mathbb{R}$
is the neutral mode.

By the \textit{center manifold theorem}~\cite{Carr1981,Guckenheimer1983}, there
exists a locally invariant manifold $W_c$, tangent to $E_c$ at the
origin, such that all nearby trajectories approach $W_c$ and the
dynamics reduces to the flow on $W_c$.

We therefore seek a reduced description in which the full state $\mvec{X}$
is parameterized by the center variables $y \in E_c$. Since $W_c$ is
tangent to $E_c$ at the origin, any point on the center manifold can be
written as
\begin{equation}
\mvec{X} = y + h(y,\gamma),
\qquad y \in E_c,
\end{equation}
where $h : E_c \to E_c^\perp$ describes the components along the
stable directions. The tangency condition implies
\begin{equation}
h(0,\gamma)=0, \qquad D_y h(0,\gamma)=0,
\end{equation}
so that $h = \mathcal{O}(|y|^2)$.

Thus, to leading order, the dynamics is confined to $E_c$, while the
stable modes are slaved to $(w,\bar{w},z)$ through the nonlinear
correction $h$.

\subsection{Invariance Condition and Determination of $h$}

We seek the dynamics on $W_c$. The full system is
\begin{equation}\label{si:eq:full}
  \partial_t \mvec{X} = J\mvec{X} + f(\mvec{X},\gamma) + D\,\partial_x^2 \mvec{X}.
\end{equation}
We look for a reduced flow of the form
\begin{equation}\label{si:eq:reduced}
  \partial_t y = Jy + g(y,\gamma),
  \qquad g = \mathcal{O}(|y|^2),
\end{equation}
where $g$ is to be determined. To find both $h$ and $g$, we enforce
that $\mvec{X} = y + h(y,\gamma)$ satisfies~\eqref{si:eq:full} whenever $y$
satisfies~\eqref{si:eq:reduced}. Differentiating $\mvec{X} = y + h$ with respect
to $t$,
\begin{equation}
  \partial_t \mvec{X}
  = \partial_t y + \frac{\partial h}{\partial y}\,\partial_t y
  = \left(I + \frac{\partial h}{\partial y}\right)\partial_t y.
\end{equation}
Substituting $\partial_t y$ from~\eqref{si:eq:reduced} on the left and
$\mvec{X} = y + h$ into the right-hand side of~\eqref{si:eq:full},
\begin{equation}\label{si:eq:invariance}
  \left(I + \frac{\partial h}{\partial y}\right)
  \bigl(Jy + g(y,\gamma)\bigr)
  = J(y+h) + f(y+h,\gamma) + D\,\partial_x^2(y+h).
\end{equation}
The linear terms $Jy$ cancel from both sides. Expanding all remaining
functions in multi-index Taylor series~\cite{Ipsen2000},
\begin{equation}
  f(y{+}h,\gamma) = \sum_{\mathbf{p},q} f_{\mathbf{p}q}\,
  y^{\mathbf{p}}\gamma^q,
  \quad
  h = \sum_{\mathbf{p},q} h_{\mathbf{p}q}\, y^{\mathbf{p}}\gamma^q,
  \quad
  g = \sum_{\mathbf{p},q} g_{\mathbf{p}q}\, y^{\mathbf{p}}\gamma^q,
\end{equation}
where $\mathbf{p}=(p_1,p_2,p_3)\in\mathbb{N}_0^3$,
$y^{\mathbf{p}} = w^{p_1}\bar{w}^{p_2}z^{p_3}$, and
$|\mathbf{p}|+|q|\geq 2$ throughout. Collecting~\eqref{si:eq:invariance}
at each order $(\mathbf{p},q)$ yields~\cite[Eqs.~(12)--(13)]{Ipsen2000},
\begin{equation}\label{si:eq:gpq}
  g_{\mathbf{p}q}
  = \left(J - \sum_j p_j\lambda_j\,I\right)h_{\mathbf{p}q}
  + \Phi_{\mathbf{p}q},
\end{equation}
where
\begin{equation}\label{si:eq:Phi}
  \Phi_{\mathbf{p}q}
  = f_{\mathbf{p}q}
  - \sum_{\mathbf{p}',q'} h_{\mathbf{p}'q'}
    \sum_j p'_j\,\mathbf{u}_j^*\cdot
    g_{(\mathbf{p}-\mathbf{p}'+\bm{\delta}_j)(q-q')}
\end{equation}
here the sum over $\mathbf{p}',q'$ runs over all indices for which
$\mathbf{p} - \mathbf{p}' + \bm{\delta}_j \geq 0$ componentwise and
$q - q' \geq 0$; since $|\mathbf{p}'| \geq 2$ throughout, the
$g$-coefficients appearing on the right always carry total order
$|\mathbf{p}| - |\mathbf{p}'| + 1 \leq |\mathbf{p}| - 1$, making
\eqref{si:eq:gpq}--\eqref{si:eq:Phi} a well-posed order-by-order recurrence.

\subsection{Solvability and Resonance}

Projecting~\eqref{si:eq:gpq} onto the $i$-th adjoint eigenvector gives
\begin{equation}
  \textbf{u}_i^*\cdot g_{\mathbf{p}q}
  =
  \Bigl(\sum_j p_j\lambda_j - \lambda_i\Bigr)\,
  \textbf{u}_i^*\cdot h_{\mathbf{p}q}
  + \textbf{u}_i^*\cdot\Phi_{\mathbf{p}q}.
\end{equation}
Two cases arise.

\medskip
\noindent\textbf{Non-resonant} ($\sum_j p_j\lambda_j \neq \lambda_i$):
the prefactor is nonzero, so we choose
\begin{equation}
  \textbf{u}_i^*\cdot h_{\mathbf{p}q}
  = \frac{-\textbf{u}_i^*\cdot\Phi_{\mathbf{p}q}}
         {\lambda_i - \sum_j p_j\lambda_j},
\end{equation}
setting $\textbf{u}_i^*\cdot g_{\mathbf{p}q}=0$ and eliminating this term
from the reduced dynamics. For stable-subspace components of
$h_{\mathbf{p}q}$, the operator $J - \sum_j p_j\lambda_j I$ is always
invertible since $\mathrm{Re}(\lambda_{4,5})<0$.

\medskip
\noindent\textbf{Resonant} ($\sum_j p_j\lambda_j = \lambda_i$):
the prefactor vanishes for any $h_{\mathbf{p}q}$. We set
$\textbf{u}_i^*\cdot h_{\mathbf{p}q}=0$ and the term contributes
irreducibly,
\begin{equation}
  \textbf{u}_i^*\cdot g_{\mathbf{p}q} = \textbf{u}_i^*\cdot\Phi_{\mathbf{p}q}.
\end{equation}
This is the standard normal-form resonance
condition~\cite{Guckenheimer1983}. In compact form,
$h_{\mathbf{p}q}$ is determined by~\cite[Eq.~(19)]{Ipsen2000},
\begin{equation}\label{si:eq:hsolve}
  \left(J - \sum_j p_j\lambda_j\,I\right)h_{\mathbf{p}q}
  = -Q_{\mathbf{p}}\cdot\Phi_{\mathbf{p}q},
  \qquad
  R_{\mathbf{p}}\cdot h_{\mathbf{p}q} = 0,
\end{equation}
where $R_{\mathbf{p}}$ projects onto resonant center directions and
$Q_{\mathbf{p}} = I - R_{\mathbf{p}}$.

\subsection{Slow-Fast Decomposition and Projection}

To incorporate diffusion and isolate slow spatial modulation, we
follow~\cite{Ipsen1999} and separate fast oscillation
$\theta = \omega_c t$ from slow dynamics $\tau$. On $E_c$,
\begin{equation}
  y(x,t) = e^{J\theta}\,z(x,\tau),
  \qquad
  \partial_t y = \partial_\theta y + \partial_\tau y,
\end{equation}
where $z=(w,\bar{w},z)$ are slowly varying envelopes. Substituting
$\mvec{X} = y + h(y,\gamma)$ into the governing equation gives
\begin{equation}
  \partial_\tau \mvec{X}
  = (J\mvec{X} - \partial_\theta \mvec{X}) + f(\mvec{X},\gamma)
  + D\,\partial_x^2 \mvec{X}.
\end{equation}
Multiplying by $e^{-\lambda_i\theta}\textbf{u}_i^*$ and averaging over one
period $T = 2\pi/\omega_c$, two identities
eliminate the bracketed terms~\cite[App.~A--B]{Ipsen1999},
\begin{equation}
  \frac{1}{T}\int_0^T e^{-\lambda_i\theta}\,
  \textbf{u}_i^*\cdot(J\mvec{X}-\partial_\theta \mvec{X})\,d\theta = 0,
  \qquad
  \frac{1}{T}\int_0^T e^{-\lambda_i\theta}\,
  \textbf{u}_i^*\cdot\partial_\tau h\,d\theta = 0,
\end{equation}
the first because $e^{J\theta}$ exactly cancels the fast rotation, the
second because $h\in E_c^{\perp}$. Using
$y^{\mathbf{p}} = e^{\mathbf{p}\cdot\bm{\lambda}\,\theta}z^{\mathbf{p}}$,
the nonlinear average selects only resonant orders,
\begin{equation}
  \frac{1}{T}\int_0^T e^{-\lambda_i\theta}\,
  \textbf{u}_i^*\cdot f(\mvec{X},\gamma)\,d\theta
  =
  \sideset{}{^i}\sum_{\mathbf{p},q}
  (\textbf{u}_i^*\cdot f_{\mathbf{p}q})\,z^{\mathbf{p}}\gamma^q,
  \qquad
  \mathbf{p}\cdot\bm{\lambda} = \lambda_i,
\end{equation}
since all non-resonant integrals
$\frac{1}{T}\int_0^T e^{(\mathbf{p}\cdot\bm{\lambda}-\lambda_i)\theta}
d\theta$ vanish. The diffusion term contributes
$\sum_j d_{ij}\nabla^2 z_j$ with
$d_{ij}=\textbf{u}_i^*\cdot D\cdot\textbf{u}_j$~\cite[Eqs.~(17)--(20)]{Ipsen1999}.

\subsection{Amplitude Equation}
Combining the above, the reduced dynamics on $W_c$ is
\begin{equation}
  \partial_t z_i
  = \sideset{}{^i}\sum_{\mathbf{p},q}
    (\textbf{u}_i^*\cdot\Phi_{\mathbf{p}q})\,z^{\mathbf{p}}\gamma^q
  + \sum_j d_{ij}\nabla^2 z_j + \cdots,
\end{equation}
where only resonant $\Phi_{\mathbf{p}q}$ contribute. Applying the resonance condition to the spectrum ${\pm i\omega_c, 0}$: for $\lambda_1 = i\omega_c$, the condition $p_1 - p_2 = 1$ selects terms with one excess $w$ over $\bar{w}$; for $\lambda_3 = 0$, the condition $p_1 = p_2$ selects combinations of the form $|w|^{2k} z^m$. Although the expansion is, in principle, an infinite series, we truncate it at cubic order in $w$, which is sufficient to capture the leading-order dynamics. We have also verified that the coefficients of higher-order terms are smaller in magnitude compared to the retained terms. While these higher-order contributions may influence the dynamics, their effects are beyond the scope of the present work and will be explored in future studies. Thus, the amplitude equation can be written as, 

\begin{eqnarray}
  \partial_t w &=&d_w \lap{w}+ a_1\gamma w + a_2zw+ a_3z^2 w+a_4\,z^3 w+a_5\,w|w|^2
    + \cdots,\label{si:eq:amps_w} \\[4pt]
  \partial_t z &=& \lap{\left(d_z z + d_{ww}|w|^2 + \cdots\right)}.
    \label{si:eq:amps_z}
\end{eqnarray}
All coefficients are computed from
$\textbf{u}_i^*\cdot\Phi_{\mathbf{p}q}$ at the corresponding resonant orders
via~\eqref{si:eq:hsolve}, and their explicit values are given below.

\medskip

\begin{table}[H]
\caption{Coefficients of the reduced amplitude equations.}
\label{si:tab:coeff}
\begin{ruledtabular}
\renewcommand{\arraystretch}{1.5}
\begin{tabular}{lll}
Coefficient & Term & Expression \\
\hline
$a_1$ & $\gamma w$ 
& $-12.7864 - 2.24842\,i$ \\

$a_2$ & $z w$ 
& $\dfrac{(1.42469 + 0.250524\,i)\,\beta}
{\sqrt{1 + (0.777515 + 0.49562\,\alpha^2)\beta^2}}$ \\

$a_3$ & $z^2 w$ 
& $-\dfrac{(0.64333 - 0.437904\,i)\,\beta^2}
{1.28615 + (1 + 0.637441\,\alpha^2)\beta^2}$ \\

$a_4$ & $z^3 w$ 
& $-\dfrac{(0.284787 - 0.535231\,i)\,\beta^2}
{1.28615 + (1 + 0.637441\,\alpha^2)\beta^2}$ \\

$a_5$ & $w|w|^2$ 
& $-\dfrac{0.563631 - 0.383654\,i}{1.68608 + \alpha^2}$ \\

$d_w$ & $w$  
& $(-0.152283 + 0.380418\,i)\,\alpha\beta$ \\

$d_z$ & $z$  
& $1 - 0.704003\,\alpha\beta$ \\

$d_{ww}$ & $|w|^2$  
& $-\dfrac{5.57504\,\alpha\,\sqrt{1+(0.777515+0.49562\,\alpha^2)\beta^2}}
{1.68608+\alpha^2}$ \\
\end{tabular}
\end{ruledtabular}
\end{table}

\subsection{Time-scale separation and reduction to conserved dynamics}

From Eqs.~\eqref{si:eq:amps_w} and \eqref{si:eq:amps_z} we observe that the field $z$ evolves diffusively whereas $w$ follows purely local dynamics. This indicates a separation of timescales with $w$ relaxing much faster than $z$. We therefore employ a quasi-steady-state approximation for $w$.

We write the complex amplitude in polar form as $w = R e^{i\theta}$. Retaining leading-order terms the amplitude $R$ satisfies
\begin{equation}
\partial_t R = a_1 \gamma R + a_2 z R + a_3 z^2 R + a_4 z^3 R + a_5 R^3
\end{equation}

Assuming fast relaxation we set $\partial_t R = 0$ which gives
\begin{equation}
R^2 = - \frac{a_1 \gamma + a_2 z + a_3 z^2 + a_4 z^3}{a_5}
\end{equation}

Substituting this into the equation for $z$ we obtain
\begin{equation}
\partial_t z = \nabla^2 \left( b_1 \gamma + b_2 z + b_3 z^2 + b_4 z^3 \right)
\end{equation}

where the coefficients are
\begin{equation}
b_1 = - \frac{a_1}{a_5} d_{ww}
\end{equation}
\begin{equation}
b_2 = d_z - \frac{a_2}{a_5} d_{ww}
\end{equation}
\begin{equation}
b_3 = - \frac{a_3}{a_5} d_{ww}
\end{equation}
\begin{equation}
b_4 = - \frac{a_4}{a_5} d_{ww}
\end{equation}

The evolution equation can be expressed in the form of a conserved dynamics,
\begin{equation}
\partial_t z + \nabla \cdot \mathbf{J} = 0 ,
\end{equation}
where the associated current is
\begin{equation}
\mathbf{J} = - \nabla \mu_z ,
\end{equation}
and the chemical potential is defined through the variational derivative
\begin{equation}
\mu_z = \frac{\delta \mathcal{L}[z]}{\delta z} .
\end{equation}

The corresponding Lyapunov functional is
\begin{equation}
\mathcal{L}[z]=\int dx \left[b_1 \gamma z+ \frac{b_2}{2} z^2+ \frac{b_3}{3} z^3+ \frac{b_4}{4} z^4\right]
\end{equation}

Its temporal evolution is obtained as
\begin{eqnarray}
\frac{d\mathcal{L}}{dt}&=&\int dx \frac{\delta \mathcal{L}}{\delta z}\,\partial_t z\nonumber\\
&=&\int dx \mu_z \partial_t z =-\int dx \mu_z \nabla^2 \mu_z=-\int dx |\nabla \mu_z|^2\le 0 
\end{eqnarray}

Here integration by parts has been performed assuming periodic boundary conditions. Therefore, the functional $\mathcal{L}$ decreases monotonically with time and acts as a Lyapunov functional for the dynamics.

Under periodic boundary conditions, the total mass of the real mode is conserved,
\begin{equation}
\int z \, dx = 0 .
\end{equation}

The dynamics is controlled by the local free-energy density
\begin{equation}
l(z)
=
b_1 \gamma z
+ \frac{b_2}{2} z^2
+ \frac{b_3}{3} z^3
+ \frac{b_4}{4} z^4 .
\end{equation}

As the parameters are varied, the local potential develops a double-well structure when
\begin{equation}
l''(0) < 0 ,
\end{equation}
signaling the linear instability of the homogeneous state.

Because the dynamics decreases $\mathcal{L}$ while preserving the conserved quantity
\begin{equation}
\int z \, dx = 0 ,
\end{equation}
the system cannot relax toward a spatially uniform configuration. Instead, it approaches states that minimize $\mathcal{L}$ subject to the conservation constraint.

For a non-convex free-energy density $l(z)$, the minimizing configurations correspond to coexistence between domains near the two minima of the potential, leading to phase separation.
We further observe that one branch satisfies
\begin{equation}
R^2 > 0
\end{equation}

corresponding to sustained oscillations while the other satisfies
\begin{equation}
R^2 < 0
\end{equation}

where only the solution $R = 0$ is allowed. This leads to coexistence of oscillatory and non oscillatory states characteristic of oscillation death.

\subsection{Generality of the reduced description}

Although the derivation above was carried out for the Harmonic Brusselator model, the resulting mechanism is not specific to the details of the underlying chemical kinetics. Near a Hopf bifurcation, a generic oscillatory reaction system reduces, under center manifold reduction, to the universal Stuart--Landau equation governing the complex Hopf amplitude.

The essential effect of the mechanochemical feedback is the introduction of an additional slow neutral mode associated with the conserved mechanical degree of freedom. Consequently, the dynamics is governed by coupled amplitude equations for the oscillatory mode $w$ and the conserved field $z$. The conserved field locally modulates the effective distance from the Hopf bifurcation, leading to phase separation between oscillatory and non-oscillatory states.

We therefore expect the emergence of oscillation-death patterns to be robust across a broad class of Hopf oscillators provided the symmetry and conservation structure of the coupling are preserved.

\subsection{Phase diagram construction}
The phase diagram in Fig.~3(C) is constructed from the reduced amplitude equations by combining the amplitude condition $R^2(z)$ with the effective potential $l(z)$. For each set of parameters, the classification is determined using the following criteria:

\begin{itemize}
\item \textbf{Amplitude condition:}  
We evaluate $R^2(z)$ over the range of $z$. If $R^2>0$ for all $z$, the system is classified as oscillatory. If $R^2<0$ for all $z$, only the solution $R=0$ is allowed, corresponding to amplitude death. If $R^2$ changes sign, coexistence between oscillatory and non-oscillatory states is possible.

\item \textbf{Energetic selection:}  
In the coexistence regime, we restrict attention to the regions of $z$ where each branch is allowed and evaluate the local Lyapunov density $l(z)$ within those regions. The corresponding states are retained only if $l(z)$ attains minima at finite $z$ with values lower than $l(0)$, ensuring that they are energetically favorable.

\item \textbf{Instability condition:}  
We additionally require $l''(0)<0$, so that the homogeneous state is linearly unstable and small perturbations grow.
\end{itemize}

Together, these criteria determine the boundaries between oscillatory, amplitude death, and oscillation death regimes shown in Fig.~3(C).

\begin{figure}[!htbp]
  \centering
  \includegraphics[width=0.52\linewidth]{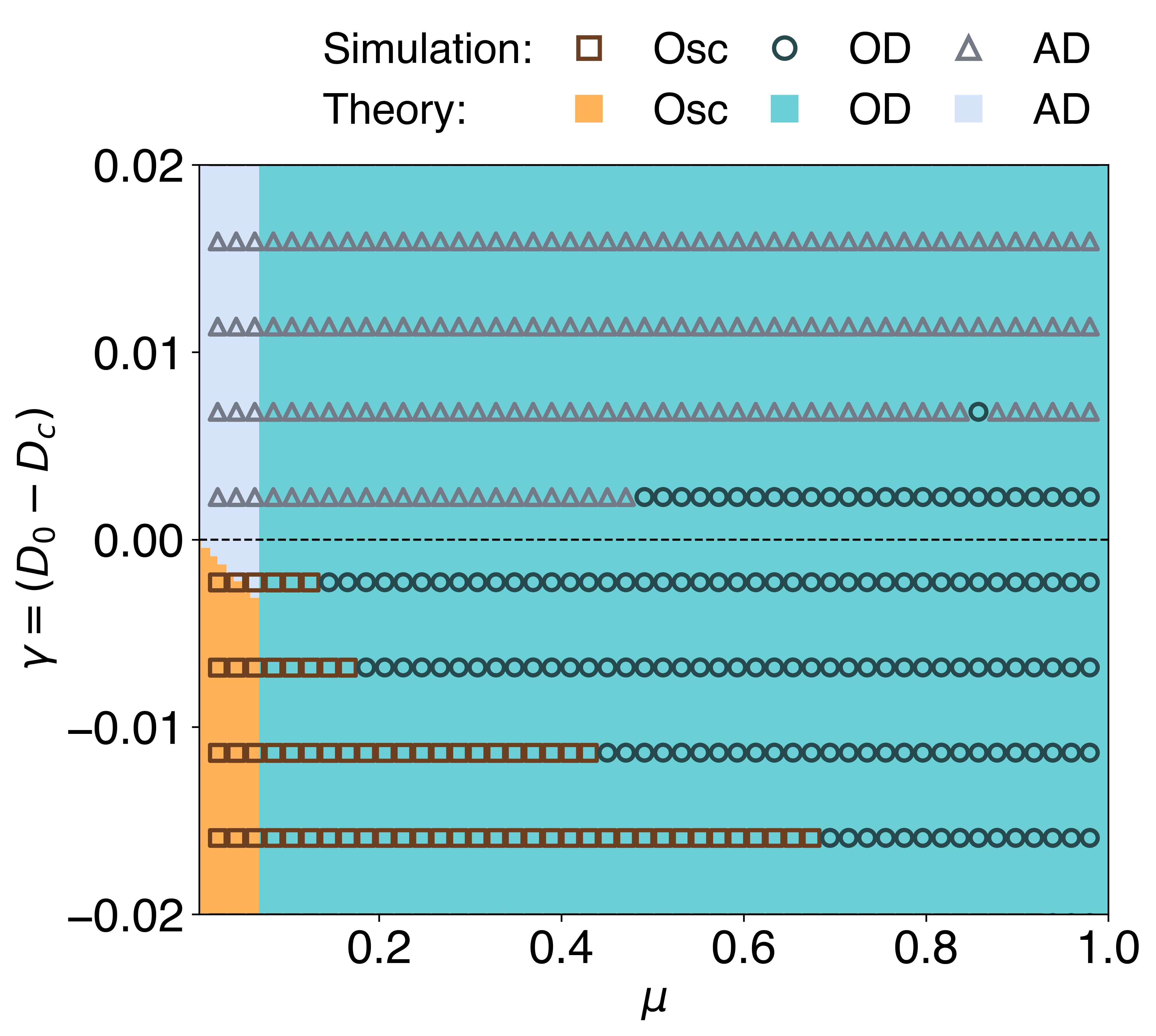}
  \caption{Comparison between the
analytical prediction (heatmap) obtained from the amplitude
equation and the corresponding simulation results (markers)}
  \label{si:fig:phase_plane_all_mu}
\end{figure}

\section{Phase turbulence}

To analyze the weakly nonlinear oscillatory regime, we consider the
long-wavelength dynamics of the Hopf amplitude $w$ obtained from the
reduced amplitude equations derived in the previous section. In the
weak-coupling regime, fluctuations of the conserved mode $z$ remain
small, and the leading-order dynamics of the oscillatory mode reduces
to the complex Ginzburg--Landau equation
\begin{equation}
\partial_t w
=
d_w \nabla^2 w
+
(\lambda_r+i\lambda_i)w
+
(g_r+i g_i)|w|^2w,
\label{si:eq:cgle_generic}
\end{equation}
where
\begin{equation}
d_w=d_r+i d_i,
\qquad
g=g_r+i g_i .
\end{equation}

Equation~\eqref{si:eq:cgle_generic} is the universal amplitude equation
governing oscillatory media near a Hopf bifurcation
\cite[Chs.~4--6]{kuramoto1984chemical}. Following the standard phase
reduction procedure~\cite[Ch.~7]{kuramoto1984chemical}, we write the
complex amplitude in polar form,
\begin{equation}
w(x,t)=R(x,t)e^{i\Theta(x,t)}.
\end{equation}

For weak spatial modulation, the amplitude $R$ relaxes rapidly to its
local limit-cycle value, while the phase $\Theta$ evolves on a much
slower timescale. Eliminating the fast amplitude fluctuations yields
the nonlinear phase diffusion equation
\begin{equation}
\partial_t \Theta
=
\omega_0
+
D_\Theta \nabla^2 \Theta
+
N_\Theta (\nabla \Theta)^2
+\cdots,
\label{si:eq:phase_eq}
\end{equation}
where
\begin{equation}
\omega_0
=
\lambda_i
-
\lambda_r\frac{g_i}{g_r},
\end{equation}
and
\begin{equation}
D_\Theta
=
d_r
+
\frac{g_i}{g_r}d_i,
\qquad
N_\Theta
=
-d_i
+
\frac{g_i}{g_r}d_r.
\end{equation}

Higher-order gradient terms appear at next order in the long-wave
expansion and regularize short-wavelength modes, but are not required
for determining the onset of the instability.

The stability of the spatially homogeneous oscillatory state is
controlled by the sign of the phase diffusion coefficient
$D_\Theta$. When
\begin{equation}
D_\Theta<0,
\label{si:eq:BF_condition}
\end{equation}
long-wavelength phase perturbations grow, corresponding to the
Benjamin--Feir instability and the onset of phase turbulence
\cite[Ch.~7]{kuramoto1984chemical}.

For sufficiently weak coupling $\mu$, the system remains globally
oscillatory but develops irregular spatial fluctuations in the phase of
the Hopf mode. To characterize this regime, we project the full
dynamics onto the critical Hopf eigenspace using the adjoint
eigenvectors obtained in the center manifold reduction and extract the
slow phase field $\theta(x,t)$ after subtracting the uniform Hopf
rotation.

Following Ref.~\cite[Ch.~7]{kuramoto1984chemical}, we define the
stationary phase fluctuation spectrum
\begin{equation}
S(k)=\left\langle |\psi_k|^2 \right\rangle,
\end{equation}
where
\begin{equation}
\psi_k(t)
=
\frac{1}{L}
\int_0^L
dx\,\theta(x,t)e^{ikx},
\end{equation}
and $\langle\cdot\rangle$ denotes a long-time average over the
statistically stationary state.

\begin{figure}[!htbp]
    \centering
    \includegraphics[width=0.49\linewidth]{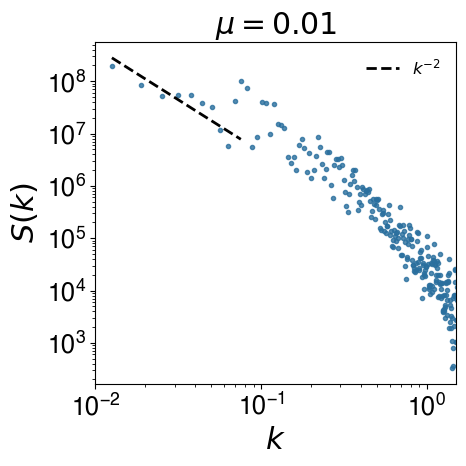}
    \includegraphics[width=0.49\linewidth]{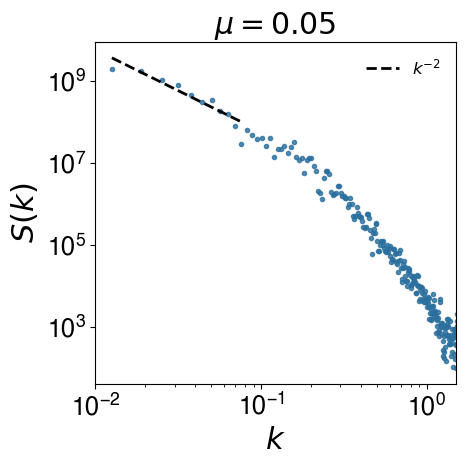}
    \caption{Stationary phase fluctuation spectrum $S(k)$ for weak
    coupling strengths $\mu=0.01$ and $\mu=0.05$. The dashed line
    indicates the scaling $S(k)\sim k^{-2}$ at small wavenumbers.}
    \label{si:fig:SI_phase_turbulence}
\end{figure}

The spectra obtained from direct numerical simulations are shown in
Fig.~\ref{si:fig:SI_phase_turbulence}. At low wavenumbers we observe the
approximate scaling
\begin{equation}
S(k)\sim k^{-2},
\end{equation}
consistent with the long-wavelength phase turbulence spectrum of the
Kuramoto--Sivashinsky phase equation
\cite[Ch.~7]{kuramoto1984chemical}.

To interpret this scaling, we introduce the phase-gradient field
\begin{equation}
v(x,t)=\partial_x\theta(x,t).
\end{equation}
In Fourier space,
\begin{equation}
v_k = ik\psi_k,
\end{equation}
so that
\begin{equation}
\left\langle |v_k|^2 \right\rangle
=
k^2S(k)
\sim \mathrm{const.}
\end{equation}
for small $k$. Thus, the phase-gradient fluctuations are approximately
white in space,
\begin{equation}
\langle v(x)v(x')\rangle
\sim
\delta(x-x').
\end{equation}

The phase field may therefore be expressed as
\begin{equation}
\theta(x)-\theta(0)
=
\int_0^x v(s)\,ds,
\end{equation}
from which the mean-square phase fluctuation becomes
\begin{align}
\left\langle
|\theta(x)-\theta(0)|^2
\right\rangle
&=
\int_0^x\int_0^x
\langle v(s)v(s')\rangle
\,ds\,ds' \\
&\sim
\int_0^x\int_0^x
\delta(s-s')
\,ds\,ds' \\
&\sim x.
\end{align}

Thus, the phase undergoes random-walk-like spatial wandering,
resulting in the loss of long-range phase coherence characteristic of
phase turbulence.

\section{Generality of the result: Testing the predicted patterns using the Fitzhugh-Nagumo model}
In the main text, we have explored the pattern formation using a Brusselator as a chemical oscillator. However, it is a well-established result that for a 1D periodic system consisting of coupled Hopf oscillators, the emergent patterns do not depend on the details of the Hopf oscillator as long as the symmetry of the coupling is preserved. This result implies that as long as we have a Hopf oscillator that can be driven to oscillation death by a degrader molecule, we will reproduce the observed collective behavior. To test the robustness and universality of our conclusions, we have repeated our analysis using the FitzHugh–Nagumo (FHN) model, a canonical representative of Hopf-type oscillators. The mechanochemical feedback-modified FHN system is described by,
\begin{eqnarray}
    \partial_t M &=& M - \frac{M^3}{3} -E + R I, \label{si:FHN0:1}\\
    \tau_E \partial_t E &=& M + a -bE, \label{si:FHN0:2}\\
    \tau_B \partial_t b &=& - \{(b-b_0) + (b-b_0)^3\} - \beta b (\partial_x R - 1), \\
    \tau_L \partial_t L_0 &=& - \{(L_0 - 1)+(L_0 - 1)^3\} - \alpha (E - E_0), \\
    \tau_R \partial_t R &=& \partial_x^2 R - \partial_x L_0,
\end{eqnarray}
with $\tau_E = 12.5, a = 0.7, R = 0.1$. $I$, $b_0$, $\tau_B$, $\tau_L$, and $\tau_R$ are free parameters. We set $\tau_R = 1$ to set the timescale of our simulations. The simulation was conducted for $L = 200$ and a timescale of 25,000$\tau_R$ using periodic boundary conditions.

\begin{figure}[!htbp]
    \centering
    \includegraphics[width=\textwidth]{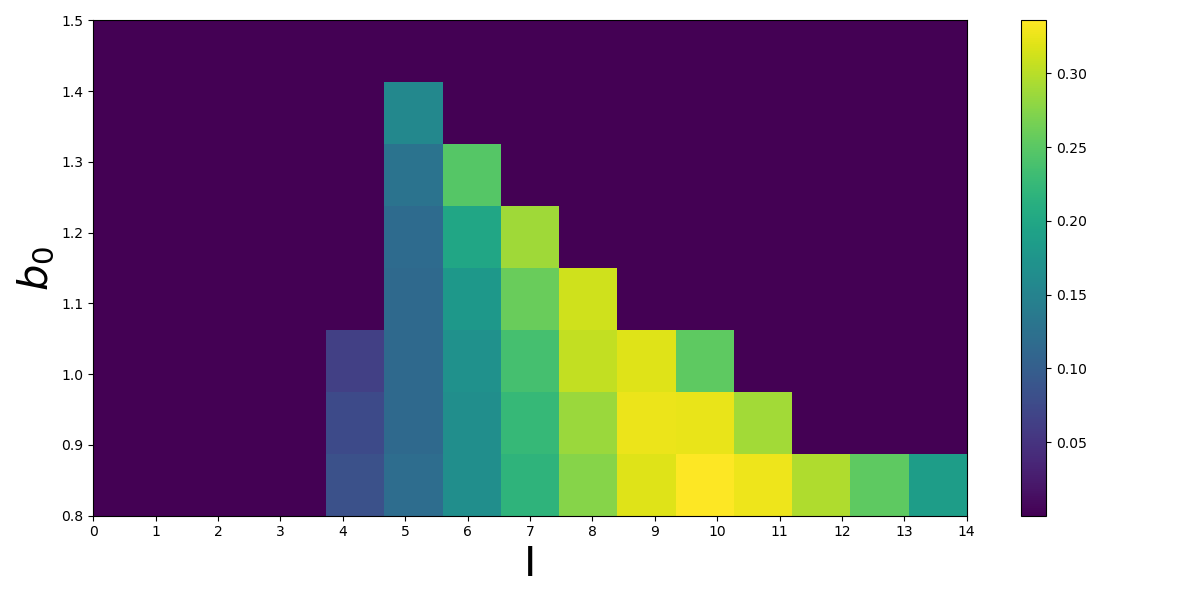}
    \caption{Variance map of $E$ for $b_0$ versus $I$ without MCFL ($\alpha =\beta =0$). The regions of high variance correspond to strong oscillatory activity in the FitzHugh–Nagumo system}
    \label{si:fig:SI_FHN_Varience}
\end{figure}
We first analyzed the system without the mechanochemical feedback to identify the boundaries of Hopf bifurcation, and to choose representative values of $b_0$ and $I$ (Eqs.~\ref{si:FHN0:1} and \ref{si:FHN0:2}). To do so, we set $\alpha = \beta = 0$, $b = b_0$, and $L_0 = 1$, and note that in the absence of noise, a stable fixed point will have zero variance, but an oscillatory state will have finite variance. We use this fact to draw the locations of the Hopf bifurcation. Specifically, we compute the variance of $E$ for different values of $b_0$ and $I$ (Fig.~\ref{si:fig:SI_FHN_Varience}). From this variance landscape, representative values of $b_0$ and $I$ are identified and subsequently used to explore the full feedback-coupled dynamics by scanning $\alpha$ and $\beta$. In line with our choice of $D_0$, we picked $b_0 = 1.3$ and $I = 5$ to ensure that the system remains close to a bifurcation point. However, we observe the same phenomenology for other $b_0$ and $I$ combinations.

\begin{figure}[!htbp]
    \centering
    \includegraphics[width=\textwidth,keepaspectratio]{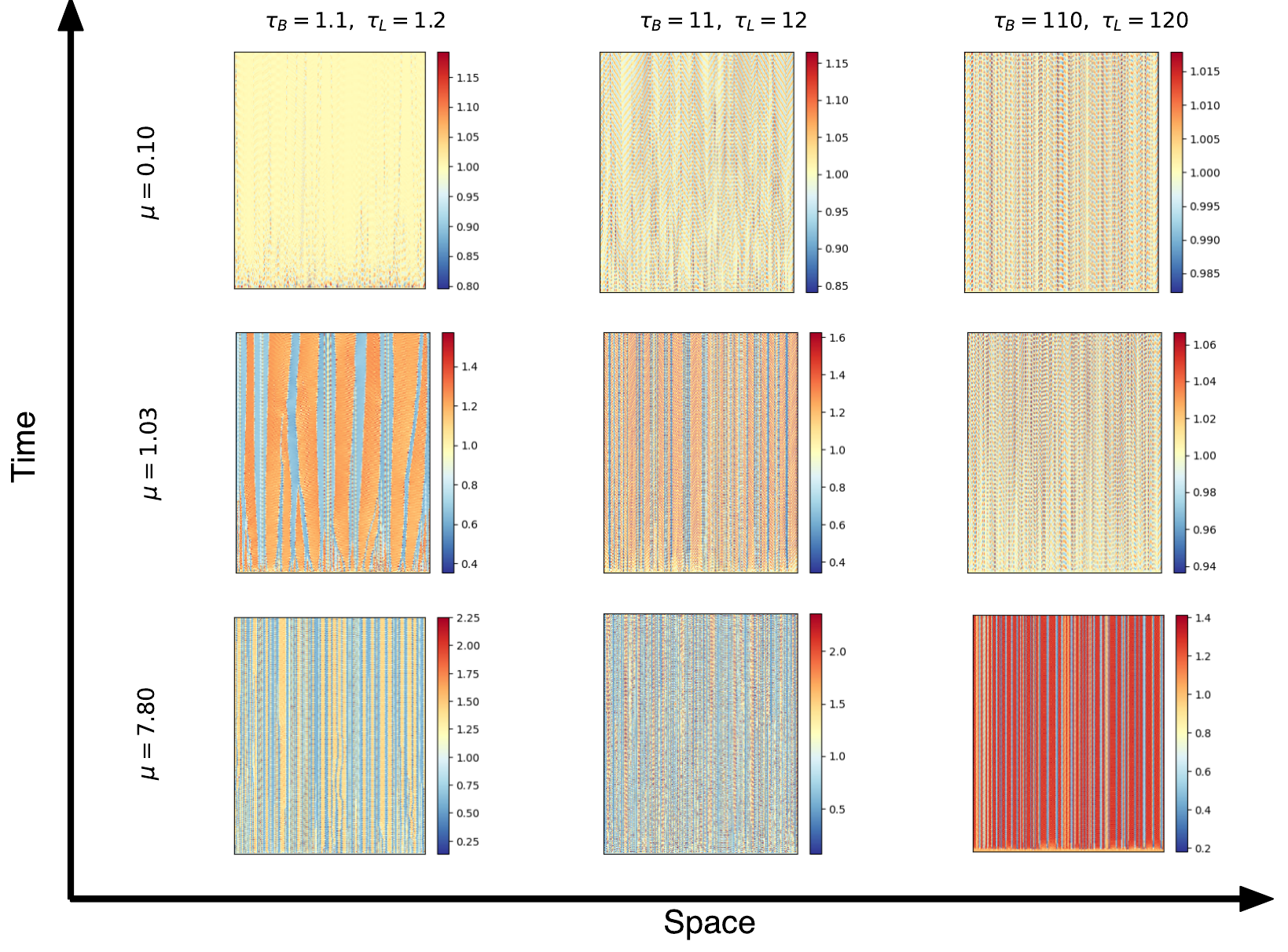}
    \caption{The kymograph of the spring length ($\partial_x R$) for the FHN model is shown for different values of $\mu=0.1,1.03,7.80$ (top to bottom) and $\tau_B = 1.1$, $\tau_L = 1.2$ (left column); $\tau_B = 11$, $\tau_L = 12$ (middle column); $\tau_B = 110$, $\tau_L = 120$ (right column). The X axis represents the space, and the Y axis represents the time in the figure.}
    \label{si:fig:Rev_FHN_Kymo_X}
\end{figure}
\begin{figure}[!htbp]
    \centering
    \includegraphics[width=\textwidth,keepaspectratio]{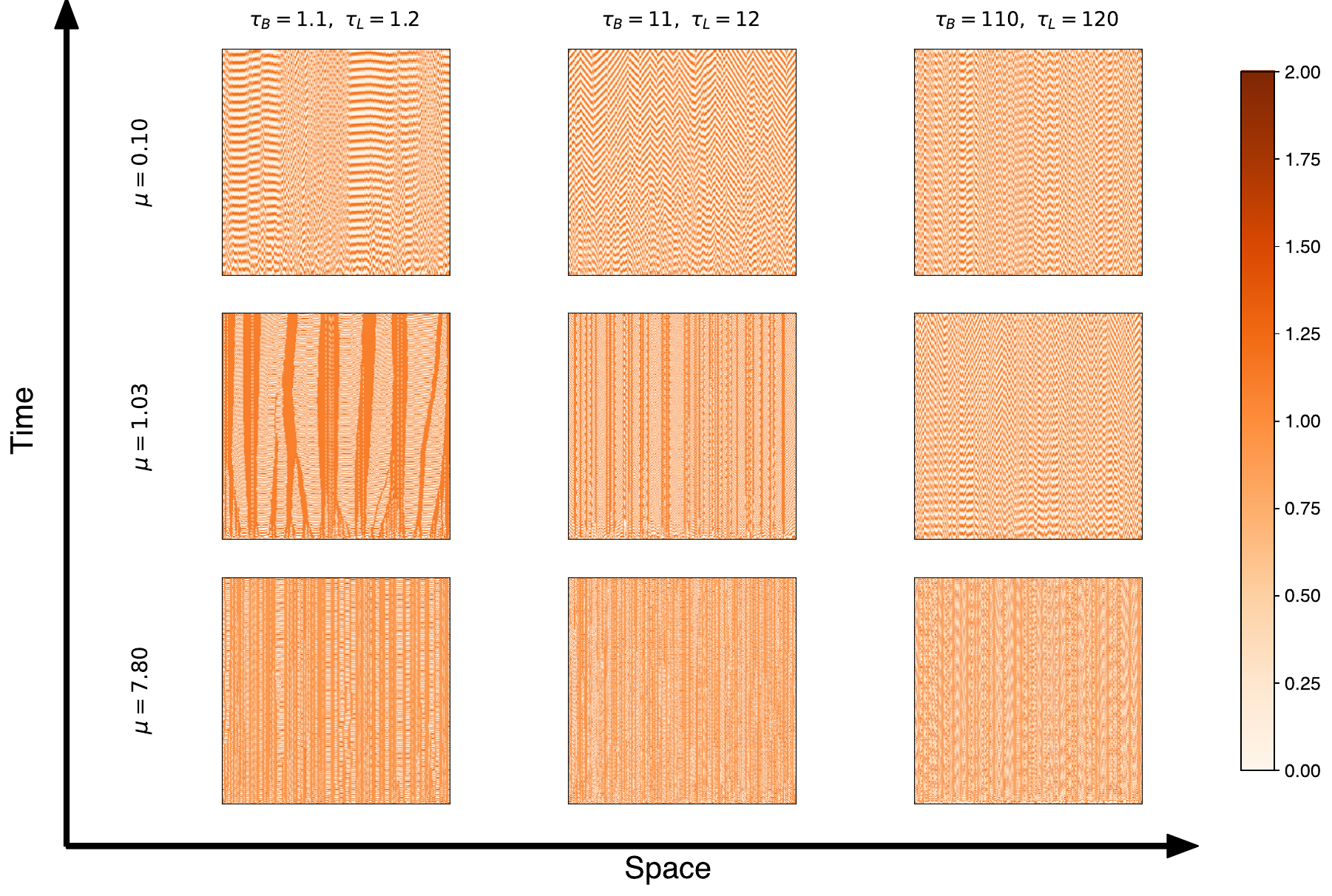}
    \caption{The kymograph of $E$ for the FHN model is shown for different values of $\mu=0.1,1.03,7.80$ (top to bottom) and $\tau_B = 1.1$, $\tau_L = 1.2$ (left column); $\tau_B = 11$, $\tau_L = 12$ (middle column); $\tau_B = 110$, $\tau_L = 120$ (right column). The X axis represents the space, and the Y axis represents the time in the figure.}
    \label{si:fig:Rev_FHN_Kymo_E}
\end{figure}

\begin{figure}[htbp]
    \centering
\includegraphics[width=0.9\paperwidth,height=0.9\paperheight,keepaspectratio]{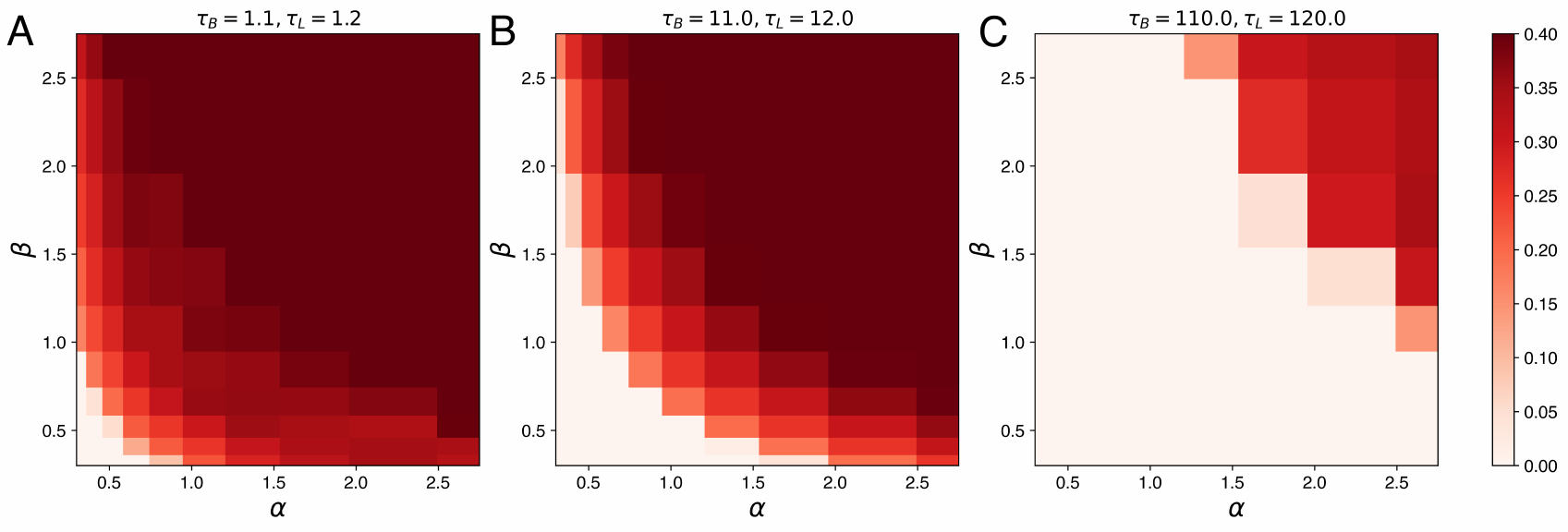}
    \caption{The total size of regions with collective oscillation death as the fraction of system size with FHN model (A) $\tau_B = 1.1$, $\tau_L = 1.2$; (B) $\tau_B = 11$, $\tau_L = 12$; (C) $\tau_B = 110$, $\tau_L = 120$.}
    \label{si:fig:Rev_Static_region}
\end{figure}

\begin{figure}[htbp]
    \centering
\includegraphics[width=0.9\paperwidth,height=0.9\paperheight,keepaspectratio]{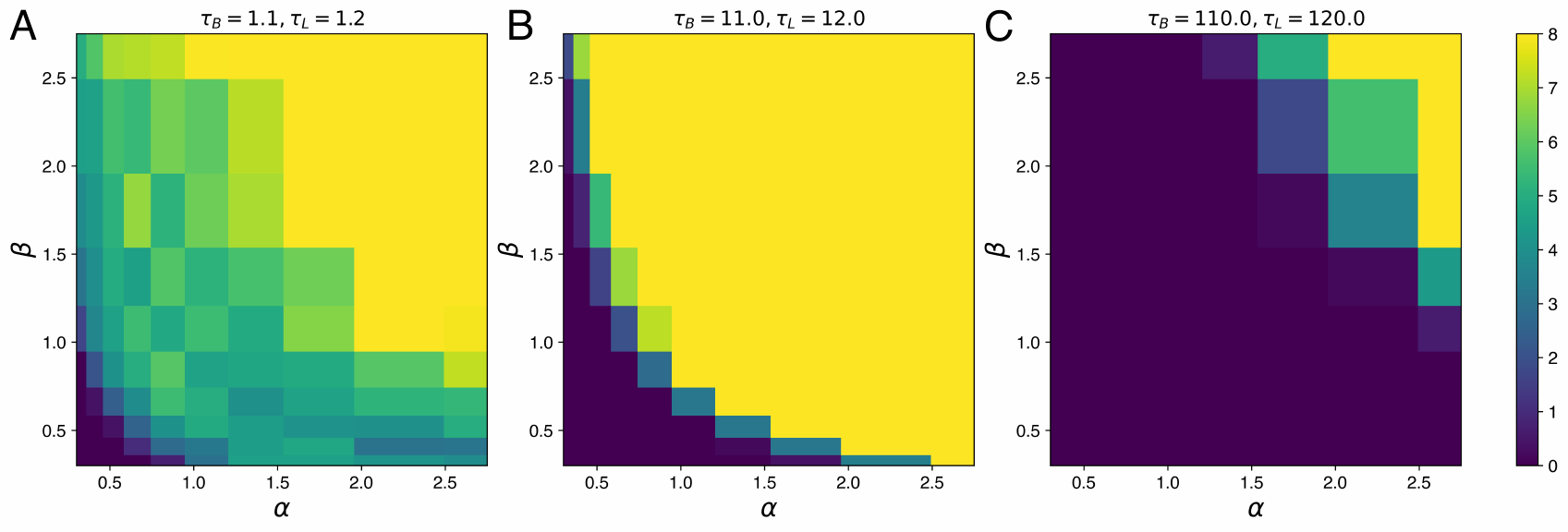}
    \caption{The total number of regions with collective oscillation death for the FHN model.  (A) $\tau_B = 1.1$, $\tau_L = 1.2$; (B) $\tau_B = 11$, $\tau_L = 12$; (C) $\tau_B = 110$, $\tau_L = 120$.}
    \label{si:fig:Rev_Static_number}
\end{figure}

Below, we summarize the behavior of the model for different $\tau_B$ and $\tau_L$ values. Briefly, we reproduce all the phases observed in the Harmonic Brusselator model. From the kymograph of the spring length ($\partial_x R$,  Fig~\ref{si:fig:Rev_FHN_Kymo_X}) and $E$ (Fig.~\ref{si:fig:Rev_FHN_Kymo_E}), it is clear that the oscillation death and other patterns are present when we use the FHN model as a chemical oscillator. Moreover, the size (Fig.~\ref{si:fig:Rev_Static_region}) and the number (Fig.~\ref{si:fig:Rev_Static_number}) of the regions with oscillation death changes in a manner similar to what we observe in the HHS model. This is expected from the group theoretic argument~\cite{collins_coupled_1993,collins_group-theoretic_1994}, which states that all 1D ring of Hopf oscillators have universal behavior as long as the nature of the coupling is preserved, which we have done here. Please note that, although we recover all the observed patterns in the FHN model, the patterns observed at a specific value of $\tau_L$ and $\tau_D$ differ in the two models. This may stem from a nonlinear mapping between the models, exploring which is beyond the scope of this paper.     
  

\newpage
\section{The effect of noise in the Harmonic Hopf Solid}
\begin{figure}[!htbp]
    \centering
    \includegraphics[width=\textwidth]{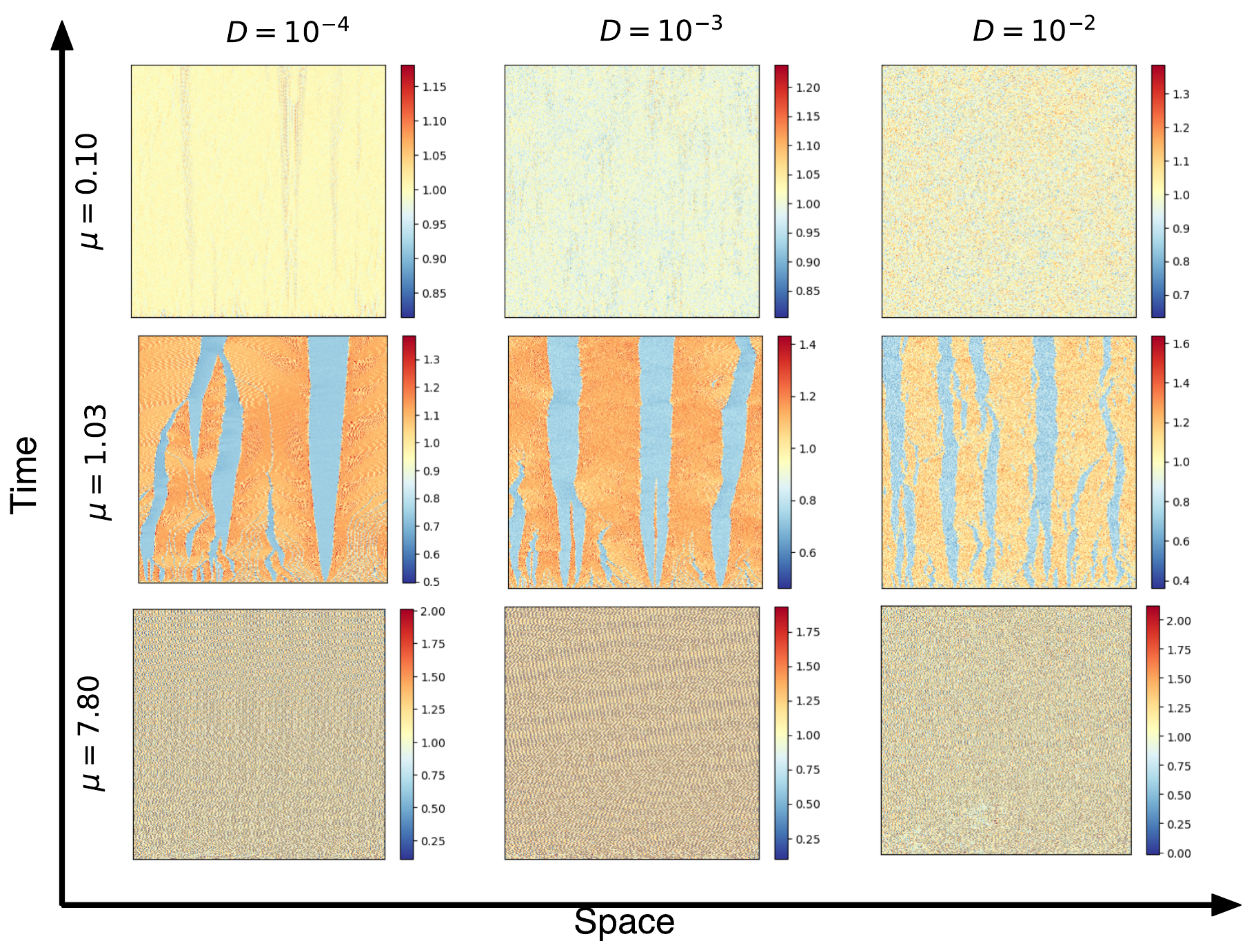}
    \caption{The kymograph of the spring length ($\partial_x R$) shows the effect of noise for different noise-strength  ($D$): $D=10^{-4}$ $D=10^{-3}$, and $D=10^{-2}$ (left to right) for different values of $\mu=0.10,1.03 \text{and} 7.80$ (top to bottom). The X-axis represents the space, and the Y-axis represents the time in the figure.}
    \label{si:fig:SI_kymo_noise_X}
\end{figure}

\begin{figure}[!htbp]
    \centering
    \includegraphics[width=\textwidth]{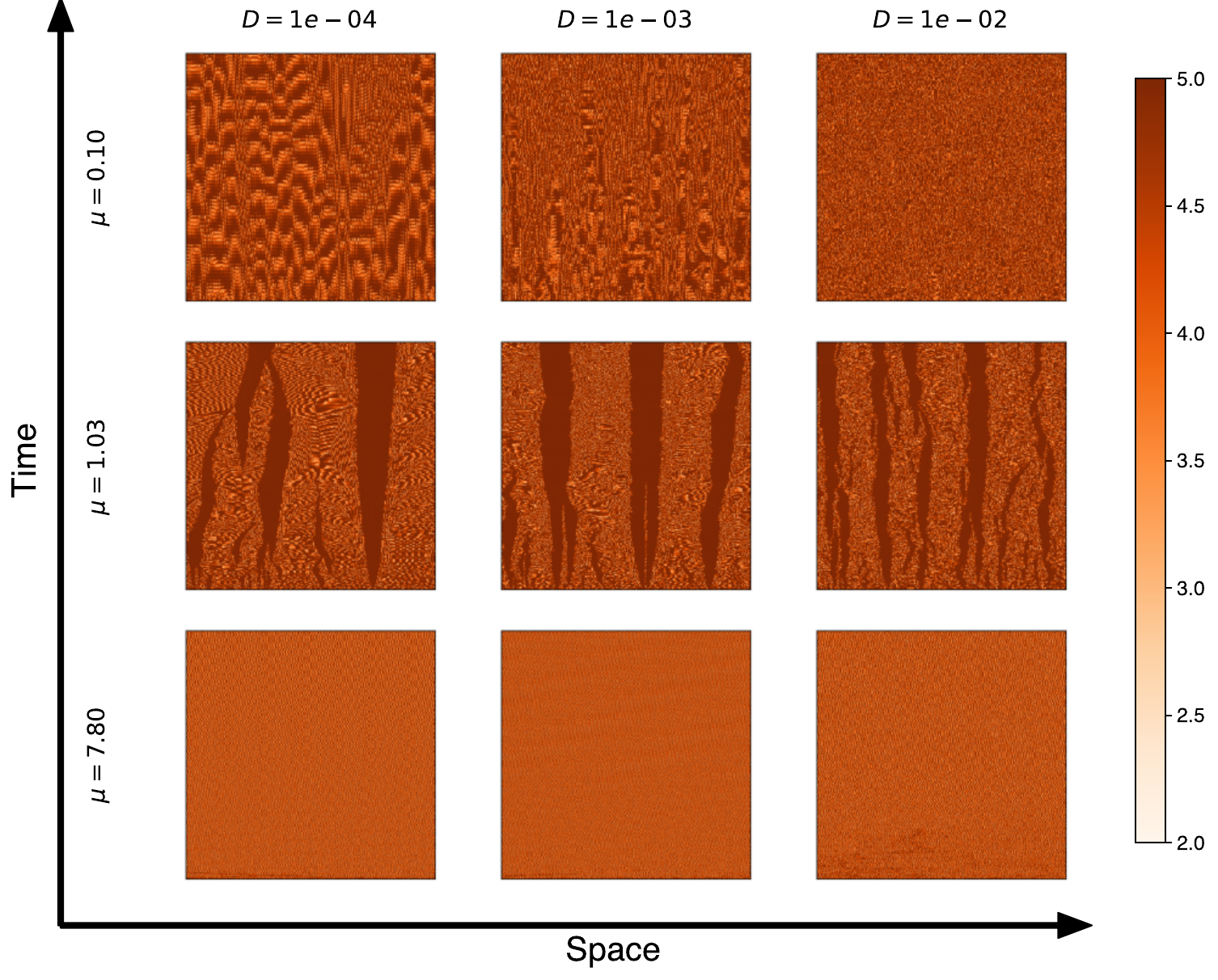}
    \caption{The kymograph of the ERK field ($E$) shows the effect of noise for different noise-strength  ($D$):   $D=10^{-4}$ $D=10^{-3}$, and $D=10^{-2}$ (left to right) for different values of $\mu=0.10,1.03 \text{and} 7.80$ (top to bottom). The X-axis represents the space, and the Y-axis represents the time in the figure.}
    \label{si:fig:SI_kymo_noise_E}
\end{figure}

Below we provide the kymograph of the spring length ($\partial_x R$, Fig.~\ref{si:fig:SI_kymo_noise_X}) and ERK ($E$, Fig.~\ref{si:fig:SI_kymo_noise_E}) to show the effect of the noise in the system. As is apparent, the non-Hermitian patterns, such as COD is robust even at extremely high noise values ($2D/\mbox{var}(\partial_xR) > 100$), whereas system-spanning traveling waves are easily affected by noise. 


\section{Trimodal cell length}
\begin{figure}[!htbp]
    \centering
    \includegraphics[width=\textwidth,height=0.6\paperheight,keepaspectratio]{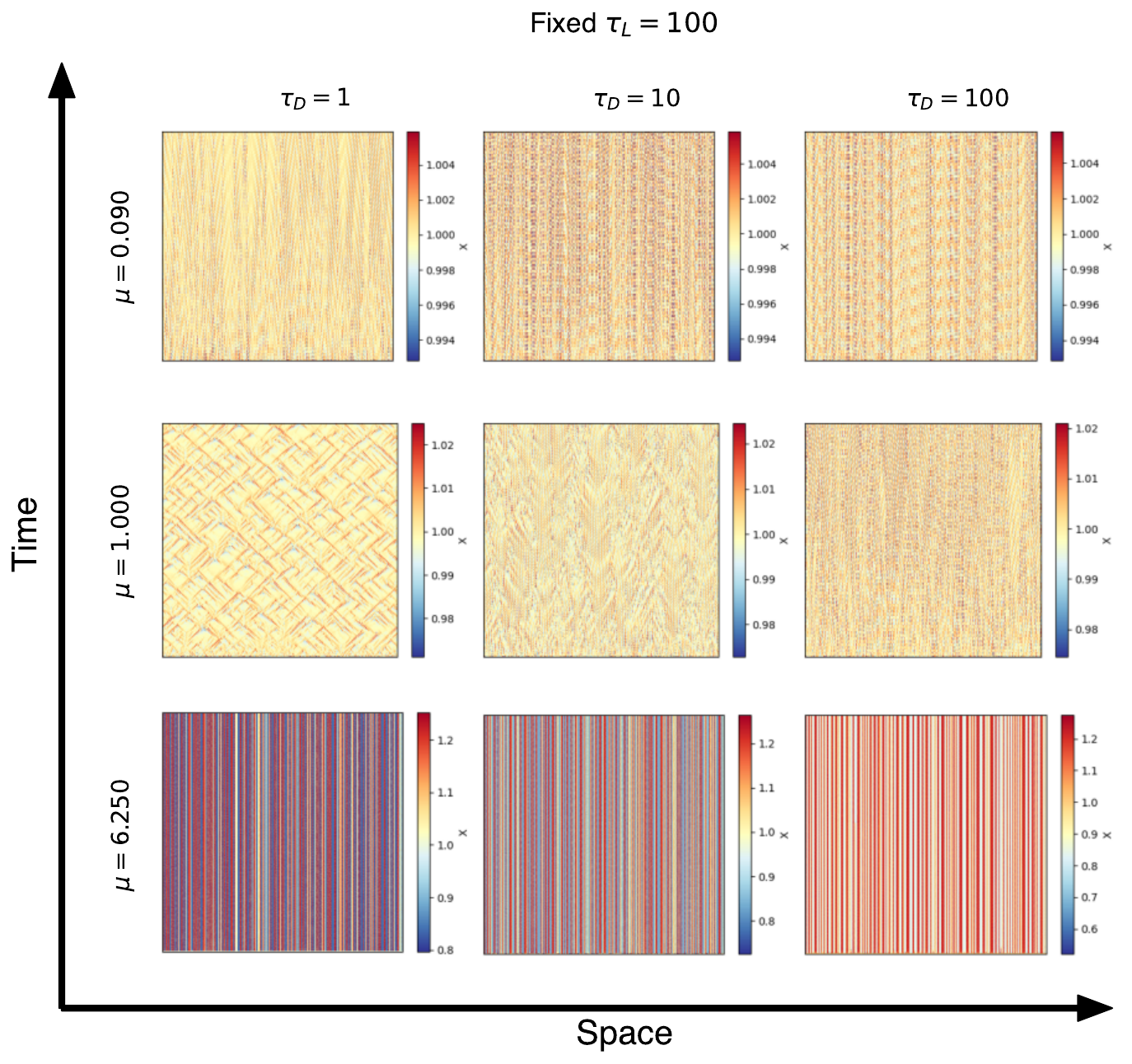}
    \caption{Kymographs of spring length ($\partial_x R$) for $\tau_L = 100$, for varying $\tau_D$ values ($1, 10, 100$) (left to right) and $\mu = 0.09, 1, 6.25$ (top to bottom). The X-axis represents the space and Y-axis represents the time in the figure.}
    \label{si:fig:SI_higher_tauL_X}
\end{figure}

In our manuscript, we have analyzed the system over a wide range of $\tau_D$ and $\tau_L$ values. The phase boundaries will change for higher values of $\tau_D$ and $\tau_L$, as shown in Fig.\ref{si:fig:SI-total-number}. The trimodal cell length ($\partial_x R$) distribution observed at large $\tau_L$ reflects the coexistence of multiple dynamical states resulting from delayed mechanical relaxation. At higher values of $\tau_L$ and $\mu$, as seen in the kymographs (Fig.~\ref{si:fig:SI_higher_tauL_X}), this coexistence of phases becomes evident. In fact, some of the kymographs show patterns reminiscent of X, V, and W-mode traveling waves described in \cite{yin_emergence_2024}. 

\begin{figure}[!htbp]
    \centering
    \includegraphics[width=\textwidth,height=0.6\paperheight,keepaspectratio]{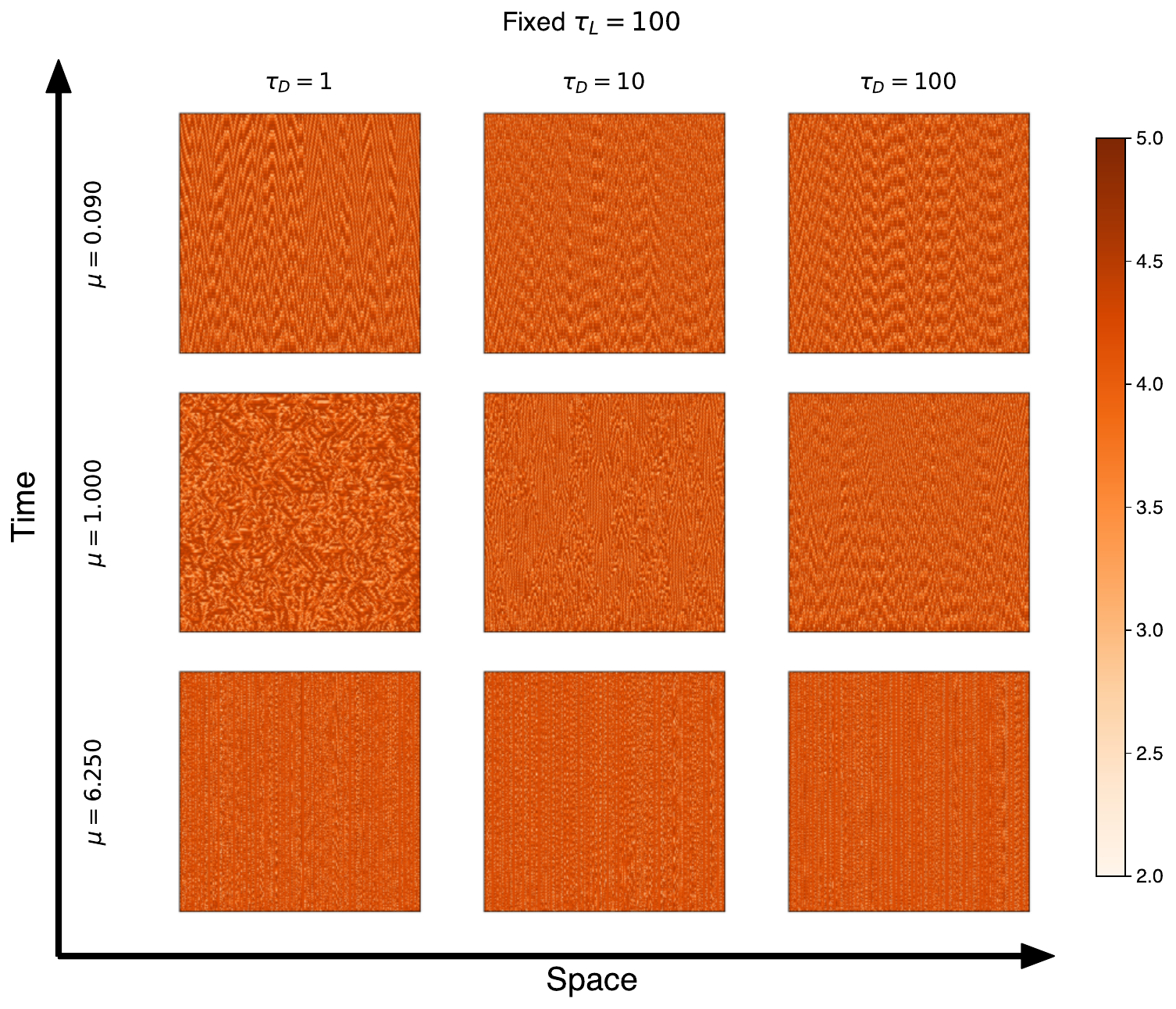}
    \caption{Kymographs of $E$ for $\tau_L = 100$, for varying $\tau_D$ values ($1, 10, 100$) (left to right) and $\mu = 0.09, 1, 6.25$ (top to bottom). The X-axis represents the space and Y-axis represents the time in the figure.}
    \label{si:fig:SI_higher_tauL_E}
\end{figure}

\section{Chimera State Characterization}

To characterize chimera states, we will compute \emph{local order parameter}  \cite{abrams2004chimera} and  \emph{coherence fraction} \cite{Sethia2008}. The local order parameter gives a local measure which changes over time whereas the coherence fraction here is a global measure which can either be shown to evolve with time or taken as an average over time. These measures require the phase of each oscillator, which is obtained from the ERK signal $E(t)$ by converting it into an analytic signal \cite{synchronizationArkady2001}.

The local order parameter tells us whether an oscillator is in a coherent neighbourhood or not. For an oscillator $i$, it is defined as:

\begin{equation}
    R_i(t) = \left|\frac{1}{2\delta} \sum_{|i-j|\leq \delta}e^{i\theta_i(t)} \right|
\end{equation}

where, 

\begin{align*}
    \theta_i(t) &= \text{phase of oscillator $i$}\\
    \delta &= \text{neighbourhood of oscillator $i$}\\
\end{align*}

And $R_i(t)\in[0,1]$, where

\begin{align*}
    R_i \approx 1 &: \text{oscillator $i$ is in coherent domain}\\
    R_i \ll 1 &: \text{oscillator $i$ is in incoherent domain}
\end{align*}

The coherence fraction is the fraction of oscillators that belong to coherent domains at a given time:

\begin{equation}
    g_0(t) = \frac{1}{N}\sum_{i=1}^N \Theta(R_i(t) > \eta)
\end{equation}

where $\Theta$ is the Heaviside step function and $\eta$ is the threshold for $R_i$ above which we consider oscillators to be coherent. We will use $\eta = 0.95$ in our calculations.

This can also be calculated as an average over time:

\begin{equation}
    g_0 = \left< g_0(t) \right>_t
\end{equation}

where, we can infer that:

\begin{align*}
    g_0 \approx 1 &: \text{global synchrony}\\
    g_0 \approx 0 &: \text{global incoherence}\\
    0 < g_0 <1 &: \text{chimera state}
\end{align*}

\textbf{Mesh size:} To check the robustness of $g_0$ with mesh size $\delta$, we calculated it with different $\delta$ for three $(\tau_D,\tau_L)$ pairs.

\begin{figure}[!htbp]
    \centering
    \includegraphics[width=0.85\linewidth]{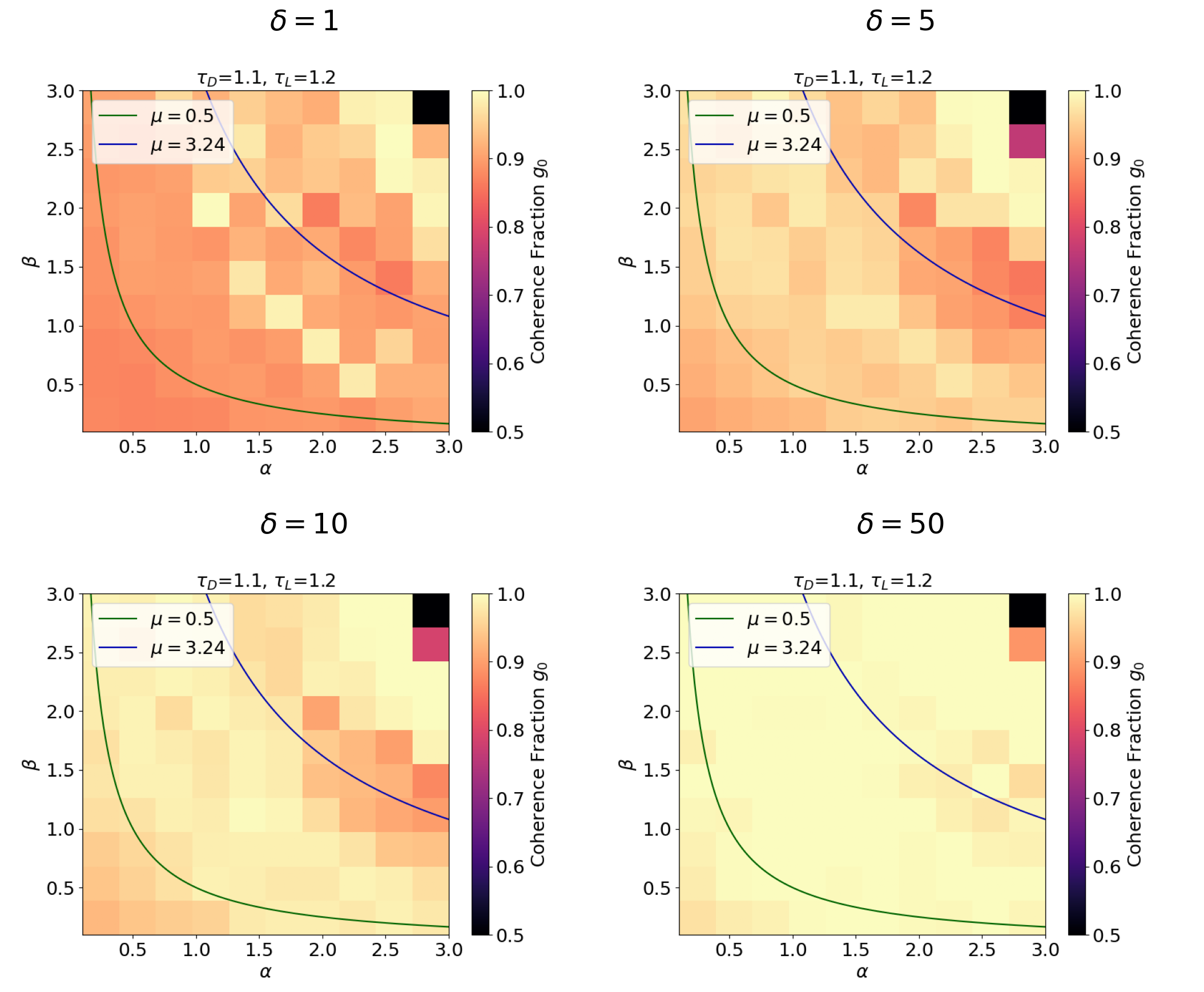}
    \caption{Coherence fraction for different $\delta$ for $(\tau_D,\tau_L)=(1.1,1.2)$.}
    \label{si:fig:placeholder}
\end{figure}

The main observations are as follows:

\begin{itemize}
    \item For low $\mu=\alpha\beta$, $g_0$ goes below 1, showing that there are chimera states.
    \item For intermediate $\mu$, $g_0\approx1$, which is the in-phase state where we get oscillation death.
    \item At high $\mu$, $g_0<1$ again where we get bistable limit cycles and traveling waves.
    \item The dependence of $g_0$ on $\alpha$ and $\beta$ does change with $\delta$ particularly for $(\tau_D,\tau_L)=(1.1,1.2)$, but the general behaviour of $g_0$ at low, intermediate and high $\mu$ are seen in all the $(\tau_D,\tau_L)$ values shown.
    \item For higher $(\tau_D,\tau_L)$ we see that changes in $g_0$ with $\mu$ becomes more pronounced across phases as we take larger $\delta$.
\end{itemize}

\textbf{Random initial conditions:} To check robustness of the measure to initial conditions, we have calculated $g_0$ for different initial random seeds.

\begin{figure}[!htbp]
    \centering
    \includegraphics[width=0.9\linewidth]{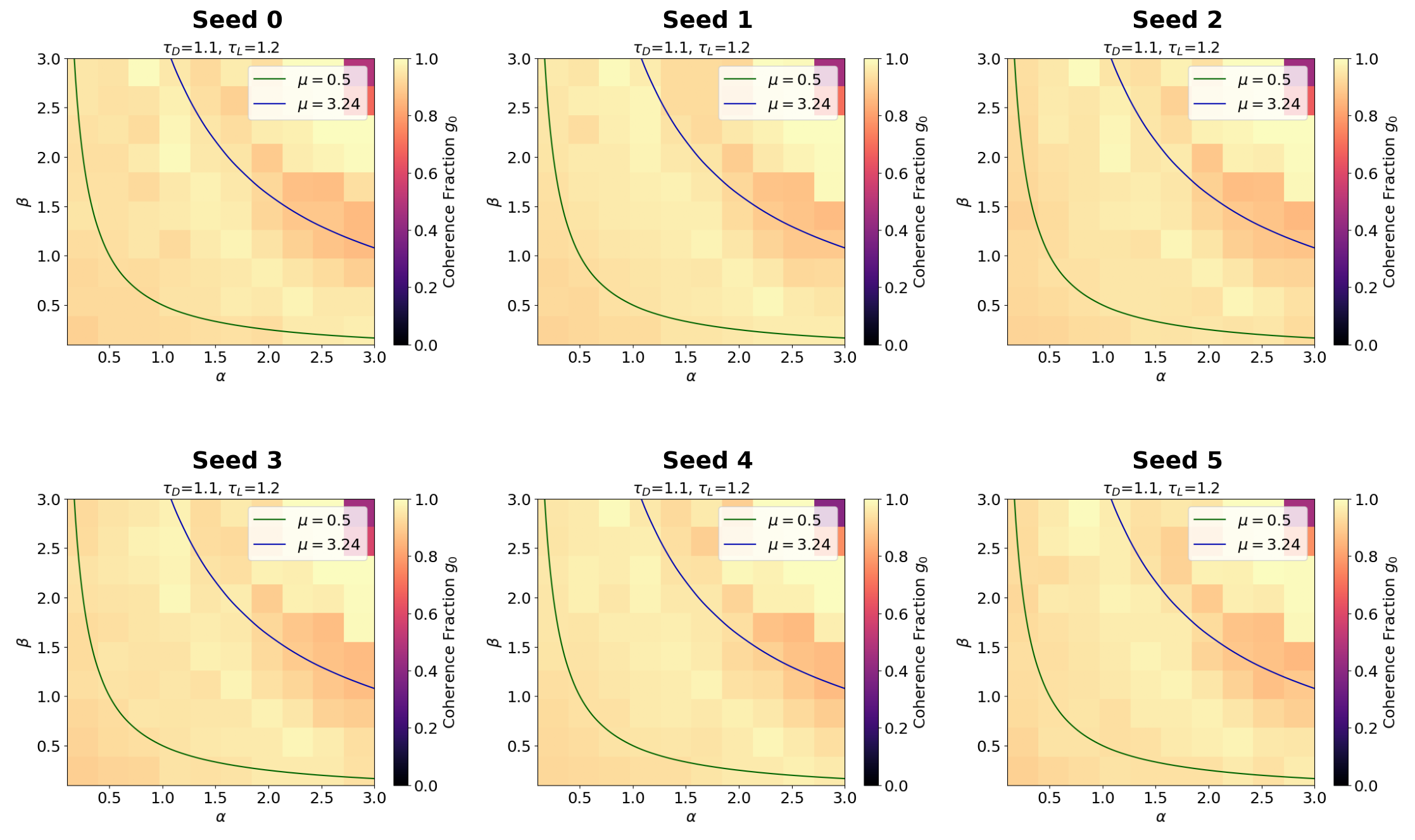}
    \caption{Coherence fraction for different initial seeds for $(\tau_D,\tau_L)=(1.1,1.2)$ for $\delta=5$.}
    \label{si:fig:placeholder}
\end{figure}

We observe that the dependence of $g_0$ on $\alpha$ and $\beta$ does not change with random initial conditions.

\textbf{System size:} We do not need to check the robustness to system size as we have been using periodic boundary conditions in our simulations. However, to check, we systematically calculated $g_0$ for $n_X = \{200,300,500\}$.

\begin{figure}[!htbp]
    \centering
    \includegraphics[width=0.9\linewidth]{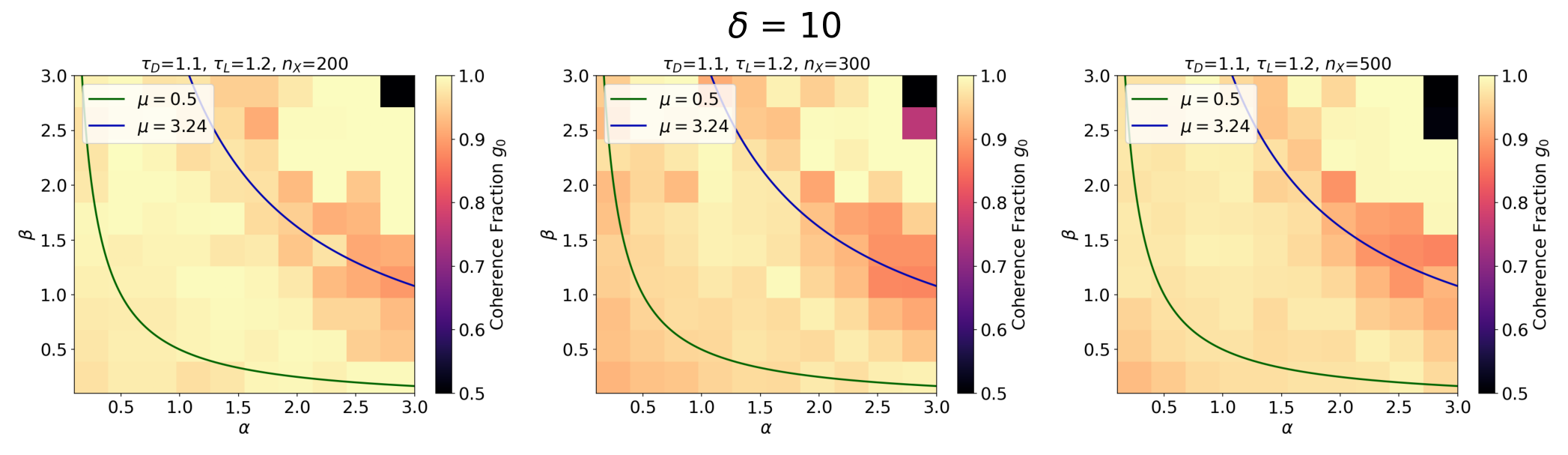}
    \caption{Coherence fraction for different system sizes for $(\tau_D,\tau_L)=(1.1,1.2)$ for $\delta=10$.}
    \label{si:fig:placeholder}
\end{figure}

Clearly, the $g_0$ dependence on $\alpha,\beta$ does not change appreciably on system size.

\section{Lyapunov Exponents}

Following Ref.~\cite{JanakiRangarajan}, for an n-dimensional continuous-time dynamical system, 
\begin{equation*}
    \dot{\mathbf{x}} = \mathbf{F}(\mathbf{x},t) \;\;\; \text{where,} \;\;\; \mathbf{x}=(x_1,x_2,...,x_n) \in \mathbb{R}^n
\end{equation*}

The Jacobian of the system can be calculated as:
\begin{equation*}
    \mathbf{J} = \frac{\partial \mathbf{F}}{\partial \mathbf{x}} \in \mathbb{R}^{n \times n}
\end{equation*}

The infinitesimal perturbations in the system grow according to:
\begin{equation*}
    \dot{\mathbf{\xi}} = \mathbf{J}(\mathbf{x},t) \mathbf{\xi}
\end{equation*}

Now, if we start with orthogonal initial perturbations for the variables in the system, $\mathbf{\xi}(0) = (1 0 0 ... 0)$, etc. and then write out the full equation as:
\begin{equation}
    \dot{\mathbf{\Xi}} = \mathbf{J}  \mathbf{\Xi} 
\end{equation}
where,
\begin{equation}
   \mathbf{\Xi} = \begin{pmatrix}
1 & 0 & \cdots & 0 \\\\
0 & 1 & \cdots & 0 \\\\
\vdots & \vdots & \ddots & \vdots \\\\
0 & 0 & \cdots & 1
\end{pmatrix}
\end{equation}

Now, if we write $\mathbf{\Xi} = \mathbf{Q}  \mathbf{R}$, the Lyapunov exponents for each component of $\mathbf{x}$ are the corresponding diagonal elements of $\mathbf{R}$.

We can then calculate the Lyapunov spectra for the different $\alpha,\beta$ values and observed that $\lambda_i>0 , i\in(1,2,3,4), \forall \alpha,\beta$. $\lambda_5$ (corresponding to the mechanical degrees of freedom), however is entirely $>0$ only for a small region of the $\alpha - \beta$ space. This matches well with the second phase in Fig. 1B.

\begin{figure}[!htbp]
    \centering
    \includegraphics[width=0.6\linewidth]{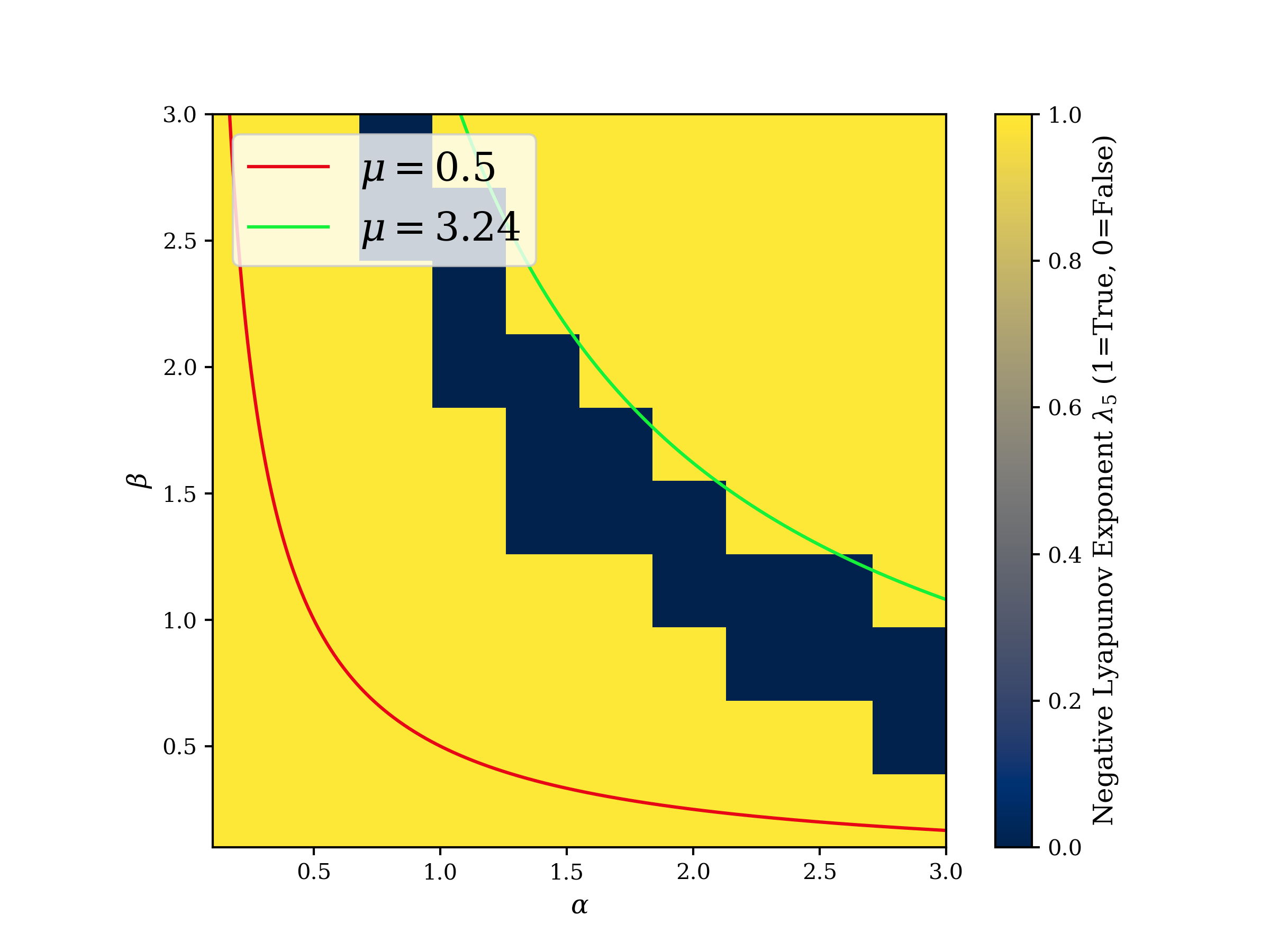}
    \caption{Binary map for $\lambda_5 < 0 \text{ or } > 0$ for $(\tau_D,\tau_L)=(1.1,1.2)$.}
    \label{si:fig:placeholder}
\end{figure}

\section{Supplementary movies}

Please note that you don't need to sign in to see these movies. Just ignore the pop up message. 

\subsection{Kymograph and time-trace of $E$}
\begin{itemize}
    \item Chimera state and spatially-localized traveling waves: \href{https://indianinstituteofscience-my.sharepoint.com/:v:/g/personal/sumantra_iisc_ac_in/EbcbQzlf_bhNkZpIrKFyfWEBbH84G9L727Kwt3Fe4cxBkA?e=WPhjGj}{$\alpha = 0.1, \beta = 0.1, \tau_D = 1.1, \tau_L = 1.2, L_x = 200, t_{max} = 50000$}

    \item Coexisting COD and spatially-localized traveling waves: \href{https://indianinstituteofscience-my.sharepoint.com/:v:/g/personal/sumantra_iisc_ac_in/EW9IXUtxHnNEg5tZVctS5AYBwEQ4B_5k4gSICyzCkoHLkw?e=lo7lYa}{$\alpha = 1, \beta = 1, \tau_D = 1.1, \tau_L = 1.2, L_x = 200, t_{max} = 50000$}

    \item Spatially-extended traveling waves: \href{https://indianinstituteofscience-my.sharepoint.com/:v:/g/personal/sumantra_iisc_ac_in/Ebk6PImLZA1GiykGIfByD2gBr9r_pcvLZQSaip458m96Cg?e=EVV4dr}{$\alpha = 2.5, \beta = 2.5, \tau_D = 1.1, \tau_L = 1.2, L_y = 200, t_{max} = 50000$}
\end{itemize}

\subsection{Time-trace of $E$ and Phase plot of $E$ vs. $M$}

The red line is the attractor of the Brusselator without mechanochemical coupling. 

\begin{itemize}
    \item \href{https://indianinstituteofscience-my.sharepoint.com/:v:/g/personal/sumantra_iisc_ac_in/ESxhF3MeR7lDqsHnVqLUlO8BNCyI41otWhMBuZPhfnEqYg?e=y2aJbU}{$\alpha = 0.1, \beta = 0.1, \tau_D = 1.1, \tau_L = 1.2, L_x = 200, t_{max} = 50000$}. Shows that at small enough coupling, the HHS follows the Brusselator limit cycle.

    \href{https://indianinstituteofscience-my.sharepoint.com/:v:/g/personal/sumantra_iisc_ac_in/EWn64eT-vE9GhkcOu16jqyYBgLslB8i_92BIWaWzjmZCgQ?e=bxD6Ph}{$\alpha = 0.7, \beta = 0.7, \tau_D = 1.1, \tau_L = 1.2, L_x = 200, t_{max} = 50000$} The location of the first dynamical phase transition (DPT). 
    
    \item \href{https://indianinstituteofscience-my.sharepoint.com/:v:/g/personal/sumantra_iisc_ac_in/Ee7qQi6hLfZLqiJNhVgrHhwBu8UoU5t3TaY19n8BX68wNQ?e=fIrvLV}{$\alpha = 1, \beta = 1, \tau_D = 1.1, \tau_L = 1.2, L_x = 200, t_{max} = 50000$}. Shows the expansion of the available phase space from the original limit cycle

    \item \href{https://indianinstituteofscience-my.sharepoint.com/:v:/g/personal/sumantra_iisc_ac_in/Edn3rvgBjjlLutvbQWMQ3bEBfpM2m9fSAk0bjLmBiikwRA?e=hkOAm5}{$\alpha = 1.8, \beta = 1.8, \tau_D = 1.1, \tau_L = 1.2, L_x = 200, t_{max} = 50000$} There are large collective fluctuations in $E$ and $M$ values, suggesting a dynamic phase transition. This is the location of the second DPT.

    \item \href{https://indianinstituteofscience-my.sharepoint.com/:v:/g/personal/sumantra_iisc_ac_in/EXOP6ubw9iVJg5Khzj9xC4kBL7T9GkvgdmJDTBh7mNvzqw?e=z7J82P}{$\alpha = 2, \beta = 2, \tau_D = 1.1, \tau_L = 1.2, L_x = 200, t_{max} = 50000$} Shows the coexistence of two limit cycles. One corresponds to the chemical oscillation (larger), and one corresponds to traveling waves (smaller). 

    \item \href{https://indianinstituteofscience-my.sharepoint.com/:v:/g/personal/sumantra_iisc_ac_in/Ed_uk7I6S-dPhuxgUoiVnQcBTXR2aAi93TnqWL1_HM8tng?e=dXS2pF}{$\alpha = 2.36, \beta = 2.36, \tau_D = 1.1, \tau_L = 1.2, L_x = 200, t_{max} = 50000$} It is the location of the third DPT. 
   
    \item \href{https://indianinstituteofscience-my.sharepoint.com/:v:/g/personal/sumantra_iisc_ac_in/Ec-ISPWzeLxDsNNIvC9vTwgBgEm5hWgFse9Dgi2l4K_N5w?e=dGVgjM}{$\alpha = 2.5, \beta = 2.5, \tau_D = 1.1, \tau_L = 1.2, L_x = 200, t_{max} = 50000$}. Shows the limit cycle of traveling waves. 
\end{itemize}

\subsection{Time trace of E under the influence of conservative noise}
\begin{itemize}

    \item {Chimera states in the presence of noise:} \href{https://indianinstituteofscience-my.sharepoint.com/:v:/g/personal/sumantra_iisc_ac_in/ETApHZfhZ_lKqLRZj6CEWfkBlvMzTCx3RjsS898b6JF6Cg?e=PJZ9by}{$\alpha = 0.3, \beta = 0.3, \tau_D = 1.1, \tau_L = 1.2, L_x = 200, t_{max} = 50000$}. 

    \item {Coexisting COD and spatially-localized traveling waves in the presence of noise:} \href{https://indianinstituteofscience-my.sharepoint.com/:v:/g/personal/sumantra_iisc_ac_in/Eenyy2K-32lFnw8Vfetb1oMBNoCCIT9edG3dTySpBUJIdA?e=rIHhnS}{$\alpha = 1, \beta = 1, \tau_D = 1.1, \tau_L = 1.2, L_x = 200, t_{max} = 50000$}. 

    \item {Spatially-extended traveling waves in the presence of noise:} \href{https://indianinstituteofscience-my.sharepoint.com/:v:/g/personal/sumantra_iisc_ac_in/EYW_-55NObVOhdieVrfc_RYBp9TYdrvKXAq4iTVF_SZqyw?e=4zmlft}{$\alpha = 2.8, \beta = 2.8, \tau_D = 1.1, \tau_L = 1.2, L_x = 200, t_{max} = 50000$}. 
  
\end{itemize}

\subsection{Time trace of E in the MCFL-Fitzhugh-Nagumo (FHN) model}
For all the videos presented in this section $I = 5, b_0 = 1.3$. 
\begin{itemize}
    \item {Chimera states in the FHN model:} \href{https://indianinstituteofscience-my.sharepoint.com/:v:/g/personal/sumantra_iisc_ac_in/ET_9TBLjl0FAgJex-zhgYTUBnBZPwb1TE_gi8FKmiuF9Uw?e=SgUscj}{$\alpha = 0.5, \beta = 0.5, \tau_B = 1.1, \tau_L = 1.2$}

    \item {Coexisting COD and chimera states in the FHN model:} \href{https://indianinstituteofscience-my.sharepoint.com/:v:/g/personal/sumantra_iisc_ac_in/EeLj52x1V0pGm9DwrXtbGk4B1TDUJvEUxS1AMc85-be7Jg?e=QVjkZM}{$\alpha = 1, \beta = 1, \tau_B = 1.1, \tau_L = 1.2$}

    \item {Coexisting COD and spatially-localized traveling waves in the FHN model:} \href{https://indianinstituteofscience-my.sharepoint.com/:v:/g/personal/sumantra_iisc_ac_in/EQljHw17q-hOgCYKXeA6o34BauI0HQp_PYAw-EFEIccPjg?e=jf5sCd}{$\alpha = 1, \beta = 1, \tau_B = 11, \tau_L = 12$}
\end{itemize}

\newpage
\section{Supplementary figures}
\begin{figure}[htbp]
    \centering
    \includegraphics[width=\linewidth]{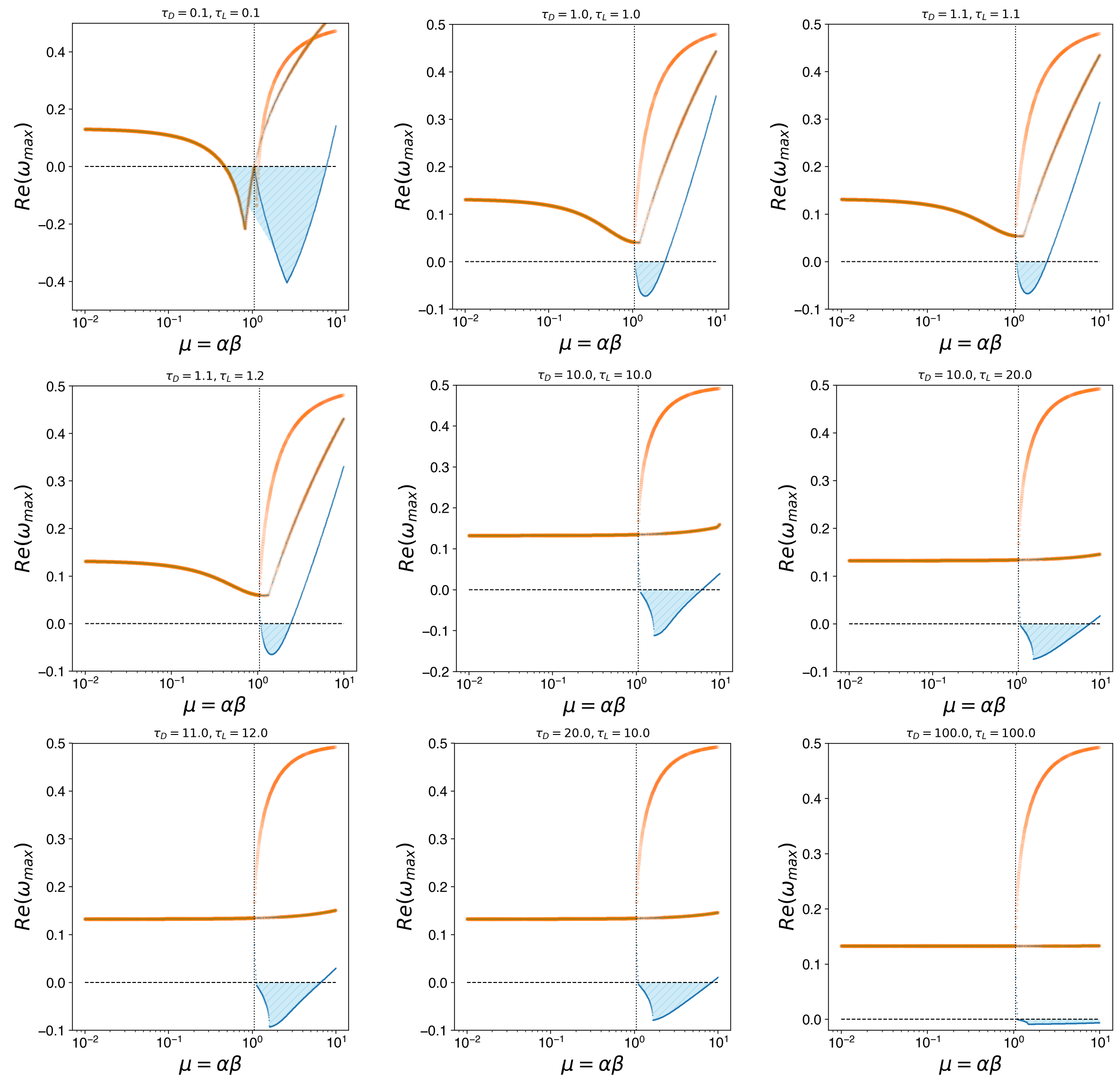}
    \caption{The real part of the eigenvalue with the maximum real part. Plotted for different $\tau_D$ and $\tau_L$ values. The orange marker shows $L \geq 1$ and the blue markers show $L < 1$. The maximum real part is negative in the shaded area and demonstrates the existence of stable nodes with $L < 1$. }
    \label{si:fig:SI-maxEV}
\end{figure}
\newpage

\begin{figure}[htbp]
    \centering
    \includegraphics[width=\linewidth]{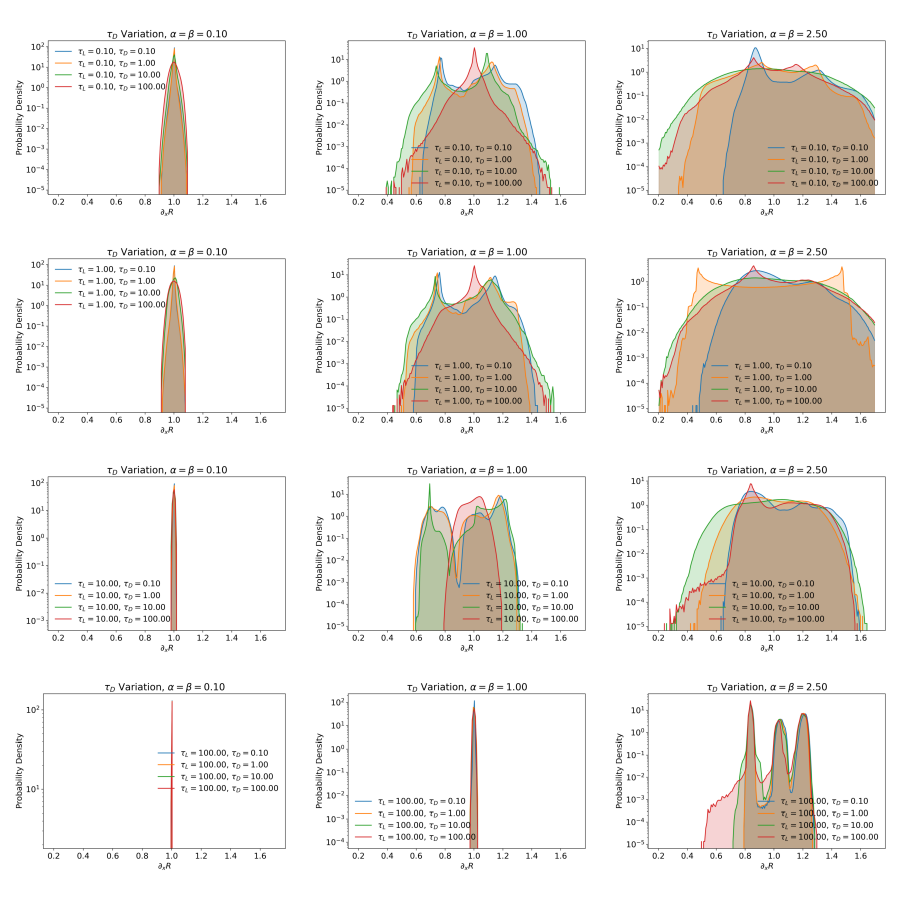}
    \caption{$\del_x R$ distributions for $\alpha=\beta = \{0.1, 1, 2.5\}$ with different $\tau_D$ but fixed $\tau_L$.}
    \label{si:fig:SI-VaryTauD}
\end{figure}
\newpage

\begin{figure}[htbp]
    \centering
    \includegraphics[width=\linewidth]{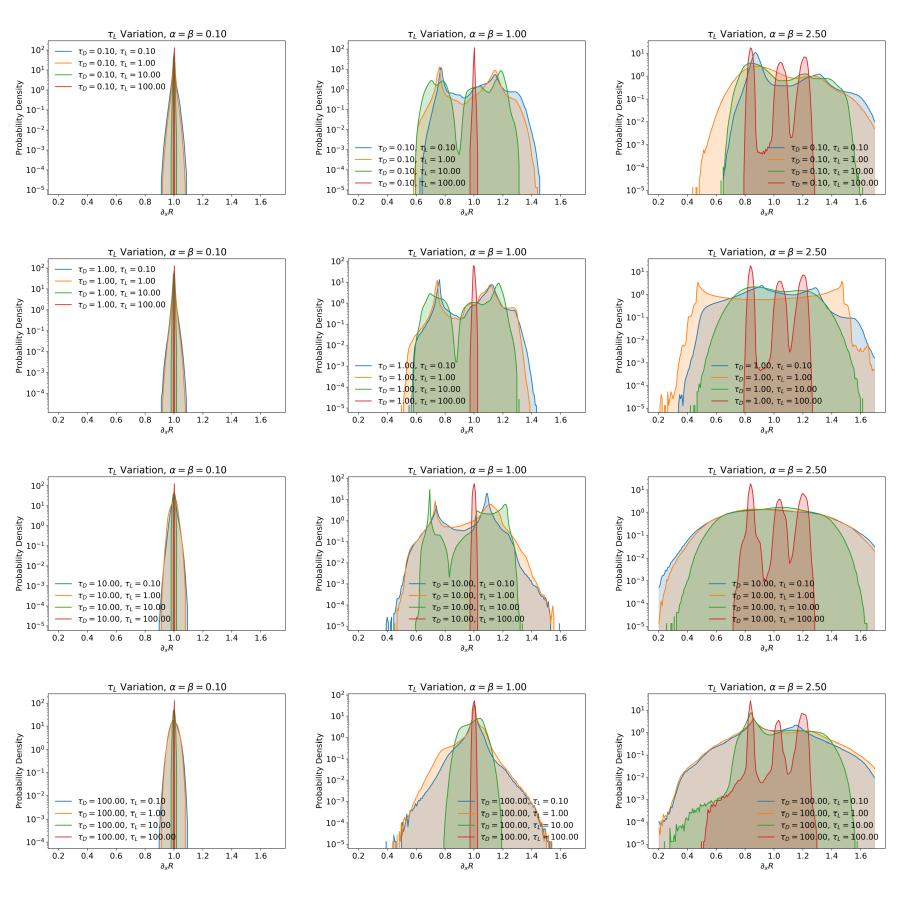}
    \caption{$\del_x R$ distributions for $\alpha=\beta = \{0.1, 1, 2.5\}$ with different $\tau_L$ but fixed $\tau_D$.}
    \label{si:fig:SI-VaryTauL}
\end{figure}
\newpage

\begin{figure}[htbp]
    \centering
    \includegraphics[width=\linewidth]{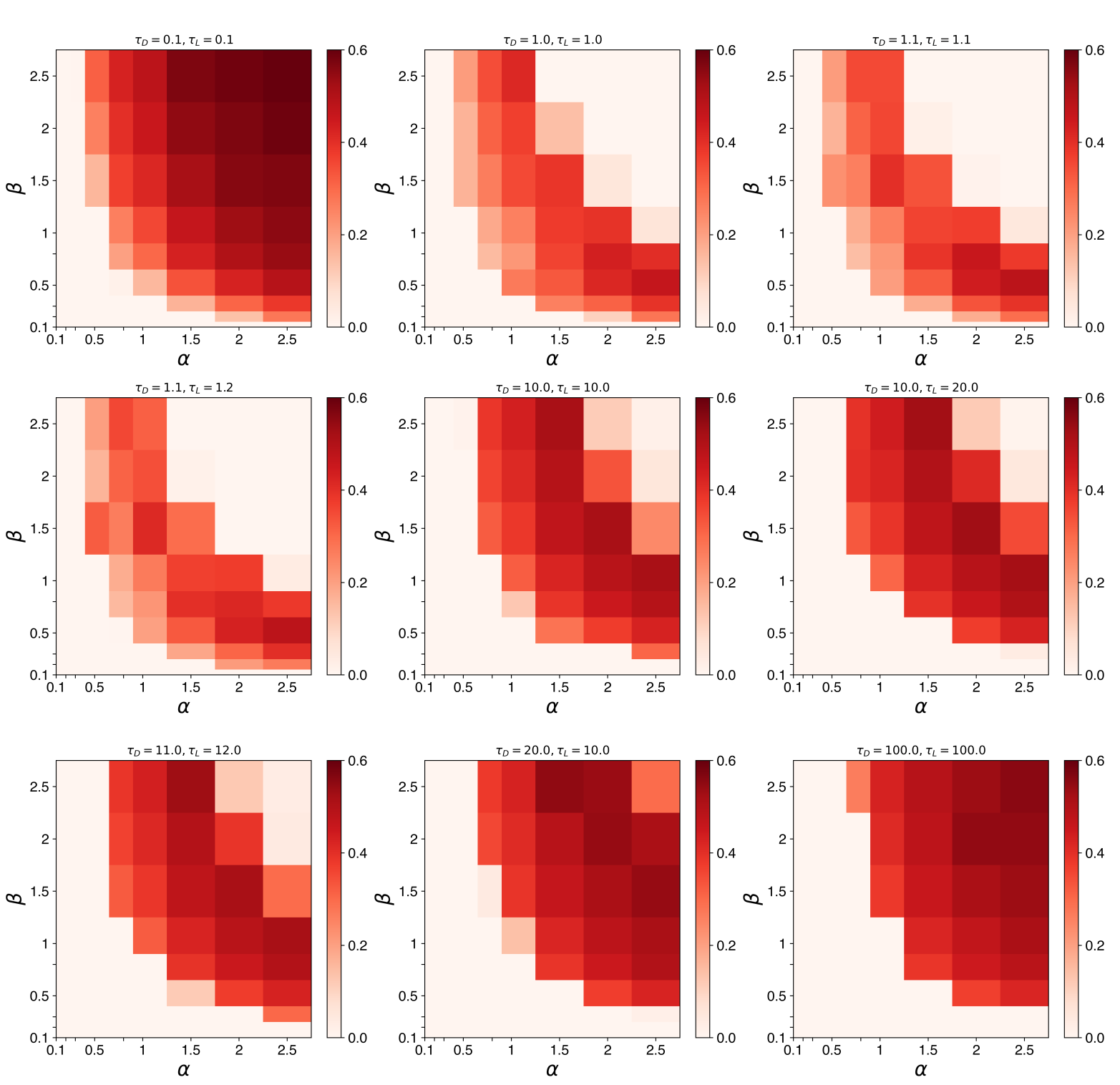}
    \caption{The total size of regions with collective oscillation death, expressed as the fraction of system size. Note the correlation with the shaded area in Fig.~\ref{si:fig:SI-maxEV}. }
    \label{si:fig:SI-total-size}
\end{figure}
\newpage

\begin{figure}[htbp]
    \centering
    \includegraphics[width=\linewidth]{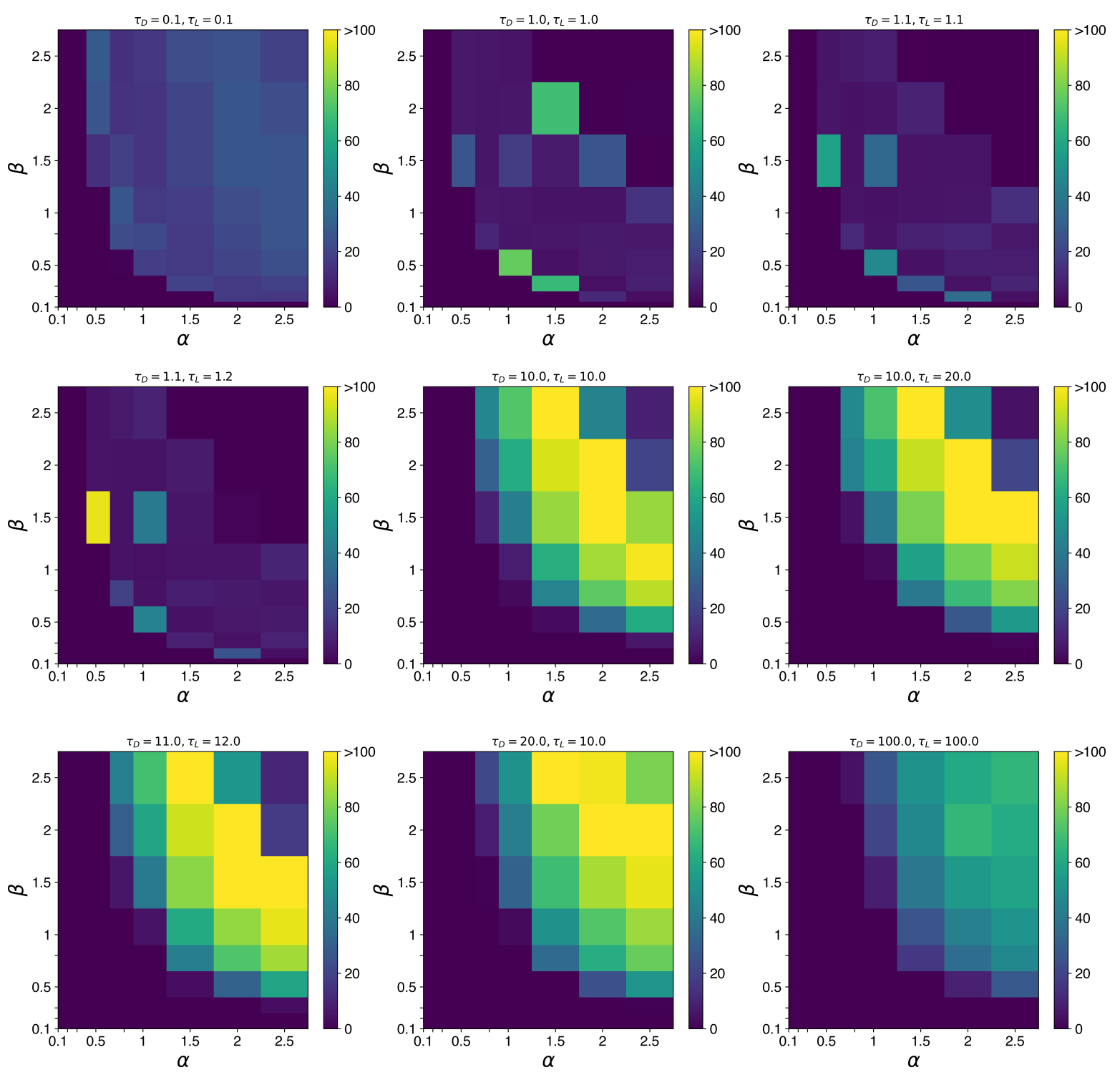}
    \caption{The total number of regions with collective oscillation death. Note the correlation with the shaded area in Fig.~\ref{si:fig:SI-maxEV}. }
    \label{si:fig:SI-total-number}
\end{figure}

\bibliography{mechano-chemical-feedback}

@article{hannezo_mechanochemical_2019,
	title = {Mechanochemical {Feedback} {Loops} in {Development} and {Disease}},
	volume = {178},
	issn = {0092-8674, 1097-4172},
	url = {https://www.cell.com/cell/abstract/S0092-8674(19)30623-3},
	doi = {10.1016/j.cell.2019.05.052},
	number = {1},
	urldate = {2025-02-23},
	journal = {Cell},
	author = {Hannezo, Edouard and Heisenberg, Carl-Philipp},
	month = jun,
	year = {2019},
	pmid = {31251912},
	note = {Publisher: Elsevier},
	pages = {12--25},
}

@article{ioratim-uba_mechanochemical_2023,
	title = {Mechanochemical {Active} {Feedback} {Generates} {Convergence} {Extension} in {Epithelial} {Tissue}},
	volume = {131},
	issn = {0031-9007, 1079-7114},
	url = {https://link.aps.org/doi/10.1103/PhysRevLett.131.238301},
	doi = {10.1103/PhysRevLett.131.238301},
	
	number = {23},
	urldate = {2025-02-23},
	journal = {Physical Review Letters},
	author = {Ioratim-Uba, Aondoyima and Liverpool, Tanniemola B. and Henkes, Silke},
	month = dec,
	year = {2023},
	pages = {238301},
}

@article{boocock_interplay_2023,
	title = {Interplay between {Mechanochemical} {Patterning} and {Glassy} {Dynamics} in {Cellular} {Monolayers}},
	volume = {1},
	doi = {10.1103/PRXLife.1.013001},
	number = {1},
	journal = {PRX Life},
	author = {Boocock, Daniel},
	year = {2023},
url={https://journals.aps.org/prxlife/abstract/10.1103/PRXLife.1.013001}
}

@article{parmar2025spontaneous,
  title={Spontaneous and Induced Oscillations in Confined Epithelia},
  author={Parmar, Toshi and Dow, Liam P and Pruitt, Beth L and Marchetti, M Cristina},
  journal={PRX Life},
  volume={3},
  number={1},
  pages={013002},
  year={2025},
  publisher={APS},
url={https://journals.aps.org/prxlife/abstract/10.1103/PRXLife.3.013002}
}

@article{perez-verdugo_excitable_2024,
	title = {Excitable dynamics driven by mechanical feedback in biological tissues},
	volume = {7},
	copyright = {2024 The Author(s)},
	issn = {2399-3650},
	url = {https://www.nature.com/articles/s42005-024-01661-2},
	doi = {10.1038/s42005-024-01661-2},
	
	number = {1},
	urldate = {2025-02-25},
	journal = {Communications Physics},
	author = {Pérez-Verdugo, Fernanda and Banks, Samuel and Banerjee, Shiladitya},
	month = may,
	year = {2024},
	note = {Publisher: Nature Publishing Group},
	pages = {1--8},
}

@article{moreno_competition_2019,
	title = {Competition for {Space} {Induces} {Cell} {Elimination} through {Compaction}-{Driven} {ERK} {Downregulation}},
	volume = {29},
	issn = {0960-9822},
	url = {https://www.cell.com/current-biology/abstract/S0960-9822(18)31473-8},
	doi = {10.1016/j.cub.2018.11.007},
	
	number = {1},
	urldate = {2025-02-27},
	journal = {Current Biology},
	author = {Moreno, Eduardo and Valon, Léo and Levillayer, Florence and Levayer, Romain},
	month = jan,
	year = {2019},
	pmid = {30554899},
	note = {Publisher: Elsevier},
	pages = {23--34.e8},
}

@article{bruckner_tissue_2024,
	title = {Tissue {Active} {Matter}: {Integrating} {Mechanics} and {Signaling} into {Dynamical} {Models}},
	issn = {, 1943-0264},
	shorttitle = {Tissue {Active} {Matter}},
	url = {http://cshperspectives.cshlp.org/content/early/2024/07/01/cshperspect.a041653},
	doi = {10.1101/cshperspect.a041653},
	
	urldate = {2025-03-01},
	journal = {Cold Spring Harbor Perspectives in Biology},
	author = {Brückner, David B. and Hannezo, Edouard},
	month = jul,
	year = {2024},
	pmid = {38951023},
	note = {Company: Cold Spring Harbor Laboratory Press
Distributor: Cold Spring Harbor Laboratory Press
Institution: Cold Spring Harbor Laboratory Press
Label: Cold Spring Harbor Laboratory Press
Publisher: Cold Spring Harbor Lab},
	pages = {a041653},
}

@article{yin_emergence_2024,
	title = {Emergence, {Pattern}, and {Frequency} of {Spontaneous} {Waves} in {Spreading} {Epithelial} {Monolayers}},
	volume = {24},
	issn = {1530-6984},
	url = {https://doi.org/10.1021/acs.nanolett.3c04876},
	doi = {10.1021/acs.nanolett.3c04876},
	number = {12},
	urldate = {2025-03-01},
	journal = {Nano Letters},
	author = {Yin, Xu and Liu, Yong-Quan and Zhang, Li-Yuan and Liang, Dong and Xu, Guang-Kui},
	month = mar,
	year = {2024},
	note = {Publisher: American Chemical Society},
	pages = {3631--3637},
}

@article{martens_chimera_2013,
	title = {Chimera states in mechanical oscillator networks},
	volume = {110},
	url = {https://www.pnas.org/doi/abs/10.1073/pnas.1302880110},
	doi = {10.1073/pnas.1302880110},
	
	number = {26},
	urldate = {2025-03-04},
	journal = {Proceedings of the National Academy of Sciences},
	author = {Martens, Erik Andreas and Thutupalli, Shashi and Fourrière, Antoine and Hallatschek, Oskar},
	month = jun,
	year = {2013},
	pages = {10563--10567},
}

@article{boocock_theory_2021,
	title = {Theory of mechanochemical patterning and optimal migration in cell monolayers},
	volume = {17},
	copyright = {2020 The Author(s), under exclusive licence to Springer Nature Limited},
	issn = {1745-2481},
	url = {https://www.nature.com/articles/s41567-020-01037-7},
	doi = {10.1038/s41567-020-01037-7},
	number = {2},
	urldate = {2025-03-06},
	journal = {Nature Physics},
	author = {Boocock, Daniel and Hino, Naoya and Ruzickova, Natalia and Hirashima, Tsuyoshi and Hannezo, Edouard},
	month = feb,
	year = {2021},
	note = {Publisher: Nature Publishing Group},
	pages = {267--274},
}

@article{collins_group-theoretic_1994,
	title = {A group-theoretic approach to rings of coupled biological oscillators},
	volume = {71},
	issn = {1432-0770},
	url = {https://doi.org/10.1007/BF00197312},
	doi = {10.1007/BF00197312},
	
	number = {2},
	urldate = {2025-03-08},
	journal = {Biological Cybernetics},
	author = {Collins, J. J. and Stewart, I.},
	month = jun,
	year = {1994},
	pages = {95--103},
}

@article{collins_coupled_1993,
	title = {Coupled nonlinear oscillators and the symmetries of animal gaits},
	volume = {3},
	issn = {1432-1467},
	url = {https://doi.org/10.1007/BF02429870},
	doi = {10.1007/BF02429870},
	
	number = {1},
	urldate = {2025-03-08},
	journal = {Journal of Nonlinear Science},
	author = {Collins, J. J. and Stewart, I. N.},
	month = dec,
	year = {1993},
	pages = {349--392},
}

@article{epstein_symmetric_1993,
	title = {Symmetric patterns in linear arrays of coupled cells},
	volume = {3},
	issn = {1054-1500},
	url = {https://doi.org/10.1063/1.165974},
	doi = {10.1063/1.165974},
	number = {1},
	urldate = {2025-03-10},
	journal = {Chaos: An Interdisciplinary Journal of Nonlinear Science},
	author = {Epstein, Irving R. and Golubitsky, Martin},
	month = jan,
	year = {1993},
	pages = {1--5},
}

@article{shankaran_rapid_2009,
	title = {Rapid and sustained nuclear–cytoplasmic {ERK} oscillations induced by epidermal growth factor},
	volume = {5},
	copyright = {http://creativecommons.org/licenses/by-nc-sa/3.0/},
	issn = {1744-4292, 1744-4292},
	url = {https://www.embopress.org/doi/10.1038/msb.2009.90},
	doi = {10.1038/msb.2009.90},
	
	number = {1},
	urldate = {2025-04-12},
	journal = {Molecular Systems Biology},
	author = {Shankaran, Harish and Ippolito, Danielle L and Chrisler, William B and Resat, Haluk and Bollinger, Nikki and Opresko, Lee K and Wiley, H Steven},
	month = jan,
	year = {2009},
	pages = {332},
}

@article{brody2013biorthogonal,
  title={Biorthogonal quantum mechanics},
  author={Brody, Dorje C},
  journal={Journal of Physics A: Mathematical and Theoretical},
  volume={47},
  number={3},
  pages={035305},
  year={2013},
  publisher={IOP Publishing},
url={https://iopscience.iop.org/article/10.1088/1751-8113/47/3/035305}
}

@article{kholodenko2010signalling,
  title={Signalling ballet in space and time},
  author={Kholodenko, Boris N and Hancock, John F and Kolch, Walter},
  journal={Nature reviews Molecular cell biology},
  volume={11},
  number={6},
  pages={414--426},
  year={2010},
  publisher={Nature Publishing Group UK London},
url={https://www.nature.com/articles/nrm2901}
}

@article{arkun2018dynamics,
  title={Dynamics and control of the ERK signaling pathway: Sensitivity, bistability, and oscillations},
  author={Arkun, Yaman and Yasemi, Mohammadreza},
  journal={PloS one},
  volume={13},
  number={4},
  pages={e0195513},
  year={2018},
  publisher={Public Library of Science San Francisco, CA USA},
url={https://pubmed.ncbi.nlm.nih.gov/29630631/}
}

@article{logue2015erk,
  title={Erk regulation of actin capping and bundling by Eps8 promotes cortex tension and leader bleb-based migration},
  author={Logue, Jeremy S and Cartagena-Rivera, Alexander X and Baird, Michelle A and Davidson, Michael W and Chadwick, Richard S and Waterman, Clare M},
  journal={elife},
  volume={4},
  pages={e08314},
  year={2015},
  publisher={eLife Sciences Publications, Ltd},
url={https://pmc.ncbi.nlm.nih.gov/articles/PMC4522647/}
}

@article{lavoie2020erk,
  title={ERK signalling: a master regulator of cell behaviour, life and fate},
  author={Lavoie, Hugo and Gagnon, Jessica and Therrien, Marc},
  journal={Nature reviews Molecular cell biology},
  volume={21},
  number={10},
  pages={607--632},
  year={2020},
  publisher={Nature Publishing Group UK London},
url={https://www.nature.com/articles/s41580-020-0255-7}
}

@article{tanimura2017erk,
  title={ERK signalling as a regulator of cell motility},
  author={Tanimura, Susumu and Takeda, Kohsuke},
  journal={The Journal of Biochemistry},
  volume={162},
  number={3},
  pages={145--154},
  year={2017},
  publisher={Oxford University Press},
url={https://academic.oup.com/jb/article/162/3/145/3954057}
}

@article{mendoza2015erk,
  title={ERK reinforces actin polymerization to power persistent edge protrusion during motility},
  author={Mendoza, Michelle C and Vilela, Marco and Juarez, Jesus E and Blenis, John and Danuser, Gaudenz},
  journal={Science signaling},
  volume={8},
  number={377},
  pages={ra47--ra47},
  year={2015},
  publisher={American Association for the Advancement of Science},
url={https://pubmed.ncbi.nlm.nih.gov/25990957/}
}

@article{barros2005activation,
  title={Activation of either ERK1/2 or ERK5 MAP kinase pathways can lead to disruption of the actin cytoskeleton},
  author={Barros, Joana Castro and Marshall, Christopher J},
  journal={Journal of cell science},
  volume={118},
  number={8},
  pages={1663--1671},
  year={2005},
  publisher={Company of Biologists},
url={https://journals.biologists.com/jcs/article/118/8/1663/28734/Activation-of-either-ERK1-2-or-ERK5-MAP-kinase}
}

@article{mendoza2011erk,
  title={ERK-MAPK drives lamellipodia protrusion by activating the WAVE2 regulatory complex},
  author={Mendoza, Michelle C and Er, E Emrah and Zhang, Wenjuan and Ballif, Bryan A and Elliott, Hunter L and Danuser, Gaudenz and Blenis, John},
  journal={Molecular cell},
  volume={41},
  number={6},
  pages={661--671},
  year={2011},
  publisher={Elsevier},
url={https://pmc.ncbi.nlm.nih.gov/articles/PMC3078620/}
}

@article{de2022interplay,
  title={Interplay between mechanics and signalling in regulating cell fate},
  author={De Belly, Henry and Paluch, Ewa K and Chalut, Kevin J},
  journal={Nature Reviews Molecular Cell Biology},
  volume={23},
  number={7},
  pages={465--480},
  year={2022},
  publisher={Nature Publishing Group UK London},
url={https://www.nature.com/articles/s41580-022-00472-z}
}

@article{hino2020erk,
  title={ERK-mediated mechanochemical waves direct collective cell polarization},
  author={Hino, Naoya and Rossetti, Leone and Mar{\'\i}n-Llaurad{\'o}, Ariadna and Aoki, Kazuhiro and Trepat, Xavier and Matsuda, Michiyuki and Hirashima, Tsuyoshi},
  journal={Developmental cell},
  volume={53},
  number={6},
  pages={646--660},
  year={2020},
  publisher={Elsevier},
url={https://www.sciencedirect.com/science/article/pii/S1534580720304007}
}

@book{strogatz2024nonlinear,
  title={Nonlinear dynamics and chaos: with applications to physics, biology, chemistry, and engineering},
  author={Strogatz, Steven H},
  year={2024},
  publisher={Chapman and Hall/CRC},
url={https://www.taylorfrancis.com/books/mono/10.1201/9780429398490/nonlinear-dynamics-chaos-steven-strogatz}
}

@article{lan2015biomechanical,
  title={A biomechanical model for cell polarization and intercalation during Drosophila germband extension},
  author={Lan, Haihan and Wang, Qiming and Fernandez-Gonzalez, Rodrigo and Feng, James J},
  journal={Physical biology},
  volume={12},
  number={5},
  pages={056011},
  year={2015},
  publisher={IOP Publishing},
url={https://pubmed.ncbi.nlm.nih.gov/26356256/}
}

@article{banerjee2015propagating,
  title={Propagating stress waves during epithelial expansion},
  author={Banerjee, Shiladitya and Utuje, Kazage JC and Marchetti, M Cristina},
  journal={Physical review letters},
  volume={114},
  number={22},
  pages={228101},
  year={2015},
  publisher={APS},
url={https://journals.aps.org/prl/abstract/10.1103/PhysRevLett.114.228101}
}

@article{sknepnek2023generating,
  title={Generating active T1 transitions through mechanochemical feedback},
  author={Sknepnek, Rastko and Djafer-Cherif, Ilyas and Chuai, Manli and Weijer, Cornelis and Henkes, Silke},
  journal={Elife},
  volume={12},
  pages={e79862},
  year={2023},
  publisher={eLife Sciences Publications Limited},
url={https://elifesciences.org/articles/79862}
}

@article{dierkes2014spontaneous,
  title={Spontaneous oscillations of elastic contractile materials with turnover},
  author={Dierkes, Kai and Sumi, Angughali and Solon, J{\'e}r{\^o}me and Salbreux, Guillaume},
  journal={Physical review letters},
  volume={113},
  number={14},
  pages={148102},
  year={2014},
  publisher={APS},
url={https://journals.aps.org/prl/abstract/10.1103/PhysRevLett.113.148102}
}

@article{staddon2019mechanosensitive,
  title={Mechanosensitive junction remodeling promotes robust epithelial morphogenesis},
  author={Staddon, Michael F and Cavanaugh, Kate E and Munro, Edwin M and Gardel, Margaret L and Banerjee, Shiladitya},
  journal={Biophysical Journal},
  volume={117},
  number={9},
  pages={1739--1750},
  year={2019},
  publisher={Elsevier},
url={https://www.sciencedirect.com/science/article/pii/S0006349519308148}
}

@article{cavanaugh2020adaptive,
  title={Adaptive viscoelasticity of epithelial cell junctions: from models to methods},
  author={Cavanaugh, Kate E and Staddon, Michael F and Banerjee, Shiladitya and Gardel, Margaret L},
  journal={Current opinion in genetics \& development},
  volume={63},
  pages={86--94},
  year={2020},
  publisher={Elsevier},
url={https://www.sciencedirect.com/science/article/pii/S0959437X20300691}
}

@article{etournay2015interplay,
  title={Interplay of cell dynamics and epithelial tension during morphogenesis of the Drosophila pupal wing},
  author={Etournay, Rapha{\"e}l and Popovi{\'c}, Marko and Merkel, Matthias and Nandi, Amitabha and Blasse, Corinna and Aigouy, Beno{\^\i}t and Brandl, Holger and Myers, Gene and Salbreux, Guillaume and J{\"u}licher, Frank and others},
  journal={Elife},
  volume={4},
  pages={e07090},
  year={2015},
  publisher={eLife Sciences Publications, Ltd},
url={https://elifesciences.org/articles/07090}
}

@article{noll2017active,
  title={Active tension network model suggests an exotic mechanical state realized in epithelial tissues},
  author={Noll, Nicholas and Mani, Madhav and Heemskerk, Idse and Streichan, Sebastian J and Shraiman, Boris I},
  journal={Nature physics},
  volume={13},
  number={12},
  pages={1221--1226},
  year={2017},
  publisher={Nature Publishing Group UK London},
url={https://www.nature.com/articles/nphys4219}
}

@article{saw2017topological,
  title={Topological defects in epithelia govern cell death and extrusion},
  author={Saw, Thuan Beng and Doostmohammadi, Amin and Nier, Vincent and Kocgozlu, Leyla and Thampi, Sumesh and Toyama, Yusuke and Marcq, Philippe and Lim, Chwee Teck and Yeomans, Julia M and Ladoux, Benoit},
  journal={Nature},
  volume={544},
  number={7649},
  pages={212--216},
  year={2017},
  publisher={Nature Publishing Group UK London},
url={https://www.nature.com/articles/nature21718}
}

@article{streichan2018global,
  title={Global morphogenetic flow is accurately predicted by the spatial distribution of myosin motors},
  author={Streichan, Sebastian J and Lefebvre, Matthew F and Noll, Nicholas and Wieschaus, Eric F and Shraiman, Boris I},
  journal={Elife},
  volume={7},
  pages={e27454},
  year={2018},
  publisher={eLife Sciences Publications, Ltd},
url={https://elifesciences.org/articles/27454}
}

@article{ibrahimi2023deforming,
  title={Deforming polar active matter in a scalar field gradient},
  author={Ibrahimi, Muhamet and Merkel, Matthias},
  journal={New Journal of Physics},
  volume={25},
  number={1},
  pages={013022},
  year={2023},
  publisher={IOP Publishing},
url={https://iopscience.iop.org/article/10.1088/1367-2630/acb2e5#:~:text=Active%20matter%20with%20local%20polar,anisotropic%20deformation%20during%20animal%20morphogenesis.}
}

@article{clement2017viscoelastic,
  title={Viscoelastic dissipation stabilizes cell shape changes during tissue morphogenesis},
  author={Cl{\'e}ment, Rapha{\"e}l and Dehapiot, Beno{\^\i}t and Collinet, Claudio and Lecuit, Thomas and Lenne, Pierre-Fran{\c{c}}ois},
  journal={Current biology},
  volume={27},
  number={20},
  pages={3132--3142},
  year={2017},
  publisher={Elsevier},
url={https://www.sciencedirect.com/science/article/pii/S0960982217311697}
}

@article{khalilgharibi2019stress,
  title={Stress relaxation in epithelial monolayers is controlled by the actomyosin cortex},
  author={Khalilgharibi, Nargess and Fouchard, Jonathan and Asadipour, Nina and Barrientos, Ricardo and Duda, Maria and Bonfanti, Alessandra and Yonis, Amina and Harris, Andrew and Mosaffa, Payman and Fujita, Yasuyuki and others},
  journal={Nature physics},
  volume={15},
  number={8},
  pages={839--847},
  year={2019},
  publisher={Nature Publishing Group UK London},
url={https://www.nature.com/articles/s41567-019-0516-6}
}

@book{lim2014cell,
  title={Cell signaling},
  author={Lim, Wendell and Mayer, Bruce and Pawson, Tony},
  year={2014},
  publisher={Garland Science},
url={https://www.taylorfrancis.com/books/mono/10.1201/9780429258893/cell-signaling-wendell-lim-tony-pawson-bruce-mayer}
}

@article{wurthner2023geometry,
  title={Geometry-induced patterns through mechanochemical coupling},
  author={W{\"u}rthner, Laeschkir and Goychuk, Andriy and Frey, Erwin},
  journal={Physical Review E},
  volume={108},
  number={1},
  pages={014404},
  year={2023},
  publisher={APS},
url={https://journals.aps.org/pre/abstract/10.1103/PhysRevE.108.014404}
}

@article{banerjee2024hydrodynamics,
  title={Hydrodynamics of pulsating active liquids},
  author={Banerjee, Tirthankar and Desaleux, Thibault and Ranft, Jonas and Fodor, {\'E}tienne},
  journal={arXiv preprint arXiv:2407.19955},
  year={2024},
url={https://arxiv.org/html/2407.19955v1}
}

@article{kaouri2017mathematical,
  title={Mathematical modelling of calcium signalling taking into account mechanical effects},
  author={Kaouri, Katerina and Maini, Philip K and Chapman, S Jonathan},
  journal={arXiv preprint arXiv:1703.00540},
  year={2017},
url={https://arxiv.org/abs/1703.00540}
}

@article{gross2017active,
  title={How active mechanics and regulatory biochemistry combine to form patterns in development},
  author={Gross, Peter and Kumar, K Vijay and Grill, Stephan W},
  journal={Annual review of biophysics},
  volume={46},
  number={1},
  pages={337--356},
  year={2017},
  publisher={Annual Reviews},
url={https://pubmed.ncbi.nlm.nih.gov/28532214/}
}

@article{sartori2019effect,
  title={Effect of curvature and normal forces on motor regulation of cilia},
  author={Sartori, Pablo},
  journal={arXiv preprint arXiv:1905.04138},
  year={2019},
url={https://arxiv.org/abs/1905.04138}
}

@article{rautu2024active,
  title={Active morphodynamics of intracellular organelles in the trafficking pathway},
  author={Rautu, S Alex and Morris, Richard G and Rao, Madan},
  journal={arXiv preprint arXiv:2409.19084},
  year={2024},
url={https://arxiv.org/html/2409.19084v1}
}

@article{dhanuka2025excitability,
  title={Excitability and travelling waves in renewable active matter},
  author={Dhanuka, Ankit and Banerjee, Deb Sankar and Rao, Madan and others},
  journal={arXiv preprint arXiv:2503.19687},
  year={2025},
url={https://arxiv.org/abs/2503.19687}
}

@article{zhai2004amplitude,
  title={Amplitude death through a Hopf bifurcation in coupled electrochemical oscillators: Experiments and simulations},
  author={Zhai, Yumei and Kiss, Istv{\'a}n Z and Hudson, John L},
  journal={Physical Review E},
  volume={69},
  number={2},
  pages={026208},
  year={2004},
  publisher={APS},
url={https://journals.aps.org/pre/abstract/10.1103/PhysRevE.69.026208}
}

@article{tompkins2014testing,
  title={Testing Turing’s theory of morphogenesis in chemical cells},
  author={Tompkins, Nathan and Li, Ning and Girabawe, Camille and Heymann, Michael and Ermentrout, G Bard and Epstein, Irving R and Fraden, Seth},
  journal={Proceedings of the National Academy of Sciences},
  volume={111},
  number={12},
  pages={4397--4402},
  year={2014},
  publisher={National Academy of Sciences},
url={https://www.pnas.org/doi/10.1073/pnas.1322005111#:~:text=We%20quantitatively%20test%20Turing's%20ideas,differentiation%2C%20which%20drives%20physical%20morphogenesis}
}

@book{epstein1998introduction,
  title={An introduction to nonlinear chemical dynamics: oscillations, waves, patterns, and chaos},
  author={Epstein, Irving R and Pojman, John A},
  year={1998},
  publisher={Oxford university press},
url={https://academic.oup.com/book/40727}
}

@article{das2024order,
  title={From Order to Chimeras: Unraveling Dynamic Patterns in Active Fluids with Nonlinear Growth},
  author={Das, Joydeep and Chaudhuri, Abhishek and Sinha, Sudeshna},
  journal={arXiv preprint arXiv:2412.14729},
  year={2024},
url={https://arxiv.org/pdf/2412.14729}
}

@article{abrams2004chimera,
  title={Chimera states for coupled oscillators},
  author={Abrams, Daniel M and Strogatz, Steven H},
  journal={Physical review letters},
  volume={93},
  number={17},
  pages={174102},
  year={2004},
  publisher={APS},
url={https://journals.aps.org/prl/abstract/10.1103/PhysRevLett.93.174102}
}

@article{kuramoto2002coexistence,
  title={Coexistence of coherence and incoherence in nonlocally coupled phase oscillators},
  author={Kuramoto, Yoshiki and Battogtokh, Dorjsuren},
  journal={arXiv preprint cond-mat/0210694},
  year={2002},
url={https://arxiv.org/abs/cond-mat/0210694}
}

@article{arora2024shape,
  title={A shape-driven reentrant jamming transition in confluent monolayers of synthetic cell-mimics},
  author={Arora, Pragya and Sadhukhan, Souvik and Nandi, Saroj Kumar and Bi, Dapeng and Sood, AK and Ganapathy, Rajesh},
  journal={Nature Communications},
  volume={15},
  number={1},
  pages={5645},
  year={2024},
  publisher={Nature Publishing Group UK London},
url = {https://pmc.ncbi.nlm.nih.gov/articles/PMC11226658/}
}

@article{baconnier2022selective,
  title={Selective and collective actuation in active solids},
  author={Baconnier, Paul and Shohat, Dor and L{\'o}pez, C Hern{\'a}ndez and Coulais, Corentin and D{\'e}mery, Vincent and D{\"u}ring, Gustavo and Dauchot, Olivier},
  journal={Nature Physics},
  volume={18},
  number={10},
  pages={1234--1239},
  year={2022},
  publisher={Nature Publishing Group UK London},
url = {https://pmc.ncbi.nlm.nih.gov/articles/PMC11226658/}
}

@article{kumar2024emergent,
  title={Emergent dynamics due to chemo-hydrodynamic self-interactions in active polymers},
  author={Kumar, Manoj and Murali, Aniruddh and Subramaniam, Arvin Gopal and Singh, Rajesh and Thutupalli, Shashi},
  journal={Nature Communications},
  volume={15},
  number={1},
  pages={4903},
  year={2024},
  publisher={Nature Publishing Group UK London},
url = {https://pubmed.ncbi.nlm.nih.gov/38851777/}
}

@book{davidson2015turbulence,
  title={Turbulence: an introduction for scientists and engineers},
  author={Davidson, Peter},
  year={2015},
  publisher={Oxford university press},
url ={https://www.google.co.in/books/edition/Turbulence/VblDCQAAQBAJ?hl=en&gbpv=0}
}

@misc{international_centre_for_theoretical_sciences_dynamical_2019,
	title = {Dynamical phase transitions in {Markov} processes by {Hugo} {Touchette}},
	url = {https://www.youtube.com/watch?v=_rBzf3JqBpc},
	urldate = {2025-04-20},
	author = {{International Centre for Theoretical Sciences}},
	month = jul,
	year = {2019},
}

@article{nyawo2017minimal,
  title={A minimal model of dynamical phase transition},
  author={Nyawo, Pelerine Tsobgni and Touchette, Hugo},
  journal={Europhysics Letters},
  volume={116},
  number={5},
  pages={50009},
  year={2017},
  publisher={IOP Publishing},
url={https://iopscience.iop.org/article/10.1209/0295-5075/116/50009}
}

@article{wechselberger2007canards,
  title={Canards},
  author={Wechselberger, Martin},
  journal={Scholarpedia},
  volume={2},
  number={4},
  pages={1356},
  year={2007}
}

@article{mora2011biological,
  title={Are biological systems poised at criticality?},
  author={Mora, Thierry and Bialek, William},
  journal={Journal of Statistical Physics},
  volume={144},
  pages={268--302},
  year={2011},
  publisher={Springer},
url = {https://link.springer.com/article/10.1007/s10955-011-0229-4}
}

@article{munoz2018colloquium,
  title={Colloquium: Criticality and dynamical scaling in living systems},
  author={Munoz, Miguel A},
  journal={Reviews of Modern Physics},
  volume={90},
  number={3},
  pages={031001},
  year={2018},
  publisher={APS},
url = {https://journals.aps.org/rmp/abstract/10.1103/RevModPhys.90.031001}
}

@article{roshan2023multiscale,
  title={Multiscale studies of delayed afterdepolarizations I: A comparison of two biophysically realistic mathematical models for human ventricular myocytes},
  author={Roshan, Navneet and Pandit, Rahul},
  journal={arXiv preprint arXiv:2307.10134},
  year={2023},
  url ={https://arxiv.org/abs/2307.10134}
}

@article{andrenvsek2023wrinkling,
  title={Wrinkling instability in unsupported epithelial sheets},
  author={Andren{\v{s}}ek, Ur{\v{s}}ka and Ziherl, Primo{\v{z}} and Krajnc, Matej},
  journal={Physical Review Letters},
  volume={130},
  number={19},
  pages={198401},
  year={2023},
  publisher={APS},
  url={https://journals.aps.org/prl/abstract/10.1103/PhysRevLett.130.198401}
}

@article{guru2022microtubule,
  title={The microtubule end-binding proteins EB1 and Patronin modulate the spatiotemporal dynamics of myosin and pattern pulsed apical constriction},
  author={Guru, Anwesha and Saravanan, Surat and Sharma, Deepanshu and Narasimha, Maithreyi},
  journal={Development},
  volume={149},
  number={22},
  pages={dev199759},
  year={2022},
  publisher={The Company of Biologists Ltd},
  url={https://pubmed.ncbi.nlm.nih.gov/36440630/}
}

@article {muthukrishnan2025glassy,
	author = {Muthukrishnan, Sindhu and Dewan, Phanindra and Tejaswi, Tanishq and Sebastian, Michelle B and Chhabra, Tanya and Mondal, Soumyadeep and Kolya, Soumitra and Sarkar, Sumantra and Vishwakarma, Medhavi},
	title = {Glassy dynamics in active epithelia emerge from an interplay of mechanochemical feedback and crowding},
	elocation-id = {2025.11.08.687351},
	year = {2025},
	doi = {10.1101/2025.11.08.687351},
	publisher = {Cold Spring Harbor Laboratory},
	URL = {https://www.biorxiv.org/content/early/2025/11/11/2025.11.08.687351},
    journal = {bioRxiv}
}

@article{cross1993pattern,
  title={Pattern formation outside of equilibrium},
  author={Cross, Mark C and Hohenberg, Pierre C},
  journal={Reviews of modern physics},
  volume={65},
  number={3},
  pages={851},
  year={1993},
  publisher={APS},
  url="https://journals.aps.org/rmp/abstract/10.1103/RevModPhys.65.851"
}

@article{Sethia2008,
  title = {Clustered Chimera States in Delay-Coupled Oscillator Systems},
  volume = {100},
  ISSN = {1079-7114},
  DOI = {10.1103/physrevlett.100.144102},
  number = {14},
  journal = {Physical Review Letters},
  publisher = {American Physical Society (APS)},
  author = {Sethia,  Gautam C. and Sen,  Abhijit and Atay,  Fatihcan M.},
  year = {2008},
  month = apr, 
  url = {http://dx.doi.org/10.1103/PhysRevLett.100.144102}
}

@article{synchronizationArkady2001,
author="Arkady, Pikovsky and Michael, Rosenblum and Jürgen, Kurths",
title="Synchronization",
journal="Cambridge University Press",
publisher="Cambridge University Press",
year="2001",
month="10",
volume="12",
DOI="10.1017/cbo9780511755743",
URL="https://cir.nii.ac.jp/crid/1361981469325365376"
}

@article{serra2012mechanical,
  title={Mechanical waves during tissue expansion},
  author={Serra-Picamal, Xavier and Conte, Vito and Vincent, Romaric and Anon, Ester and Tambe, Dhananjay T and Bazellieres, Elsa and Butler, James P and Fredberg, Jeffrey J and Trepat, Xavier},
  journal={Nature Physics},
  volume={8},
  number={8},
  pages={628--634},
  year={2012},
  publisher={Nature Publishing Group UK London},
  url="https://www.nature.com/articles/nphys2355"
}

@article{Ipsen2000,
  author  = {Ipsen, M. and Hynne, F. and S{\o}rensen, P. G.},
  title   = {Amplitude equations for reaction--diffusion systems with a {H}opf bifurcation and slow real modes},
  journal = {Physica D: Nonlinear Phenomena},
  volume  = {136},
  number  = {1--2},
  pages   = {66--92},
  year    = {2000},
  month   = feb,
  doi     = {10.1016/S0167-2789(99)00149-9},
  url     = {https://doi.org/10.1016/S0167-2789(99)00149-9}
}

@article{Ipsen1999,
  author  = {Ipsen, M. and Hynne, F. and Sorensen, P. G.},
  title   = {Systematic Derivation of Amplitude Equations and Normal Forms for Dynamical Systems},
  journal = {Chaos},
  volume  = {8},
  number  = {4},
  pages   = {834--852},
  year    = {1998},
  month   = dec,
  doi     = {10.1063/1.166370},
  pmid    = {12779791},
  url     = {https://pubmed.ncbi.nlm.nih.gov/12779791/}
}

@article{PhysRevE.66.026102,
  title = {Amplitude equations for systems with long-range interactions},
  author = {Kassner, Klaus and Misbah, Chaouqi},
  journal = {Phys. Rev. E},
  volume = {66},
  issue = {2},
  pages = {026102},
  numpages = {9},
  year = {2002},
  month = {Aug},
  publisher = {American Physical Society},
  doi = {10.1103/PhysRevE.66.026102},
  url = {https://link.aps.org/doi/10.1103/PhysRevE.66.026102}
}

@book{Carr1981,
  author    = {Carr, J.},
  title     = {Applications of Centre Manifold Theory},
  publisher = {Springer},
  address   = {New York},
  year      = {1981}
}

@book{Guckenheimer1983,
  author    = {Guckenheimer, J. and Holmes, P.},
  title     = {Nonlinear Oscillations, Dynamical Systems, and
               Bifurcations of Vector Fields},
  publisher = {Springer},
  address   = {New York},
  year      = {1983}
}

@article{bailles2022mechanochemical,
  title={Mechanochemical principles of spatial and temporal patterns in cells and tissues},
  author={Bailles, Ana{\"\i}s and Gehrels, Emily W and Lecuit, Thomas},
  journal={Annual review of cell and developmental biology},
  volume={38},
  number={1},
  pages={321--347},
  year={2022},
  publisher={Annual Reviews},
  URL = "https://www.annualreviews.org/content/journals/10.1146/annurev-cellbio-120420-095337"
}

@article{turing1952chemical,
  author  = {Turing, Alan Mathison},
  title   = {The Chemical Basis of Morphogenesis},
  journal = {Philosophical Transactions of the Royal Society of London. Series B, Biological Sciences},
  volume  = {237},
  number  = {641},
  pages   = {37--72},
  year    = {1952},
  publisher = {Royal Society},
  doi     = {10.1098/rstb.1952.0012},
  url     = {https://royalsocietypublishing.org}
}

@incollection{kuramoto1984chemical,
  title={Chemical turbulence},
  author={Kuramoto, Yoshiki},
  booktitle={Chemical oscillations, waves, and turbulence},
  pages={111--140},
  year={1984},
  publisher={Springer}, 
  url={https://link.springer.com/chapter/10.1007/978-3-642-69689-3_7}
}

@article{lin2017activation,
  title={Activation and synchronization of the oscillatory morphodynamics in multicellular monolayer},
  author={Lin, Shao-Zhen and Li, Bo and Lan, Ganhui and Feng, Xi-Qiao},
  journal={Proceedings of the National Academy of Sciences},
  volume={114},
  number={31},
  pages={8157--8162},
  year={2017},
  publisher={National Academy of Sciences},
  url = "https://www.pnas.org/doi/full/10.1073/pnas.1705492114"
}

@Inbook{Sherwood2013,
author="Sherwood, William Erik",
editor="Jaeger, Dieter
and Jung, Ranu",
title="FitzHugh--Nagumo Model",
bookTitle="Encyclopedia of Computational Neuroscience",
year="2013",
publisher="Springer New York",
address="New York, NY",
pages="1--11",
isbn="978-1-4614-7320-6",
doi="10.1007/978-1-4614-7320-6_147-1",
url="https://doi.org/10.1007/978-1-4614-7320-6_147-1"
}

@misc{JanakiRangarajan,
	author = {T. M. Janaki and Govindan Rangarajan},
	title = {Lyapunov Exponents for Continuous-Time Dynamical Systems},
	howpublished = {\url{https://math.iisc.ac.in/~rangaraj/wp-content/uploads/2020/07/jiisc\_lyap.pdf}},
	year = {},
}

@article{Senthil2015,
  title = {Symmetry-Protected Topological Phases of Quantum Matter},
  volume = {6},
  ISSN = {1947-5462},
  url = {http://dx.doi.org/10.1146/annurev-conmatphys-031214-014740},
  DOI = {10.1146/annurev-conmatphys-031214-014740},
  number = {1},
  journal = {Annual Review of Condensed Matter Physics},
  publisher = {Annual Reviews},
  author = {Senthil,  T.},
  year = {2015},
  month = Mar,
  pages = {299–324}
}

@misc{SI,
  note = {See Supplemental Material at [URL will be inserted by publisher] for additional figures, numerical methods, and supporting calculations.}
}
\end{document}